\documentclass[twocolumn,preprintnumbers,aps,prd,superscriptaddress,footinbib,tightenlines, hidelinks]{revtex4-1}

\usepackage{mathtools}
\usepackage{booktabs}
\usepackage[english]{babel}
\usepackage{amsmath,amssymb,amsbsy,amstext, amsthm, simplewick, amsfonts}
\usepackage{hyperref}
\usepackage{graphicx}
\usepackage{siunitx}
\usepackage{upgreek}
\usepackage{framed}
\usepackage{wrapfig}
\usepackage{multirow}
\usepackage{bbm}
\usepackage[svgnames,dvipsnames,x11names]{xcolor}
\usepackage[utf8x]{inputenc}
\usepackage{selinput}
\usepackage{bm}
\usepackage{float}
\usepackage{geometry}
\usepackage{yfonts}
\usepackage{sidecap}
\usepackage{anyfontsize}
\usepackage{tcolorbox}
\usepackage{bbm}
\usepackage{placeins}
\usepackage{setspace}

\usepackage{tikz}
\tikzset{every picture/.style={line width=0.75pt}} 
\usepackage{tikz-cd, tikz-feynman} 
\tikzfeynmanset{warn luatex=false}

\definecolor{blue3}{RGB}{31, 119, 180}
\definecolor{red3}{RGB}{	214, 39, 40}
\definecolor{orange3}{RGB}{255, 127, 14}
\definecolor{green3}{RGB}{44, 160, 44}

\newcommand{\bfx}{\mathbf{x}}
\newcommand{\bfk}{\mathbf{k}}

\begin{document}
\raggedbottom

\title{Propagator positivity bounds for cosmological correlators}

\author{Mang Hei Gordon Lee}
\affiliation{Leung Center for Cosmology and Particle Astrophysics, National Taiwan University,
Taipei 10617, Taiwan}

\author{Scott Melville}
\affiliation{Department of Physics and Astronomy, \\ Queen Mary University of London, Mile End Road, London E1 4NS, United Kingdom}


\begin{abstract}
\noindent
Using unitarity and causality, we derive an infinite tower of two-sided positivity bounds on the effective field theory coefficients which describe the propagation of heavy fields on de Sitter spacetime.
We design this EFT to describe propagators with the in-in 
boundary conditions that are relevant for cosmological correlators.
Our positivity bounds therefore identify EFT correlators
that can never emerge from a consistent underlying model of inflation.  
This implies non-trivial constraints on primordial non-Gaussianity; for instance the cosmological collider oscillations in the squeezed bispectrum from the exchange of a heavy scalar are tied to the shape of its EFT background.
\end{abstract}

\maketitle

\section{Introduction}

\noindent What are the possible effective field theories in de Sitter spacetime?
EFTs have played a central role in how we understand quantum field theory on Minkowski spacetime and how we search for new physics in particle colliders \cite{Isidori:2023pyp}. 
An analogous framework for de Sitter---the maximally symmetric spacetime most relevant for cosmology---would similarly shed light on QFT in curved spacetimes and improve the analysis of cosmological collider signals from the early inflationary Universe \cite{Arkani-Hamed:2015bza, Baumann:2022jpr}.


In particular, alongside the comoving curvature perturbations responsible for the Cosmic Microwave Background and Large Scale Structure, the early inflationary Universe is expected to contain a number of massive modes (e.g. from degrees of freedom beyond the inflaton). 
Properties of these modes---such as their mass, spin and interaction with the curvature perturbations---are encoded as characteristic features in the primordial non-Gaussianities produced by inflation. 
At their most extreme, these signals are within reach of current and upcoming surveys, e.g. DESI, EUCLID, LSST, SPHEREx and Simons Observatory \cite{Achucarro:2022qrl}. 
The most promising signal---the scalar bispectrum---can be produced by a massive mode propagating from one curvature perturbation to another two. 
An EFT description of how massive modes propagate on quasi-de Sitter spacetimes, particularly when they interact strongly with other degrees of freedom, would therefore allow these sky surveys to analyse the inflationary particle spectrum in the same spirit as conventional collider experiments.

From a purely bottom-up perspective, to construct an EFT on dS one can follow the same procedure as on Minkowski: first identify the relevant degrees of freedom and a small power counting parameter that can be used to estimate their size; then construct all local operators compatible with the spacetime isometries up to a fixed order in this parameter; and finally build an effective action by summing these operators, each multiplied by an EFT coefficient. These EFT coefficients are undetermined \emph{a priori}, and must be fixed by matching to data or to a known UV completion of the theory. 

While pragmatically useful, this 
approach does not address several conceptual issues with de Sitter. 
It assumes a Wilsonian decoupling of the heavy degrees of freedom at low energies, which is spoiled by particle production. 
It is also not yet clear what is the correct power counting parameter for de Sitter, since ``energy'' is no longer a good quantum number.
And it treats the EFT coefficients as free parameters, when there is no guarantee that all values can arise from a consistent UV completion.  
In this paper, we address all three of these issues by building a consistent EFT for scalar fields on de Sitter and bounding its coefficients so 
that a healthy underlying UV theory can exist.

This is the latest in a series of attempts to carve out the space of consistent EFTs in cosmology. 
The earliest attempts used unitarity of $2 \to 2$ scattering on sufficiently subhorizon scales that an energy-conserving $S$-matrix can be defined \cite{Baumann:2015nta, Grall:2020tqc, Grall:2021xxm}, however the resulting positivity bounds require stronger UV assumptions about analyticity which can be violated in causal theories \cite{Aoki:2021ffc, Hui:2023pxc, Creminelli:2023kze}. 
More recent work has developed Kramers-Kronig dispersion relations 
for two-point functions with broken Lorentz invariance \cite{Creminelli:2022onn, Serra:2024tmz, Hui:2025aja, Creminelli:2025rxj}: these have the advantage that they apply to a wide range of systems---including condensed matter \cite{Heller:2022ejw, Heller:2023jtd, Creminelli:2024lhd}---however they do not exploit the (approximate) de Sitter symmetries which characterise the early Universe. 
Another strategy has been to use causality classically by imposing positivity of scattering time delays on cosmological spacetime backgrounds \cite{deRham:2020zyh,Bittermann:2022hhy, CarrilloGonzalez:2023emp, CarrilloGonzalez:2025fqq, McLoughlin:2025shj} 
(see also \cite{deRham:2019ctd, Chen:2021bvg, deRham:2021bll, Bellazzini:2021shn, CarrilloGonzalez:2022fwg, CarrilloGonzalez:2023cbf, Chen:2023rar, Melville:2024zjq} for asymptotically flat spacetimes), though on Minkowski
these causality constraints often differ from the positivity bounds of unitarty/microcausality \cite{CarrilloGonzalez:2023cbf}.
The most recent dS positivity bounds have used unitarity and causality in the form of a Kallen-Lehmann spectral decomposition \cite{Hogervorst:2021uvp} 
to bound anomalous dimensions \cite{Green:2023ids, Chakraborty:2025mhh}
and to numerically explore the space of allowed bispectra \cite{deRham:2025mjh}.
This paper uses a similar implementation of unitarity and causality to produce the first analytic bounds on EFT coefficients using relativistic QFT in de Sitter.

\begin{figure}[t]
\begin{flushleft}
\includegraphics[width=0.98\columnwidth]{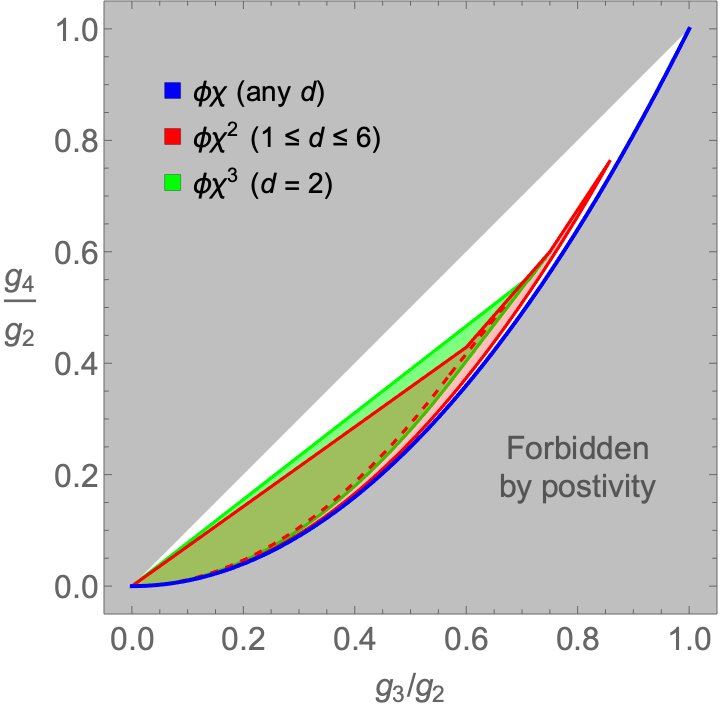}
\end{flushleft}
\caption{
Our positivity bounds \eqref{eqn:pos} forbid EFT coefficients in the grey region, for which there is no unitary/causal UV completion. 
We verify that a number of explicit UV completions---the interactions $\phi \chi$, $\phi \chi^2$ and $\phi \chi^3$ with a heavy field $\chi$ in different spatial dimensions $d$---produce EFTs that lie in the allowed region. The largest region from the $\phi \chi^2$ interaction occurs in $d=1$, and we show with a dashed red line the $d=3$ region for comparison.
}
\label{fig:main}
\end{figure}

Our main result is a tower of two-sided positivity bounds [see \eqref{eqn:pos}], which must be obeyed by cosmological correlators at every order in the EFT expansion if they were produced by unitarity/causal physics.
We demonstrate that these are satisfied in several explicit UV completions, summarised in Fig.~\ref{fig:main}. 
For instance, the first EFT interaction beyond mass and kinetic counterterms, schematically $c_2 \, \partial^4 \phi^2$, must obey $c_2 > 0$ on Minkowski.
On de Sitter, we show that $c_2$ is no longer always positive and we establish a new bound:
\begin{align}
 c_2   >  \int^\Lambda_{-\Lambda} \frac{d \mu}{\pi \mu} \frac{ \text{Im} \, \Sigma ( \mu ) }{ \mu^{4} } 
 \label{eqn:intro_c2}
\end{align}  
where $\Lambda$ is the EFT cut-off and $\text{Im} \, \Sigma$ is the decay width of $\phi$ due to the spacetime background. 
We are able to similarly map the infinite tower of EFT constraints that exist on Minkowski---the analogue of the EFThedron for two-point interactions---onto new bounds for dS EFTs at every order. 
Just like \eqref{eqn:intro_c2}, these bounds mix the conservative and dissipative parts of the EFT. 
When applied to observable signals from inflation---e.g. the squeezed limit of the scalar bispectrum---they therefore bound the decay of the cosmological collider oscillations in terms of the non-oscillating EFT background.

An important feature of the EFT that we introduce below is that it contains---in addition to the usual $c_n$ coefficients of Minkowski---another series of coefficients $\gamma_n'$ that capture the low-energy decay due to the de Sitter spacetime background:
\begin{align}
 \text{Im} \, \Sigma (\nu ) =  \pi  \nu \left( \gamma_0'  + \gamma_1' \; \nu^{2} + ... \right)  \; . 
 \label{eqn:ImSigma_from_gamma}
\end{align}
In practice, positivity on de Sitter [e.g., \eqref{eqn:intro_c2}] constrains $c_n$ in terms of $\gamma_n'$. 
A useful way to visualise these bounds is to define modified $c_n$ coefficients that include dissipative corrections. For instance 
\begin{align}
g_2  = \Lambda^{4} c_2  +    \tfrac{ 2 }{3}  \Lambda  \gamma'_0    +    2 \Lambda^{3}  \gamma'_1  + ...    
\label{eqn:g2_from_c}
\end{align}
is defined so that positivity \eqref{eqn:intro_c2} implies $g_2 > 0$.
Analogous $g_{n > 2}$ are defined in \eqref{eqn:g_from_c} and obey 
the tower of two-sided and non-linear bounds recently discovered on Minkowski (see Fig.~\ref{fig:main}). 

The relative size of the dissipative corrections depends on the background Hubble rate, which can be viewed as setting the ``temperature'' of the environment. 
When the new degrees of freedom at the EFT cut-off $\Lambda$ are ``cold,'' i.e., $\Lambda \gg H$, then the $\gamma_n'$ coefficients are exponentially suppressed and $g_n \approx c_n$ (the EFT matching is not affected by the de Sitter background). 
When the new degrees of freedom are ``hot'', i.e. $\Lambda \ll H$, then $\gamma_n' \sim \mathcal{O} (1)$ and $g_n$ can differ significantly from $c_n$ (dissipation affects the EFT matching).
Below we adopt units in which $H=1$, and indeed find 
\begin{align}
\Lambda^4 c_2 > \begin{cases}
0 + \mathcal{O} \left(  e^{- \pi \Lambda} \right)    &\text{when} \;\;  \Lambda \gg 1 \, ,   \\ 
 - \frac{2}{3} \Lambda \gamma_0' + \mathcal{O} \left(   \Lambda^3 \right)     &\text{when} \;\;  \Lambda \ll 1 \, .  
 \end{cases}
\end{align}
In the intermediate regime $\Lambda \sim H$, the first $\mathcal{O} \left( \Lambda/H \right)$ terms in \eqref{eqn:ImSigma_from_gamma} can be important and appear in the positivity bounds (see Fig.~\ref{fig:gamma}).

Our results are an important proof of principle---that consistent EFTs can be constructed and bounded despite particle production in de Sitter.
Phenomenologically, they are relevant for the cosmological collider signal from the exchange of heavy fields during inflation. 
Theoretically, they reveal a positive geometry analogous to the EFThedron on Minkowski \cite{Arkani-Hamed:2020blm, Chiang:2021ziz}, advancing the ambitious programme of elevating these mathematical structures 
to defining features of a QFT: viewing the positive geometry as the fundamental object from which unitarity/causality emerge \cite{Arkani-Hamed:2013jha,Arkani-Hamed:2017fdk,Arkani-Hamed:2024jbp}.

We organise the paper as follows. 
In Sec.~\ref{sec:prop} we introduce our observable of interest: the time-ordered two-point function with in-in boundary conditions relevant for cosmology. 
In Sec.~\ref{sec:EFT} we describe the EFT that captures its low-energy limit, and in Sec.~\ref{sec:pos} derive positivity bounds on the EFT coefficients. 
In Sec.~\ref{sec:uv} we give various concrete UV completions to demonstrate that our bounds are indeed always satisfied. 
In Appendix
we discuss phenomenological consequences of these bounds. 

Throughout we consider a fixed de Sitter spacetime, $ds^2 = (-\tau)^{-2} \left( -d \tau^2 + d \bfx^2 \right)$, in $d$ spatial dimensions and work in units where the Hubble rate $H=1$. For a real scalar field $\phi$ of mass $m$, we split the action $S = S_0 + S_{\rm int}$, where the free action is
\begin{align}
S_{0}  = \int d\tau \, d^d \bfx \; \sqrt{-g} \; \tfrac{1}{2} \left( - (  \partial_\mu \phi )^2  - m^2 \phi^2 \right) \; .
\label{eqn:Sfree}
\end{align}
We consider heavy fields to be in the \emph{principal series}, which have sufficiently large masses so that the parameter $\mu = \sqrt{  m^2 - d^2 / 4 }$ is real. 
The free equation of motion 
can be written in momentum space as 
$\Box  \phi_{\bfk} ( \tau) =  \mu^2 \phi_{\bfk} ( \tau ) $, where
\begin{align}
\Box \phi_{\bfk}  =  \tau^{d/2} \left( ( \tau \partial_\tau )^2 - k^2 \tau^2   \right)  \tau^{-d/2} \phi_{\bfk}  .
\end{align}

\section{In-in propagators}
\label{sec:prop}

\noindent The time-ordered two-point function plays a central role in defining (and calculating with) any QFT. 
It represents propagation, and therefore is constrained by the causal structure of the spacetime. It also encodes the overlap between one-particle states, and therefore is constrained by unitarity. In this section, we review the time-ordered two-point function on de Sitter with the in-in (Bunch-Davies) boundary conditions relevant for cosmology.

Recall that, on Minkowski, the time-ordered two-point function can be written as
\begin{align}
G ( p ) = \frac{-i}{m^2 - p^2 - \Sigma ( p ) }
\label{eqn:self-energy_form_Mink}
\end{align}
where $\Sigma$ is the self-energy. The free theory propagator corresponds to $\Sigma = i \epsilon$ with $\epsilon \to 0$ from above, shifting the pole into the complex plane to encode time-ordered boundary conditions. In fact, unitarity requires $\text{Im} \, \Sigma > 0$ in any interacting theory.

On de Sitter, what is the analogue of \eqref{eqn:self-energy_form_Mink}? 
There are two important differences with Minkowski. 
First, the boundary conditions are different: since the Hamiltonian depends on time, there is now a different ground state on each time-slice and so we must specify which of these vacua are used to compute the two-point function. 
Here we focus on the Bunch-Davies vacuum of the far past, as this is the correlator that underpins cosmological collider signals from the early inflationary Universe. 
The resulting ``in-in'' two-point function requires both bra- and ket- boundary conditions to be specified in the far past (while the ``in-out'' two-point function on Minkowski can be computed using one vacuum in the far past and one in the far future). 
The second difference on de Sitter is the kinematic variables. The energy which appears in $p_\mu$ (the Fourier conjugate of time) is no longer a good quantum number. Instead, the analogue of $p^2$ on Minkowski is the de Sitter Casimir $\nu^2$, defined as the eigenvalue of the free equation of motion $\Box f_\nu = \nu^2 f_\nu$. We transform to this variable using 
\begin{align}
\frac{ \langle  T \phi_{\bfk} ( \tau ) \phi_{\bfk'} ( \tau' )   \rangle'}{ ( -\tau )^{d/2} (- \tau' )^{d/2}} = \int_{-\infty}^{\infty} d \nu \, N_\nu \,  f^+_\nu ( k \tau ) f^+_\nu ( k \tau' ) \,  G ( \nu )
\label{eqn:Gn_def}
\end{align}
where $N_\nu = \nu \sinh ( \pi \nu ) /\pi$ and $f^+_\nu ( k \tau )$ is the Hankel mode function of the free theory: the analogue of $e^{+i \omega t}$ on Minkowski \footnote{
Explicitly, our mode functions are 
\begin{align}
 f_\mu^+ ( k \tau ) &= \frac{\sqrt{\pi}}{2 i} e^{+ \pi \mu/2} H_{i \mu}^{(2)} ( - k \tau) \; ,  \\ 
 f^-_\mu ( k \tau) &=\frac{i \sqrt{\pi}}{2} e^{- \pi \mu/2} H_{i \mu}^{(1)} ( - k \tau)  \; . 
\end{align} 
and obey $f^+_\mu = (f^-_\mu )^*$. 
}. 
The prime on the correlator indicates that the universal factor of $(2\pi)^d \delta^d \left( \bfk + \bfk' \right)$ has been removed. 
%
This Bunch-Davies propagator takes the form (see e.g. \cite{DiPietro:2021sjt, Melville:2024ove})
\begin{align}
 G (\nu )  &= \frac{-i}{2 \sinh ( \pi \nu)}   \label{eqn:self-energy_form} \\ 
 &\times \left[  \frac{   e^{+ \pi \nu} }{ \mu^2 - \nu^2 - \Sigma ( \nu  )  }  -  
 \frac{ e^{- \pi \nu} }{ \mu^2 - \nu^2 - \Sigma ( - \nu  )  } 
 \right] \; .  \nonumber 
\end{align}

In the free theory, $\Sigma ( \pm \nu) = \pm i \epsilon$ with $\epsilon \to 0$ from above. 
In a general interacting theory, the self-energy obeys $\Sigma ( -\nu ) = \Sigma^* (\nu )$ and unitarity will requires $\text{Im} \, \Sigma (\nu) > 0$ for all $\nu > 0$. 
This means that the second term has poles shifted in the opposite direction: physically, this reflects the particle production that takes place on de Sitter \cite{Melville:2024ove}. 
In the flat space limit---which corresponds to taking $H\to 0$ with $\nu^2 \to p^2/H^2$ and $\mu^2 \to m^2/H^2$---the particle production term turns off exponentially and we recover the usual Minkowski propagator \eqref{eqn:self-energy_form_Mink}.

Since it is the positivity of $\Sigma$ that will lead to our 
bounds on EFT coefficients, we will now give a more detailed account of what observable this self-energy corresponds to. 
On Minkowski, it is the time-ordered two-point function of the source, $\delta S_{\rm int} / \delta \phi$. 
On de Sitter, the connection between $\Sigma$ and $J_{\bfk} ( \tau ) = \delta S_{\rm int}/\delta \phi_{\bfk} ( \tau)$ is more subtle. 
To compute the in-in two-point function, we must path integrate from the far past to the far future and back again, along the two branches of the Schwinger-Keldysh contour. 
This leads to self-energy insertions with all possible operator orderings. To capture this, we define four self-energies: 
\begin{align}
  \Sigma^{++}_k ( \tau , \tau' )  &= i \,  \langle T J_{\bfk} ( \tau )  J_{\bfk'} ( \tau' ) \rangle' \; ,  \nonumber \\ 
 \Sigma^{--}_k ( \tau , \tau' )  &= i \, \langle \bar{T} J_{\bfk} ( \tau )  J_{\bfk'} ( \tau' ) \rangle' \; , \nonumber  \\ 
\Sigma^{+-}_k ( \tau , \tau' ) &= i \, \langle J_{\bfk'} ( \tau' )  J_{\bfk} ( \tau ) \rangle' \; , \nonumber \\ 
\Sigma^{-+}_k ( \tau , \tau' ) &= i \, \langle J_{\bfk} ( \tau )  J_{\bfk'} ( \tau' ) \rangle' 
\label{eqn:JJ}
\end{align} 
where the prime again denotes that an overall $(2 \pi )^3 \delta^3 \left( \bfk + \bfk' \right)$ has been removed and $T$ ($\bar{T}$) denotes (anti-)time-ordering. 
They are related by
\begin{align}
i \, \Sigma^{-\alpha, - \beta}_k &=  \left( i \,  \Sigma_k^{\alpha \beta} \right)^* \; ,   \label{eqn:Sigma_conjugate} \\ 
\Sigma^{++}_k + \Sigma_k^{--} &= \Sigma_k^{+-} + \Sigma_k^{-+}  \; .  \label{eqn:Sigma_sum}
\end{align}
To compare with \eqref{eqn:self-energy_form}, we transform these self-energy functions to the same Casimir basis using
\begin{align}
 \Sigma^{\alpha \beta}_k (  \tau,  \tau' ) = \int_{-\infty}^{+\infty} d \nu \; N_\nu   f_\nu^\alpha ( k\tau) f_\nu^\beta ( k \tau' ) \; \Sigma^{\alpha \beta} (\nu )
\end{align}
where 
both $f^{\pm}_\nu$ mode functions have been used to preserve the conjugate relation \eqref{eqn:Sigma_conjugate}. Notice that $\Sigma^{+-} (\nu) = \Sigma^{-+} (\nu)$ since they are related by permuting $\tau \leftrightarrow \tau'$, and  \eqref{eqn:Sigma_sum} becomes 
\begin{align}
i \, \text{Im} \, \Sigma^{++} (\nu) = \cosh ( \pi \nu ) \Sigma^{+-} (\nu) 
\label{eqn:Sigma_condition}
\end{align} 
thanks to a relation between the $f^{\pm}_\nu$ integrals (see \eqref{eqn:ff_integral}). 
There is therefore only one independent complex function here (rather than the naive four if all $\Sigma^{\alpha \beta}$ were independent).
An explicit calculation using in-in perturbation theory \footnote{
Concretely, consider a diagrammatic expansion of the in-in path integral in which each vertex is labelled by $\pm$ according to its branch of the integration contour. At leading order, this gives $i G  = \Pi^{++}  +  \sum_{\alpha, \beta} \Pi^{+\alpha} \Sigma^{\alpha \beta} \Pi^{\beta +} + ... $,
where $\Pi^{\alpha \beta}$ are the free theory propagators for $\phi$ with different operator orderings. Comparing with the leading order expansion of \eqref{eqn:self-energy_form} leads to \eqref{eqn:Sigma_def}. Some combinatorics then shows that \eqref{eqn:self-energy_form} will match the in-in sum over vertices at every order in $\Sigma$.
} shows that the self-energy $\Sigma$ appearing in \eqref{eqn:self-energy_form} is given by
 \begin{align}
 \Sigma (\nu ) = \frac{ e^{+ \pi \nu} \Sigma^{++} (\nu) - e^{- \pi \nu} \Sigma^{--} (\nu )}{ 2 \cosh ( \pi \nu )  } \; . 
\label{eqn:Sigma_def}
\end{align}
This combination of time- and anti-time-ordered matrix elements again reflects particle production on de Sitter, and in the flat space limit of large $\nu$ we recover $\Sigma \to \Sigma^{++}$. 
It enjoys several nice properties: for instance causality implies that $\Sigma (\nu)$ is analytic in the lower half of the complex $\nu$ plane \footnote{
Here we refer to the causality condition that the analytic structure of $G(\nu)$, which encodes the time-ordering, remains the same in the interacting theory: namely an even combination of $e^{\pm \pi \nu}$ terms which have singularities in the upper/lower half planes respectively.
} and since the imaginary part can be written using \eqref{eqn:Sigma_condition} and \eqref{eqn:ff_integral} as 
\begin{align}
\text{Im} \, \Sigma (\nu ) &=  \sinh ( \pi \nu)  \langle  | J (\nu) |^2  \rangle'   
\end{align}
with $J (\nu ) = \int_{-\infty}^0 d \tau \, f_\nu^+ ( k \tau ) J_\bfk ( \tau )$, we see explicitly that unitarity 
implies positivity of $\text{Im} \, \Sigma$.
$\Sigma$ is also the analytic continuation from the sphere \cite{Marolf:2010zp, Higuchi:2010xt} or from EAdS \cite{Sleight:2020obc, Sleight:2021plv, MdAbhishek:2025dhx} (see Appendix C of \cite{DiPietro:2021sjt}), which is a complementary way to prove these properties.

\section{In-in EFTs}
\label{sec:EFT}

\noindent Having defined a propagator and self-energy on de Sitter, we will now construct an EFT which can describe their ``low-energy'' regime in terms of only the lightest degrees of freedom in the theory. 

On Minkowski, we would do this by expanding the action $S_{\rm EFT} [\phi ] = \int d^4 x \sum_{n=0} \mathcal{L}_n [ \phi ]$ using a two-point operator basis such as
\begin{align}
\mathcal{L}_{n} = \tfrac{1}{2} c_n \; \phi \Box^n \phi \; . 
\label{eqn:Ln_Mink}
\end{align}
This gives $\Sigma_{\rm EFT} (p ) = \sum_n c_n p^{2n}$, which can be matched onto any gapped UV completion by fixing the $c_n$ accordingly: at least up to some desired accuracy (finite order in $n$).

On de Sitter, it is tempting to use the same EFT. However, this would give $\Sigma_{\rm EFT} (\nu) = \sum_n c_n \nu^{2n}$ and such an expansion cannot capture the odd (imaginary) part of $\Sigma$. 
We will see below that this difficulty is due to the lack of a mass gap on de Sitter: heavy fields can affect $\Sigma$ at arbitrarily low $\nu$, so never fully decouple. Put another way: $\text{Im} \, \Sigma$ represents the decay width of $\phi$ particles, and while the decay of a light particle into two or more heavy particles is forbidden on Minkowski by energy conservation, there is no such restriction on de Sitter. 
The EFT must therefore capture the dissipation of $\phi$ through these decay channels.

One resolution is to build an in-in EFT by doubling the number of fields so that $\phi^{\pm}$ denotes the field on each branch of the in-in path integral contour.
We can then add to the free theory the EFT corrections $S_{\rm EFT} [\phi^+, \phi^- ] = \int d^4 x \, \sqrt{-g} \sum_n \mathcal{L}_{n} [ \phi^+ , \phi ^- ]$, where
\begin{align}
\mathcal{L}_n &= \frac{1}{2} ( c_n + i \gamma_n ) \phi^+ \Box^n \phi^+   -  i \kappa_n \phi^+ \Box^n \phi^-   \nonumber \\ 
&-  \frac{1}{2} ( c_n - i \gamma_n ) \phi^- \Box^n \phi^-   
\end{align}
is the generalisation of \eqref{eqn:Ln_Mink} which uses both branches and obeys the reality condition $\mathcal{L}_n [ \phi^- , \phi^+ ] = - \mathcal{L}_n^* [ \phi^+ , \phi^- ]$. 
This doubling of the light degrees of freedom is by now familiar, given recent progress in developing open EFTs for cosmology \cite{Collins:2012nq, Burgess:2015ajz, Hongo:2018ant,Burgess:2022rdo,Burgess:2022nwu, Salcedo:2024smn,Salcedo:2024nex, Salcedo:2025ezu, Colas:2025ind}, though here we focus on only covariant quadratic terms since these capture the de Sitter invariant effects of heavy fields on the propagation of light fields (see also \cite{Salcedo:2022aal, Green:2024cmx, Kawaguchi:2024lsw, DuasoPueyo:2025lmq} for the role played by boundary terms, which we do not include here).
Explicitly, the interactions in $\mathcal{L}_n$ can be viewed as local approximations of each self-energy
\begin{widetext}
\begin{align}
 \Sigma^{++}_{k} (  \tau,  \tau' ) &=   (  c_n  + i  \gamma_n ) \Box^n \delta (  \tau -  \tau' )  \;\; &\Rightarrow& \;\; &\Sigma^{++} (\nu ) &= 
 \left(  c_n + i \gamma_n \right) \nu^{2n}
\nonumber \\ 
 \Sigma^{+-}_k (  \tau,  \tau' ) +  \Sigma^{-+}_k ( \tau,  \tau' )  &=  2 i\,  \kappa_n \Box^n \delta (  \tau -  \tau' )  \;\; &\Rightarrow& \;\; &\Sigma^{+-} (\nu ) &= \frac{i \, \kappa_n \, \nu^{2n} }{ \cosh ( \pi \nu ) }
\end{align}
\end{widetext}
where we have used the integral identities
\begin{align}
 \delta ( z - z' ) &= \int_{-\infty}^{+\infty} d \nu \;  N_\nu \; f_\nu^+( z )  f_\nu^+ ( z' )   \label{eqn:ff_integral} \\
 &= \int_{-\infty}^{+\infty} \frac{d \nu \, N_\nu}{2 \cosh ( \pi \nu )} \left[ 
f^+_\nu ( z ) f^-_\nu ( z' ) + \text{c.c.} 
\right]   \nonumber 
\end{align}
together with $\Box f^{\pm}_\nu ( k \tau ) = \nu^2 f_\nu^{\pm} ( k \tau )$. 
In order to match a consistent UV theory, with $\Sigma^{\alpha \beta}$ given by \eqref{eqn:JJ} and hence obeying \eqref{eqn:Sigma_condition},
we must fix
\begin{align}
\kappa_n = \gamma_n  \; . 
\end{align}
With this condition, the $\Sigma$ given by \eqref{eqn:Sigma_def} is 
\begin{align}
\Sigma_{\rm EFT} (\nu ) = \sum_{n=0} \left( c_n + i \gamma_n \tanh ( \pi \nu)  \right) \nu^{2n} \;  .
\end{align}
This can now be matched to the complex $\Sigma (\nu)$ of any UV completion by fixing the real EFT coefficients $c_n, \gamma_n$ accordingly: at least up to some desired accuracy in the low $\nu$ regime. 
We will abuse terminology and refer to low $\nu$ as ``low-energy'' by analogy with Minkowski, despite the fact that $\nu^2$ represents a de Sitter mass invariant which is unrelated to the Hamiltonian or time translations. 

Finally, it is convenient to reorganise the $\gamma_n$ coefficients so that $\Sigma_{\rm EFT} = \sum_n ( c_n + i \pi \nu \gamma_n'  ) \nu^{2n}$. 
Explicitly, the reorganised coefficients are
\begin{align}
\gamma_n' = \sum_{j=0}^n \; \frac{ (-1)^j \pi^{2j} T_{j+1} }{ (2j + 1)! } \;\gamma_{n-j}
\end{align}
where $T_n$ are the tangent (or zag) numbers: the first few are $\frac{ T_{j+1} }{ (2j+1)! } = \{ 1 , \frac{1}{3} , \frac{2}{15} ,  \frac{17}{315}  \}$ for $j=0, 1, 2, 3$.

\section{In-in positivity bounds}
\label{sec:pos}

\noindent Having established an EFT that captures the low-energy behaviour of the propagator, we will now show how its coefficients are bounded by unitarity and causality. We indicated above that unitarity will imply $\text{Im} \, \Sigma (\nu ) > 0$, which immediately leads to constraints on the $\gamma_n$ coefficients; the simplest of these is $\gamma_0 > 0$. 
Here we will also show that, when combined with causality, this positivity will imply constraints on $\text{Re} \, \Sigma$ that lead to a tower of two-sided bounds on the $c_n$ coefficients.

On Minkowski, this is an old story rooted in the Analytic $S$-matrix programme \cite{Eden:1966dnq}. 
Using the Kallen-Lehmann spectral representation of $\Sigma$, 
\begin{align}
\Sigma (p ) = \int_{4m^2}^{\infty} \frac{d m'^2}{ \pi} \frac{ \rho ( m' ) }{ m'^2 - p^2 - i \epsilon }
\label{eqn:KL_Mink}
\end{align}
the EFT coefficients are moments of the positive spectral density $\rho$
\begin{align}
 c_n = \int_{4m^2}^{\infty} \frac{d m'^2}{ \pi} \frac{ \rho ( m' ) }{ ( m'^2 )^{n+1} } \; . 
\end{align}
In practice we are often interested in cut-offs much larger than $2m$, so we define the couplings
\begin{align}
 g_n (\Lambda) = \int_{\Lambda^2}^{\infty} \frac{d m'^2}{ \pi \Lambda^2} \rho (m')   \left( \frac{\Lambda^2}{  m'^2 } \right)^{n+1}  \;  
 \label{eqn:cn_moment_Mink}
\end{align}
which can be computed in the EFT by subtracting the part of the branch cut which can be reliably computed below the cut-off $\Lambda$ \cite{deRham:2017avq}. 
As a series of moments, these $g_n$ obey various bounds. They
\begin{itemize}

\item[(i)] are manifestly positive, $g_n > 0$, since $\rho$ is a positive measure \cite{Adams:2006sv},

\item[(ii)] obey convergence relations, $g_{n+1}  < g_n$, since $m'^2 > \Lambda^2$ in the domain of integration \cite{deRham:2017avq}, 

\item[(iii)] obey non-linear Cauchy-Schwartz inequalities, such as $g_{n+1} g_{n-1} - g_n^2 > 0$ \cite{Bellazzini:2020cot}. 

\end{itemize}
For illustration, let us focus on $g_2, g_3$ and $g_4$. 
The bounds can be written succinctly as
\begin{align}
0 <   \frac{ g_4  }{ g_2 } < \frac{ g_3 }{ g_2 } < 1 \qquad   \left( \frac{ g_3 }{ g_2}  \right)^2 < \frac{ g_4}{g_2}
\label{eqn:pos}
\end{align}
together with $g_2  > 0$ \footnote{
Notice that $g_2 = 0$ would require $\rho = 0$ for all energies above $\Lambda$, i.e. that the theory is free. 
These bounds are 
strict inequalities when applied to theories with non-trivial interactions. 
}. 

An important caveat to these bounds is the assumption of convergence; the bounds need not apply if the spectral density grows too quickly and the integral \eqref{eqn:cn_moment_Mink} diverges. These divergences can be renormalised by introducing counter-terms (performing ``subtractions") in the dispersion relation, but the finite part of these counter-terms is fixed by imposing a renormalisation condition which can be of either sign since it is no longer connected to the underlying UV physics. 
A renormalisable propagator in the UV implies that at most two subtractions are required, i.e. once $g_0$ and $g_1$ are fixed by the mass and wavefunction renormalisaiton, the remaining $g_{n \geq 2}$ obey the above positivity bounds. 
These $n \geq 2$ bounds are then remarkably universal: they must be satisfied in any low-energy EFT with a unitary, causal Wilsonian UV completion.

On de Sitter, since the self-energies $\Sigma^{\alpha \beta}$ are two-point functions, they admit an analogous Kallen-Lehamnn representation 
\begin{align}
\Sigma^{\alpha \beta} ( \nu ) = \int_{0}^{\infty} \frac{ d \mu'^2 }{ \pi} \, \rho ( \mu' )  \Pi^{\alpha \beta}_{\mu'} ( \nu ) 
\label{eqn:KL_Sigma}
\end{align}
where $\Pi^{\alpha \beta}_\mu$ is the free propagator for a field of mass $\mu$ with operator ordering specified by the $\alpha \beta$ indices (e.g., $\Pi^{++}$ is time-ordered, $\Pi^{+-}$ is Wightman).
This dS KL representation dates back to many earlier works \cite{Bros:1990cu,Bros:1995js,Marolf:2010zp,Hollands:2011we,Marolf:2012kh} and was recently revitalised by \cite{Hogervorst:2021uvp,DiPietro:2021sjt, Penedones:2023uqc, Loparco:2023rug}. 
The simplest proof inserts a complete set of states between the two $J$ operators in \eqref{eqn:JJ}, since $\langle J \rangle$ elements are fixed by the dS isometries to be (proportional to) their free theory values. 
Combining these via \eqref{eqn:Sigma_def} produces the analogue of \eqref{eqn:KL_Mink} for the dS self-energy \footnote{
Alternatively, notice that the analytic structure of $\Sigma$ implied by causality means we can express it as $\frac{1}{2\pi i} \oint_\nu d\mu' \, 2 \mu' \Sigma (\mu') / ( \mu'^2 - \nu^2 - i \epsilon)$ around the pole in the lower half of the complex plane, and then deform this contour so that it runs over $\int_{-\infty}^{+\infty} d \mu'$ and then change variables to $\mu'^2$ to produce \eqref{eqn:Sigma_KL_dS}.   
}
\begin{align}
 \Sigma ( \nu )   =  \int_{0}^\infty \frac{ d \mu'^2 }{ \pi} \frac{ \rho ( \mu' )  }{\mu'^2 - \nu^2 - i \epsilon } \; .
 \label{eqn:Sigma_KL_dS}
\end{align}
Physically, \eqref{eqn:KL_Sigma} represents causality (interactions do not change the support of the free theory propagators \cite{Dubovsky:2007ac}
and $\rho = \text{Im} \, \Sigma > 0$ represents unitarity (interacting states can be resolved into a complete basis of free theory states).

While tempting to match \eqref{eqn:Sigma_KL_dS} directly to the EFT coefficients  as on Minkowski, 
\begin{align}
 c_n  \overset{?}{=} \int_{0}^\infty \frac{ d \mu'^2 }{ \pi} \frac{ \rho ( \mu' )  }{ ( \mu'^2 - i \epsilon )^{n+1} }
\end{align}
notice this only captures even powers of $\nu$. 
In fact, \eqref{eqn:Sigma_KL_dS} develops a branch cut at $\nu = 0$ and is no longer an even series in $\nu^{2n}$. This is due to the lack of a mass gap, a consequence of the fact that a particle on de Sitter can decay into other particles no matter the mass difference \cite{Nachtmann:1968}.
Although $\rho$ is positive, there is therefore no guarantee that the $c_n$ are positive (in the examples below we will find they are often not positive). 
The solution is to subtract part of the spectral integral and define the couplings
\begin{align}
 g_n ( \Lambda ) =  \int_{\Lambda^2}^{\infty} \frac{d \mu'^2}{\pi \Lambda^2 }   \rho (\mu' )  \left( \frac{\Lambda^2}{  \mu'^2 } \right)^{n + 1}  \; . 
 \label{eqn:g_def}
\end{align}
These can be computed within the EFT, since
\begin{align}
 \rho (\nu ) = \sum_{j=0} \pi \gamma_j' \; \nu^{2j+1}
 \label{eqn:rho_to_gamma}
\end{align}
and we therefore have
\begin{align}
g_n  = \Lambda^{2n} c_n - \sum_{j=0}  \frac{ 2 \Lambda^{2j+1}  \gamma'_j  }{1 + 2 j - 2 n  }   \; . 
\label{eqn:g_from_c}
\end{align}
which can be used to compute $g_n$ to any desired order in the EFT expansion, providing the cut-off $\Lambda$ is sufficiently low that \eqref{eqn:rho_to_gamma} converges (more on this below). 
Since they are positive moments, these $g_n$ obey the same positivity conditions \eqref{eqn:pos} as on Minkowski.
This combination of linear and quadratic constraints carves out leaf-shaped regions in parameter space, as shown in Fig.~\ref{fig:main}.  

To recap: we have shown that the time-ordered Bunch-Davies two-point function on de Sitter can be written as \eqref{eqn:self-energy_form} in terms of a self-energy $\Sigma$, which is related to the two-point function of sources by \eqref{eqn:Sigma_def}. This can be matched to an in-in EFT which contains complex coefficients, $c_n + i \gamma_n$. 
Unitarity alone would imply bounds only on $\gamma_n$, but by combining the Kallen-Lehmann representation and unitarity we have shown that the $g_n$ combinations obey an infinite tower of two-sided bounds, for instance \eqref{eqn:pos}. 
These differ from the Minkowski bounds in that, due to the lack of a mass gap, a subtraction is essential for mapping the $c_n$ onto positive moments.

\section{Explicit UV completions}
\label{sec:uv}

\noindent We will now consider a number of concrete UV completions and show that our positivity bounds are obeyed. Of course, the main virtue of the bounds are that they apply to any EFT, regardless of the detailed UV physics. This section is a useful consistency check, and also hints at whether the bounds are optimal. 
The results are summarised in Fig.~\ref{fig:main}, and we see that there are some regions of parameter space which are allowed by positivity but which cannot be realised by any of the UV completions we consider here. 
This suggests the bounds may not be optimal, or that more elaborate UV completions (e.g., with spinning fields) are needed to access this region.

\subsection{Tree-level mixing}

\noindent First consider a linear interaction $g \phi \chi$ between the light field $\phi$ of the EFT (mass $\mu_\phi$) and a heavy field $\chi$ (mass $\mu$). 
Then $\Sigma^{++} ( \nu )$ is simply the free $\chi$ propagator, namely \eqref{eqn:self-energy_form} with $\pm i \epsilon$ in place of $\Sigma ( \pm \nu )$. 
The self-energy \eqref{eqn:Sigma_def} is then:  
\begin{align}
 \Sigma ( \nu ) =  \frac{ g^2}{\mu^2 - \nu^2 - i \epsilon} \; . 
\end{align}
Taking the imaginary part gives the positive spectral density,
\begin{align}
 \rho (\nu ) = g^2 \pi \delta ( \mu^2 - \nu^2 ) \; . 
\end{align}

In this simple example, there is a mass gap. Equivalently, we see that $\gamma_n = 0$ for all $n$ and the EFT expansion of $\Sigma (\nu)$ produces a real series with EFT coefficients  
\begin{align}
 c_n = \frac{ g^2 }{\mu^{2n+2}} 
\end{align}
which coincide exactly with the Minkowski values. 
These are positive, and saturate the bounds~\eqref{eqn:pos} since every $g_n (\Lambda)/\Lambda^2$ goes from $0$ to $g^2$ as $\Lambda$ is increased from $0$ to $\mu$, and then drops back to 0 once $\Lambda$ exceeds $\mu$.  
This is shown in Fig.~\ref{fig:main} (blue line). 

As a consistency check, notice that the corresponding $\phi$ propagator contains poles at
\begin{align}
 \mu_\phi^2 - \nu^2 -   \frac{g^2}{\mu^2 - \nu^2} = 0   \; . 
\end{align}
These are precisely the mass eigenvalues obtained from diagonalising the mass matrix 
\begin{align}
\mathcal{L} \supset ( \phi \; \chi ) \left( \begin{array}{c c}  \mu^2_\phi  &  g \\ g & \mu^2 \end{array} \right) \left( \begin{array}{c} \phi \\ \chi \end{array}  \right) \; . 
\end{align} 

\subsection{One-loop bubble}

\noindent Next, consider the cubic interaction $ \phi \chi^2$. 
It contributes to the self-energies at leading order via a one-loop bubble diagram. 
In Minkowski, this loop integral is straightforward and the resulting $\Sigma$ has a branch cut along $p^2 \geq 4m^2$, with positive discontinuity:
\begin{align}
\rho (p) =  \frac{  2^{1 - d} \pi^{3/2} }{ (4 \pi )^{ \frac{d+1}{2} }  \Gamma \left( \frac{d}{2} \right) } \; \frac{ (  p^2 - 4m^2 )^{ \frac{d-2}{2} }  }{ \sqrt{p^2} }  \Theta \left( p^2 - 4m^2 \right)  . 
\label{eqn:Mink_bubble_rho}
\end{align}
The EFT coefficients appearing in $\Sigma = \sum_{n} c_n p^{2n}$ are 
\begin{align}
(4 m^2)^n c_n = \frac{  m^{d-3} }{ ( 4 \pi )^{\frac{d+1}{2}}  }  \frac{ \sqrt{\pi} \Gamma \left( \frac{3 - d}{2} + n \right) }{ 4  \Gamma \left( \frac{3}{2} + n \right) } 
\end{align}
and indeed correspond to the moments of \eqref{eqn:Mink_bubble_rho}.
Their ratio  
\begin{align}
 \frac{ 4m^2 c_{n+1}  }{ c_n  } = \frac{  3 + 2n - d}{ 3 + 2n } 
 \label{eqn:Mink_threshold}
\end{align}
indeed obeys the positivity conditions required of a series of moments.
As we take the cut-off $\Lambda \to \infty$, the individual $g_n (\Lambda) \to 0$ as the heavy physics decouples, but their ratios approach a finite value:
\begin{align}
\lim_{\Lambda \to \infty} \frac{ g_{n+1} ( \Lambda) }{ g_n ( \Lambda ) } = \frac{  3 + 2n - d}{ 5 + 2n - d } \; ,
\label{eqn:Mink_extremal}
\end{align}
which also satisfies the positivity conditions.

On dS, the spectral density for the bubble is \cite{DiPietro:2021sjt, Sleight:2021plv}
\begin{align}
  \rho ( \nu )  =  \frac{  \sinh ( \pi \nu )  \Gamma \left( \tfrac{ \frac{d}{2} \pm i \mu \pm i \mu \pm i \nu}{2}  \right)  }{ 64 \pi^{\frac{d+4}{2}} \Gamma \left( \tfrac{d}{2} \right)  \Gamma \left( \frac{d}{2} \pm i \nu\right)}
  \label{eqn:dS_bubble_rho}
\end{align}
where the $\pm$ shorthand denotes a product of $\Gamma$ functions with these arguments (i.e. there are 8 $\Gamma$ functions in the numerator and 3 $\Gamma$ functions in the denominator of this expression). 
$\rho$ is manifestly real and positive for $\nu > 0$. 
In the limit that $\nu$ and $\mu$ are both $\gg 1$, it is exponentially suppressed for $\nu < 2\mu$, and for $\nu > 2\mu$ recovers the expected Minkowski value \eqref{eqn:Mink_bubble_rho} with $p^2 = \nu^2$ and $m^2 = \mu^2$. 

\begin{figure}[t]
\begin{flushleft}
\includegraphics[width=1.0\columnwidth]{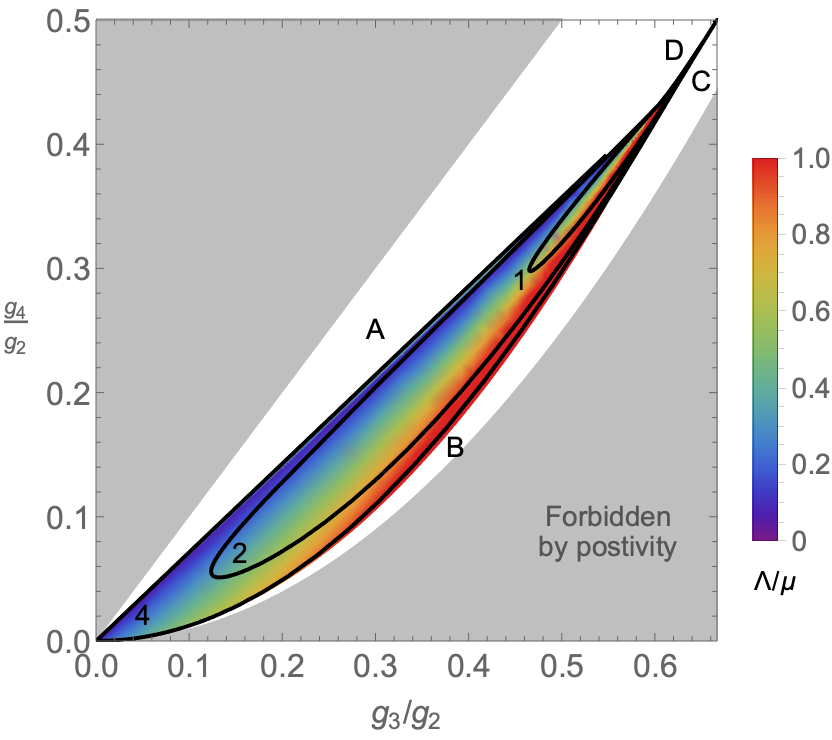}
\end{flushleft}
\caption{
The region of EFT parameter space traced out by the one-loop bubble in $d=3$ dimensions as the cut-off $\Lambda$ and 
mass $\mu$ are varied. 
Different values of $\Lambda/\mu$ are coloured so that all $\Lambda > \mu$ are displayed as red. Black lines show contours of $\mu = 1, 2$ and $4$. 
The boundary ABCD is described analytically in the text.
}
\label{fig:d3_island}
\end{figure}

The $g_n$ can be found numerically and indeed they satisfy the positivity bounds in all positive integer dimension for which the cubic interaction is renormalisable, namely $1 \leq d \leq 6$. 
The region that is traced out by varying $\Lambda$ and $\mu$ has a simple boundary which can be described analytically. 
The key observations are that (i) at large $\mu$ and $\Lambda$, we recover the Minkowski values, and (ii) at low $\mu$, the ratio becomes insensitive to $\Lambda/ \mu$,  
\begin{align}
\lim_{\mu \to 0} \left. \frac{ g_{n+1} ( \Lambda) }{ g_n (\Lambda) } \right|_{\Lambda = \alpha \mu} 
 = \frac{  2n - 1 }{  2n + 1 } \; . 
 \label{eqn:dS_kink}
\end{align}
For instance, the boundary in $\{ \frac{ g_3}{ g_2 } , \frac{ g_4 }{ g_2 } \}$ space is then made up of four segments:
\begin{itemize}

\item[A.] A straight line from \eqref{eqn:dS_kink} to the origin, from $\Lambda$ approaching zero at large $\mu$, which is dominated by the $\gamma_n$ subtraction, 

\item[B.] A quadratic curve from the origin to \eqref{eqn:Mink_threshold} from varying $0 < \Lambda \leq 2\mu$ at large $\mu$, which is captured by Minkowski limit,

\item[C.] An approximately straight line from \eqref{eqn:Mink_threshold} to \eqref{eqn:Mink_extremal} from varying $2\mu < \Lambda < \infty$ at large $\mu$, which is captured by Minkowski limit, 

\item[D.] A straight line from \eqref{eqn:Mink_extremal} to \eqref{eqn:dS_kink} , from varying $\Lambda$ at low $\mu$. 

\end{itemize}
This is shown in Fig.~\ref{fig:d3_island} \footnote{
Note that a low-energy observer can only 
determine $g_n (\Lambda)$ from the EFT coefficients when $\Lambda$ is below threshold, in this case $\Lambda < 2 \mu$. In Fig.~\ref{fig:d3_island} we have plotted all values of $\Lambda$ to illustrate that our bounds are always satisfied, regardless of the EFT series \eqref{eqn:g_from_c} converging.
}.

\begin{figure}[t]
\begin{flushleft}
\includegraphics[width=1.0\columnwidth]{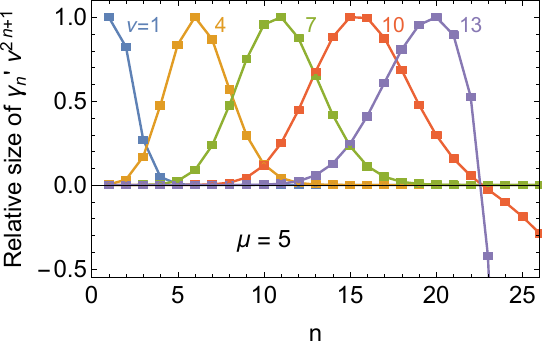}
\end{flushleft}
\caption{
The relative size of the first 25 terms appearing in the EFT expansion $\rho = \sum_n \pi \gamma_n' \nu^{2n+1}$ of the one-loop bubble in $d=2$, with $\mu=5$ and $\nu = 1, 4, 7, 10, 13$. \\
For $\nu \leq 1$, the terms strictly decrease and the EFT can be truncated at any order.
For $1 \leq \nu < 2 \mu =10$, the largest contribution comes from a finite $n$ and thus the EFT series only provides a reliable approximation if sufficiently many terms are included. For $\nu \geq \sqrt{ 4\mu^2 + 1 } \approx 10.05$, the terms become increasingly negative at large $n$ and the series does not converge.
}
\label{fig:gamma}
\end{figure}

Furthermore, we can use this one-loop example to illustrate an important point about the convergence of our EFT expansion. 
On Minkowski, the EFT expansion is controlled by the large hierarchy of scales between the mass $m$ of the light field and the cut-off $\Lambda$ which represents the characteristic mass scale of the heavy field(s). In practice, this means that the series in $c_n p^{2n} \sim ( p^2 /\Lambda^2 )^n$ not only converges for $p^2 < \Lambda^2$, but can be safely truncated at any finite order $N$ with an error of order $\left( p^2 / \Lambda^2 \right)^{N}$.
On de Sitter, we have two hierarchies: the ratio of $\nu$ to $H = 1$, and the ratio of $\nu$ to $\Lambda$.  
While the EFT expansion formally converges for $\nu^2 < \Lambda^2$, since $\Lambda$ represents the lowest-lying singularities of $\Sigma$, there can be analytic functions of $\nu/H$ that cannot be safely truncated unless $\nu \ll 1$ or sufficiently many terms are included in the EFT. 
As a toy example, consider expanding the function $e^\nu/(\Lambda - \nu) $ around $\nu =0$ and truncating after the first $N$ terms. If we take $\Lambda \gg \nu$, the error does not go to zero. The relative error is $\mathcal{O} (1)$ for all $N \lesssim \nu$, and then decays rapidly for $N \gtrsim \nu$ (faster than $e^{-\zeta \nu} \nu^{N}/N!$ for some $0< \zeta < 1$, thanks to Lagrange error bound on the exponential). 
This example captures the behaviour of $\rho (\nu)$ in \eqref{eqn:dS_bubble_rho}, which in $d=2$ dimensions simplifies to
\begin{align}
 \rho (\nu ) |_{d=2} = \frac{1}{16 \nu} \; \frac{ \cosh ( \pi \nu ) - 1 }{ \cosh (\pi \nu) +  \cosh ( 2 \pi \mu )} \; . 
\end{align}
The poles at $\nu = 2 \mu \pm i $ imply that the $\gamma_n' \nu^{2n+1}$ series will converge for $\nu^2 <   4 \mu^2 + 1 $. However, the analytic $\cosh$ functions imply truncation errors that do not vanish as $\Lambda \to \infty$.  
The relative size of each term in the EFT expansion is plotted in Fig.~\ref{fig:gamma}. 
For $\nu < 1$ they decrease uniformly as expected from Minkowski, so the series can be safely truncated at any order. But for $1 < \nu < 4\mu^2 +1$, the dominant contribution is from a finite order in the EFT series, and truncating before this order will lead to $\mathcal{O} (1)$ errors in the EFT approximation of $\rho$. For $\nu > 4 \mu^2 + 1$, the series coefficients grow more and more negative and unitarity is violated if the EFT series is truncated at any sufficiently large order. 
In contrast, if we consider the real part of $\Sigma$, which can be written as
\begin{align}
 \Sigma (\nu ) |_{d=2} = \frac{     
\psi \left( \frac{1- i \nu}{2} - i \mu \right)  
-   \psi \left( \frac{1 - i \nu}{2} + i \mu  
 \right)  }{16 \pi \nu \, \tanh ( \pi \mu ) }  + \frac{i}{16 \nu}  , 
 \label{eqn:Sigma_bubble_d2}
\end{align}
then we find that the EFT series $\sum_n^N c_n \nu^{2n}$ is always dominated by the $n = 0$ term when $\nu < 4\mu^2 + 1$ and the series can be safely truncated at any finite order.

\subsection{Two-loop sunset}

\noindent Finally, consider the interaction $\phi \chi^3$. 
This contributes to the self-energies via a two-loop sunset diagram at leading order. 
This was recently computed on de Sitter in \cite{Sleight:2021plv} using techniques from AdS \cite{Fitzpatrick:2010zm,Fitzpatrick:2011hu}. 
In $d=2$, the spectral density is
\begin{align}
&\rho (\nu ) |_{d=2}  \\
&= \frac{ \sinh ( \pi \nu ) }{16 \nu } \sum_{\mu_j = \pm \mu}  \frac{ ( \nu - \mu_{123} ) \coth \left( \frac{\pi}{2} (\nu - \mu_{123} ) \right) }{ \sinh ( \pi \mu_1 ) \sinh ( \pi \mu_2 ) \sinh ( \pi \mu_3 ) } \nonumber 
\end{align}
where $\mu_{123} = \mu_1 + \mu_2 + \mu_3$.
We have numerically integrated this for a range of $\Lambda$ and $\mu$ values and plotted the resulting $g_3/g_2$ and $g_4/g_2$ values in Fig.~\ref{fig:main}. 
This region of parameter space can be understood analytically in the same way as for the bubble. 
In the Minkowski limit of large $\nu$ and $\mu$, 
\begin{align}
\lim_{H \to 0} \rho ( \nu ) |_{d=2} =  \frac{ p - 3m }{2 p}  \; \Theta \left(  p^2 - 9 m^2 \right) \; . 
\end{align}
The lower boundary is then a quadratic curve from the origin to the Minkowski value at threshold
\begin{align}
 \left. \frac{ (3m)^2 c_{n+1}  }{ c_n  } \right|_{d=2} = \frac{( n + 1 ) ( 2 n  + 3 )}{ ( n+ 2) (2n+ 5) } \; ,
\end{align}
followed by an approximately straight line segment to the extremal value
\begin{align}
\lim_{\Lambda \to \infty} \left. \frac{ g_{n+1} ( \Lambda) }{ g_n ( \Lambda ) } \right|_{d=2} = \frac{ n + 1 }{ n + 2 } \; .
\end{align}
The upper boundary connects the origin to this extremal value via a kink at
\begin{align}
\lim_{\mu \to 0} \left. \frac{ g_{n+1} ( \Lambda) }{ g_n (\Lambda) } \right|_{\Lambda = \alpha \mu} 
 = \frac{  2n + 1 }{  2n + 3 }  \; . 
\end{align}
To sum up: in all of these examples we have verified that the EFT coefficients $\{  c_n, \gamma_n \}$ can be combined via \eqref{eqn:g_from_c} to produce a $g_n$ that closely matches the UV integral \eqref{eqn:g_def} and therefore bounded by two-sided inequalities such as \eqref{eqn:pos}.

\section{Discussion}

\noindent Given the vast space of inflationary models, it is essential that we develop an efficient EFT parameterisation with which to analyse data from the cosmological collider. 
The danger---inherent to any EFT parameterisation---is that we introduce parameter values that are secretly unphysical, i.e. which do not correspond to any consistent inflationary model. 
In this paper, we have shown that the EFT expansion of the self-energy 
must obey an infinite tower of two-sided positivity bounds if the underlying UV completion is unitary and causal. 
We have given a number of explicit UV completions to demonstrate that these bounds are indeed satisfied, and also to clarify how to estimate the error when truncating an EFT expansion on de Sitter. 
Our results open up several new questions:  \\

\noindent {\bf Phenomenology.}
Having established new constraints on the EFT description of heavy fields on de Sitter, one future direction is to explore how this positivity manifests in observables such as primordial non-Gaussianity.  
Let us now offer some first steps in this direction.

The self-energy shifts the poles in the $\phi$ propagator from the bare mass $\mu_\phi$ to the renormalised value $\tilde{\mu} - i \tilde{\gamma}$, where 
\begin{align}
\tilde{\mu} &= \mu_\phi - \frac{ \text{Re} \Sigma (\mu_\phi) }{ 2\mu_\phi  } ,  \;\;
&\tilde{\gamma} &= \frac{ \text{Im} \Sigma ( \mu_\phi) }{  2\mu_\phi }  \;  
\end{align}
at leading order in the interaction couplings. 
This has important consequences phenomenologically. 
For instance, one of the most promising observational targets is the bispectrum (equal-time three-point function of comoving curvature perturbations) \cite{Alishahiha:2004eh, Tolley:2009fg, Baumann:2011su, Achucarro:2010da, Flauger:2016idt}.
It can be calculated by acting with various differential operators on a basis of simple seed integrals that correspond to a conformally coupled scalar $\sigma$ interacting with heavy fields $\phi$ through interactions such as $\sigma^2 \phi$ \cite{Baumann:2022jpr}. 
Explicitly, the $\langle \sigma^4 \rangle$ seed integral from the $s$-channel exchange of $\phi$ is
\begin{align}
\langle \sigma^4 \rangle_s = \int_{-\infty}^0 \frac{ d \tau }{ \tau^2 } \int_{-\infty}^0 \frac{ d \tau' }{ \tau'^2 } \,  e^{i k_{12} \tau}  G_\phi ( k_s \tau, k_s \tau' )    e^{i k_{34} \tau'}
\label{eqn:sig4_s}
\end{align}  
where $G_\phi$ is the propagator for the internal $\phi$ field with spatial momentum $k_s = | \bfk_1 + \bfk_2|$ and $k_{ab} = | \bfk_a | + |\bfk_b|$ are the ``energies'' flowing into each vertex (labelled so that $k_{12} \geq k_{34}$).    
In the Appendix, we evaluate this integral and show that it naturally splits 
into ``particle production'' contributions (in which on-shell $\phi$ are produced by the spacetime background) and ``EFT'' contributions (in which off-shell $\phi$ are produced by the $\sigma$).
The particle production effects are most transparent in the collapsed limit $k_s \to 0$, where there are two contributions at leading order 
\begin{align}
 \langle \sigma^4 \rangle_s^{\rm pp} \approx  \frac{ \text{Re} \left[ 
\alpha  \left( \frac{ k_{34} }{k_{12} } \right)^{i \nu} 
+
\beta  \left( \frac{ k_s^2 }{4 k_{12} k_{34} } \right)^{i \nu} 
 \right]  }{ \sqrt{k_{12} k_{34}} } 
\,  
\label{eqn:seed_pp}
\end{align}
which exhibit logarithmic oscillations in the momenta ratios at a complex frequency $\nu = \tilde{\mu} - i \tilde{\gamma}$, i.e. $\tilde{\mu}$ sets the period and $\tilde{\gamma}$ sets the damping rate of this characteristic cosmological collider signal. 
The relative amplitudes $\alpha, \beta$ are given in \eqref{eqn:pp_ab};
both $\sim e^{- \pi  \tilde{\mu} }$ due to Boltzmann suppression and therefore vanish in the Minkowski limit. 
The EFT contribution, on the other hand, can be written as a power series in $\hat{s}^n$, where $\hat{s}$ is the differential operator \eqref{eqn:s_def} and reduces to the usual Mandelstam invariant in the Minkowski limit. 
The first few terms are:
 \begin{align}
\langle \sigma^4 \rangle_s^{\rm EFT}
=
\left[ 
\frac{1}{\mu^2_0} +  \frac{ Z_0 }{\mu_0^4}  \hat{s} + \frac{  Z_0^2 + c_2 \mu_0^2 }{\mu_0^6}  \hat{s}^2 + ... 
\right] \frac{1}{ k_{T}  } \; 
\label{eqn:sig4_EFT_s12}
\end{align}
where $1/k_T = 1/(k_{12} + k_{34})$ is the contact diagram in which the internal $\phi$ propagator has been contracted to a local interaction. 
The first term fixes the mass of the exchanged field ($\mu_0^2 = \mu^2 - c_0$), the second term fixes the normalisation of its kinetic term ($Z_0 = 1 + c_1$), 
and the third is the order at which interactions unambiguously enter. 

Our positivity bounds on the self-energy can then be translated into concrete restrictions on the bispectrum. 
The simplest of these is that, since $\text{Im} \, \Sigma > 0$, we have $\tilde{\gamma} > 0$ and therefore the particle production effects \eqref{eqn:seed_pp} are finite in the soft limits $k_s \to 0$ or $k_{ab} \to 0$. 
Our bounds on the $\text{Re} \, \Sigma$ correspond to restrictions on the EFT part of the signal, for instance $c_2 \Lambda^3 > - 2 \gamma_0'/3 + ... \approx - 4 \tilde{\gamma}/(3 \pi )$ bounds the size of the $\hat{s}^2$ shape in terms of the damping rate of the particle production oscillations. 
Any violation of the Minkowski bound $c_2 > 0$ must therefore come with a corresponding decay rate in the cosmological collider signal; conversely, a measured decay rate $\tilde{\gamma}$ implies a lower bound on corrections to the EFT background.  
Finally, at every subsequent order in the EFT expansion, our bounds place constraints on the relative size of the EFT corrections in terms of the damping rate. 
It would be interesting to compare these analytic bounds with the numerical bounds recently obtained in \cite{deRham:2025mjh}.  \\

\noindent {\bf Further bounds.} 
While our explicit UV completions all produce EFT parameters inside the allowed positivity island, they do not completely fill the island. 
One outstanding question is then: 
can our bounds be improved? 
It may be that our bounds are already optimal and more elaborate UV completions (e.g. with spinning fields) would populate the entirety of the allowed region. 
However, notice that (at least in the UV completions that we considered) the EFT parameters cluster more closely around the origin as the number of spatial dimensions is increased.
It therefore seems likely that, by adding the further requirement that $d=3$, the bounds may be strengthened (an analogous strengthening occurs in Minkowski for massive spinning fields, see e.g. \cite{Davighi:2021osh}). 
Our bounds could also be refined by including various additional UV assumptions (e.g. about spin dominance) so that they map EFT parameter regions onto UV properties beyond basic 
unitarity and causality. 

If we continue to assume only unitarity/causality in the UV, here are three directions which seem to go beyond our positivity bounds:
\begin{itemize}

\item[(i)] for $\nu = i \alpha$ purely imaginary, then we have 
\begin{align}
 \Sigma ( i \alpha ) = \int_{0}^{+\infty} \frac{d \mu'^2}{2 \pi}  \frac{ \rho (\mu' ) }{ \mu'^2 + \alpha^2 }
\end{align} 
which is real and manifestly positive \footnote{
This is analoguous to scattering with space-like momenta on Minkowski, for which the $i \epsilon$ becomes unimportant because there is no operator ordering ambiguity.
}. 
We have confirmed this
in our example UV completions, e.g. \eqref{eqn:Sigma_bubble_d2} is real and positive since $2 \alpha \,  \text{Im} \, \psi \left( \frac{1+\alpha}{2} + i \mu \right)  \leq \alpha \pi \tanh ( \pi \mu)$ for any real $\alpha$ and $\mu > 0$.
This appears to have interesting phenomenological consequences which are not obviously captured by our other bounds. For instance in the squeezed limit $k_{12} \gg k_s \sim k_{34}$ relevant for the bispectrum, 
\begin{align}
    \langle \sigma^4 \rangle_s^{\rm EFT} \approx \frac{1}{k_{12}} \;   \frac{ 1 }{\mu^2 +  1/4 - \text{Re} \, \Sigma ( i / 2 ) } 
    \label{eqn:sig4_EFT_squeezed}
\end{align}
and therefore the EFT background is always made larger by the effects of heavy physics.

\item[(ii)] Unitarity, when formulated as an optical theorem, places both lower {\it and upper} bounds on matrix elements. For instance, anomalous dimensions in AdS (analogous to $\rho$ \cite{Chakraborty:2025mhh}) are bounded by $| \gamma | < 4$ \cite{Fitzpatrick:2010zm}, and EFT couplings in Minkowski (analogous to $g_2$) are similarly bounded from above \cite[(3.35)]{Caron-Huot:2020cmc}. 
We believe that there are analogous upper bounds for EFTs on de Sitter, which have yet to be fully determined. 

\item[(iii)] Our bounds follow from the spectral representation $\Sigma$, the two-point function of the source $J$. 
Of course, the two-point function of $\phi$ itself also has a spectral representation with positive spectral density, and therefore generates an analogous series of positive moments.
The two spectral densities are related by \footnote{
\eqref{eqn:rho_phi_to_rho_J} makes it clear 
that $\rho_J$ and $\Sigma$ are related to $\rho_\phi$ and \protect{$\langle \phi \phi \rangle$} by an amputation of two propagators that closely resembles LSZ reduction: $\Sigma$ is therefore closer in spirit to an $S$-matrix, and is the natural object 
for EFT matching. 
}
\begin{align}
    \rho_\phi (\nu) =  \frac{ \rho_J (\nu) }{ | \mu_\phi^2 - \nu^2 - \Sigma ( \nu )  |^2 } \; .
    \label{eqn:rho_phi_to_rho_J}
\end{align}
It would be interesting to repeat the above using $\rho_\phi$, particularly since the recent numerical bounds of \cite{deRham:2025mjh} use $\rho_\phi$ rather than $\rho_J$.  

\end{itemize}
Furthermore, 
our bounds apply to the propagation of a massive \emph{scalar} mediator---given the recent progress in developing the analogous Kallen-Lehmann spectral representation for massive spinning fields \cite{Loparco:2023rug}, it would be interesting to repeat our construction for mediators of other spin.\\

\noindent {\bf In-in vs in-out EFTs.}
In order to match a propagator with the in-in boundary conditions relevant for inflationary cosmology,
we have used an EFT that mixes fields on different branches of the in-in path integral contour. 
The resulting in-in EFT is sufficiently similar to the in-out EFTs used for scattering on Minkowski that we can bound its coefficients. 
However, it differs in one key respect. 
Ordinarily, an in-out EFT will break down at the mass scale of the lightest state not included in the EFT description. 
An in-in EFT 
will instead break down at a scale set by the smallest {\it mass difference}.
For example, the interaction $\phi \chi_1 \chi_2$  will produce a bubble diagram that has spectral density given by \eqref{eqn:dS_bubble_rho} with $\Gamma \left( \left(  \frac{d}{2} \pm i \mu_1 \pm i \mu_2 \pm i \nu \right)/2 \right)$ in the numerator. This has singularities at points with $| \nu |^2 = | \mu_1 - \mu_2 |^2 + d^2/4$, and therefore the EFT expansion in powers of $\nu$ can only converge providing $|\nu|$ is smaller than this scale \footnote{
For the equal-mass case considered above, these singularities are cancelled by zeroes of $\sinh ( \pi \nu)$ and the cut-off is instead set by singularities at $| \nu |^2 = 4 \mu^2 + d^2/4$). 
}.
This can be much lower than the radius of convergence that an in-out observable would have, namely $| \mu_1 + \mu_2 |^2 + d^2/4$. 
In a follow-up work \cite{us}, we will present analogous positivity bounds for in-out EFTs on de Sitter, 
since while they are one step removed from cosmological correlators \footnote{
See for instance \cite{Melville:2023kgd} for a description of how to ``square'' an in-out $S$-matrix on de Sitter into an in-in Bunch-Davies correlator.
}, they enjoy a much higher cut-off and share even more similarities with the EFThedron of Minkowski.  \\

\noindent {\bf Integrating out light fields.}
We have taken the view that light fields with $m^2 < d^2 H^2/4$ (in the so-called complementary series) should be included explicitly in the EFT, while heavy fields with masses $m^2 \geq d^2 H^2/4$ (in the so-called principal series) can be integrated out to produce our EFT interactions. 
Consequently, the EFT corrections to the self-energy can be written as a Kallen-Lehmann representation involving only principal series values (real $\mu$). 
One question, which we leave open for the future, is how our positivity bounds would change if one also integrates out light fields (imaginary $\mu$). 
Mathematically, this would require deforming the integration contour in \eqref{eqn:KL_Sigma} to account for poles on the imaginary axis.
One could then split the integral into three regions: the $|\mu'| < \Lambda$ region, that can be computed explicitly within the EFT and subtracted; the $\mu > \Lambda$ region \eqref{eqn:g_def} which obeys the positivity bounds \eqref{eqn:pos}; and a new $i \mu > \Lambda$ region which must be bounded by other means. \\

\noindent {\bf Bootstrapping dS EFTs.} 
On Minkowski, unitarity and causality (analyticity) are often powerful enough to bootstrap the $S$-matrix \cite{Kruczenski:2022lot}.
In cosmology, these properties lead to various perturbative cutting rules \cite{Melville:2021lst, Goodhew:2021oqg, Tong:2021wai,AguiSalcedo:2023nds,Qin:2023bjk,Ema:2024hkj} and much progress has been made in using them to bootstrap cosmological correlators (see e.g. \cite{Baumann:2021fxj, Jazayeri:2021fvk, Meltzer:2021zin, Chowdhury:2023arc, Chowdhury:2025ohm, Jazayeri:2025vlv}).
These bootstrap strategies typically exploit the analyiticity of cosmological correlators in the complex $k$ plane \cite{Meltzer:2021bmb, Salcedo:2022aal}, which is distinct from the analyticity in $\nu$ exploited here. It would be interesting to combine these two different notions of analyticity, and also to better understand the connection with analyticity of the underlying $S$-matrix \cite{Kristiano:2025cod}. \\

\noindent {\bf Beyond de Sitter.}
The spectral representation \eqref{eqn:KL_Sigma} requires dS isometries, however we commonly consider cosmological correlators which break boosts (for instance correlators from the EFT of inflation \cite{Cheung:2007st, Senatore:2010wk, Noumi:2012vr}). In fact, since correlators which obeys full dS isometry are slow roll suppressed, these boost breaking correlators are often more important for observations \cite{Green:2020ebl}. It is therefore important to understand whether it is possible to write down a spectral representation in the absence of dS boosts.
Relatedly, it would be interesting to apply our bounds in other cosmological settings: for instance various EFTs of dark energy have been constrained using Minkowski positivity bounds \cite{Melville:2019wyy, Ye:2019oxx, Kennedy:2020ehn, Tokuda:2020mlf, deRham:2021fpu, Melville:2022ykg, Xu:2023lpq}, and it would be interesting to compare those with the dS positivity bounds developed here. \\

\section*{Acknowledgments} 

\noindent M.H.G.L. is supported by a postdoctoral fellowship at NTU funded by the Ministry of Education (114V2004-3). M.H.G.L. thanks the Yukawa Institute for Theoretical Physics at Kyoto University for hosting the workshop ``Progress of Theoretical Bootstrap", where some of the results were presented.
S.M. was supported in part by a UKRI Stephen Hawking Fellowship (EP/T017481/1), and thanks the organisers of the ``AdS/CFT meets Carrollian \& Celestial Holography" workshop at ICMS, Edinburgh, where this work was completed.

\section*{DATA AVAILABILITY}

\noindent Data are available from the authors upon request.

\section*{APPENDIX: COSMOLOGICAL COLLIDER SIGNALS}
\label{sec:pheno}

\noindent In this Appendix, we discuss some phenomenology of the de Sitter self-energy and its positivity bounds. 
In particular, we will focus on how a heavy field $\phi$ with self-energy $\Sigma$ would, if present during inflation, appear to us today in primordial non-Gaussianity. 
Thanks to recent progress in the cosmological bootstrap programme \cite{Arkani-Hamed:2018kmz, Baumann:2019oyu, Baumann:2022jpr}, these signals can be computed from simpler correlators of a conformally coupled field $\sigma$ (with mass $\mu = i/2$) on a fixed de Sitter background.
In simple terms, this procedure evaluates some $\sigma$ fields ``on the background'' by taking a soft limit $k\to0$ and then acts on the remainder with various weight-shifting operators to transform them into scalar or tensor non-Gaussianities at leading order in the slow-roll parameter (assuming de Sitter symmetries are weakly broken).
The bispectrum from the exchange of massive fields, for instance, can be computed in this way from the seeds $\langle \sigma^4 \rangle$ and $\langle \sigma^5 \rangle$ (by evaluating one or two $\sigma$ fields on the background). 
Here we focus on the single-exchange contribution $\langle \sigma^4 \rangle$, 
\begin{align}
\langle \sigma^4 \rangle_s =  \vcenter{ \hbox{ \begin{tikzpicture}
 \draw[gray!40] (-1,0.75) -- (3,0.75);
 \draw[red] (0,0) -- (2,0);
 \draw (0,0) -- (-0.5,0.75);
 \draw (0,0) -- ( 0.5,0.75);
 \draw (2,0) -- ( 1.5,0.75);
 \draw (2,0) -- ( 2.5,0.75);
 \node[below] (5) at (0.35,0) {$\phi$};
 \node[below] (6) at (1.65,0) {$\phi$};
 \filldraw[gray!40] (1,0) circle (0.25);
 \node (7) at (1,0) {$\Sigma$};
 \node[above] (1) at (-0.5,0.75) {$\sigma_1$};
 \node[above] (2) at (0.5,0.75) {$\sigma_2$};
 \node[above] (3) at (1.5,0.75) {$\sigma_3$};
 \node[above] (4) at (2.5,0.75) {$\sigma_4$};
 \end{tikzpicture} }}
\end{align}
though our results can be readily generalised to more $\sigma$ fields or more exchanges, such as 
\begin{align}
\vcenter{ \hbox{ \begin{tikzpicture}
 \draw[gray!40] (-1,0.75) -- (5,0.75);
 \draw[red] (0,0) -- (4,0);
 \draw (0,0) -- (-0.5,0.75);
 \draw (0,0) -- ( 0.5,0.75);
 \draw (2,0) -- ( 2,0.75);
 \draw (4,0) -- ( 3.5,0.75);
 \draw (4,0) -- ( 4.5,0.75);
 \node[below] (5) at (0.35,0) {$\phi$};
 \node[below] (6) at (1.65,0) {$\phi$};
  \node[below] (9) at (2.35,0) {$\phi$};
 \node[below] (10) at (3.65,0) {$\phi$};
 \filldraw[gray!40] (1,0) circle (0.25);
 \filldraw[gray!40] (3,0) circle (0.25);
 \node (7) at (1,0) {$\Sigma$};
 \node (8) at (3,0) {$\Sigma$};
 \node[above] (1) at (-0.5,0.75) {$\sigma_1$};
 \node[above] (2) at (0.5,0.75) {$\sigma_2$};
 \node[above] (3) at (2,0.75) {$\sigma_3$};
 \node[above] (4) at (3.5,0.75) {$\sigma_4$};
  \node[above] (5) at (4.5,0.75) {$\sigma_5$};
 \end{tikzpicture} }} 
\end{align}

Concretely, consider the interaction $\sigma^2 \phi$ so that two $\sigma$ can combine into an off-shell $\phi$.
At leading order in this interaction, $\langle \sigma^4 \rangle$ can be decomposed into an $s$-channel contribution ($\sigma_{\bfk_1} \sigma_{\bfk_2} \to \phi \to ... \to \phi \to \sigma_{\bfk_3} \sigma_{\bfk_4}$) plus analogous $t$- and $u$-channel contributions related by permutation. 
Dilation invariance implies that the $s$-channel contribution can be written in terms of two dimensionless ratios
\begin{align}
u &= \frac{ | \bfk_1 + \bfk_2 |}{k_1 + k_2} \; , \;\; 
&v &= \frac{|\bfk_3 + \bfk_4|}{k_3 + k_4} \; . 
\end{align}
In the time domain, $\langle \sigma^4 \rangle_s$ is given by the integral \eqref{eqn:sig4_s}. 
To perform this, use \eqref{eqn:Gn_def} to transform to the Casimir variable $\nu$, 
\begin{align}
k_s  \langle \sigma^4 \rangle_s  
 = \int_{-\infty}^{+\infty} d \nu N_\nu \,   F_\nu (u) F_\nu (v) G_\phi ( \nu ) 
 \label{eqn:O_int}
\end{align}
where $N_\nu = \nu \sinh ( \pi \nu )/\pi$, the universal vertex factors $F_\nu$ are particular Legendre functions \footnote{ 
Explicitly, they are given by:
 \begin{align}
 F_\nu^j (1/z) = \frac{1}{\sqrt{2}} \Gamma \left( j \pm i \nu \right)  P_{i \nu - \frac{1}{2}}^{\frac{1}{2} - j} (z) \left( \sqrt{z^2 - 1 } \right)^{ \frac{1}{2} - j }
 \label{eqn:F_def}
\end{align}
and are specified by a half-integer index $j = \frac{d-2}{2} ( n -1 )$ that depends on the number of spatial dimensions and the number of particles at each vertex. $j=1/2$ for \protect{$\langle \sigma^4 \rangle_s$} in $d=3$. 
}---the two-particle analogue of $f^{+}_\nu$---and the propagator $G_{\phi}$ can be written as
\begin{align}
i G_{\phi} (\nu ) = \frac{ \alpha_{+\nu} }{ \mu^2 -\nu^2 - \Sigma (\nu )  } + \frac{ \alpha_{-\nu} }{ \mu^2 -\nu^2 - \Sigma ( - \nu )  } \; . 
\end{align}
where $\alpha_{+\nu} + \alpha_{-\nu} = 1$ ensures the correct normalisation and $\alpha_{+\nu} - \alpha_{-\nu}$ depends on the observable in question (see \eqref{eqn:a_correlator} below).
To evaluate this integral, we will close the contour in the complex $\nu$ plane. The singularities of the universal $F_\nu$ functions produce the so-called ``EFT'' contribution, while the singularities of the object-dependent $G_{\phi}$ function produce the so-called ``particle production'' contribution.  \\

\noindent {\bf EFT contribution.}
To find the EFT part, note that the vertex factors can be written as,
\begin{align}
&F_\nu (u) F_\nu (v)  \\
&= \sum_{m,n=0}^\infty u^{2m+1} \left( \frac{u}{v} \right)^n \sum_{j=0}^m  \frac{ c_{mnj} }{ \nu^2 + ( n + 2 j + \frac{1}{2} )^2 } \nonumber
\end{align}
plus analytic terms that do not contribute, where the residues are given by:
\begin{align}
 c_{mnj} =  \frac{ n + 2j + \frac{1}{2} }{ j! (m-j)!}  \frac{ \Gamma \left( 1 + n + 2m \right) \Gamma \left( \frac{1}{2} + n +j \right)  }{4^{m} n! \Gamma \left( \frac{3}{2} + n +m + j  \right)} . 
\end{align} 
Closing the contour in the complex $\nu$ plane around these singularities produces
\begin{align}
k_s \langle \sigma^4 \rangle^{\rm EFT}_s  = \sum_{m,n=0}^{\infty} u^{2m + 1} \left( \frac{u}{v} \right)^n c_{mn}
\label{eqn:sig4_EFT}
\end{align} 
where
\begin{align}
 c_{mn} = \sum_{j=0}^m  \left. \frac{ (-1)^{n+j} \; c_{mnj} }{ \mu^2 - \nu^2 - \text{Re} \, \Sigma (\nu ) } \right|_{\nu = i (n + 2j + \frac{1}{2} )}
\end{align}
Since the odd (imaginary) part of $\Sigma ( \nu )$ vanishes at $i \nu = \frac{d}{2} + n$ for any integer $n$, this EFT series depends only on its real part.
This EFT contribution is therefore universal: it is the same for any observable (e.g. in-in correlator, wavefunction or $S$-matrix).
In the squeezed limit $k_{12} \gg k_s \sim k_{34}$ (i.e. $u \to 0$), this becomes \eqref{eqn:sig4_EFT_squeezed}. 

In the tree-level limit, $\text{Re} \, \Sigma (\nu )$ vanishes and 
\eqref{eqn:sig4_EFT} coincides with the result of \cite{Arkani-Hamed:2018kmz}.
At leading order in the EFT couplings, one can resum this series at large $\mu$ into the form \eqref{eqn:sig4_EFT_s12}, where 
.
the two-particle Mandelstam is \cite{Melville:2024ove} 
\begin{align}
\hat{s} = - u^2 (1 - u^2) \partial_u^2 + 2 u^3 \partial_u  - \frac{1}{4}
\label{eqn:s_def}
 \end{align}
and coincides with $-\Delta_u - \frac{1}{4}$ of \cite{Arkani-Hamed:2018kmz}. 
This resummation is to be expected, since $\hat{s} F_\nu (u) =  \nu^2 F_\nu (u)$ and therefore \eqref{eqn:sig4_EFT_s12} can also be recognised as the expansion of $G_\phi (\nu)$ in powers of $\nu^2$, together with the identity $\int_{-\infty}^{+\infty} d \nu \, N_\nu \,  F_\nu (u) F_\nu (v) = k_s/k_T$.\\

\noindent {\bf Particle production contribution.}
It is the particle production part of the integral that gives rise to distinctive oscillations in the squeezed limit of the bispectrum \cite{Arkani-Hamed:2015bza,Lee:2016vti,Chen:2016uwp,Chen:2016hrz, Wang:2019gbi, Kumar:2019ebj,Bodas:2020yho, Tong:2021wai, Pinol:2021aun,Tong:2022cdz}. 
To perform this part of the integral, notice that the vertex factors can be split
\begin{align}
 F_\nu  =    Q_{+\nu}   +   Q_{-\nu} 
\label{eqn:P_to_Q}
\end{align} 
where $Q_{+\nu}$ ($Q_{-\nu}$) decays sufficiently quickly in the lower (upper) half of the complex $\nu$ plane that we can close the integration contour. 
When $u < v$, we expand $F_\nu ( u )$ using \eqref{eqn:P_to_Q} and then close the integration contour around the singularities of $G_\phi$, 
which gives two distinct residues:
\begin{align}
k_s \langle \sigma^4 \rangle_s^{\rm pp}  
=
R_+ + R_- 
\end{align}
that are given explicitly by
\begin{align}
R_{\pm} =  \alpha_\nu Q_\nu (u ) F_\nu (v )  |_{\nu = \pm \tilde{\mu} - i \tilde{\gamma} }
\end{align}
Explicitly, $Q_\nu$ are related to Legendre functions of the second kind and have the following expansion at large argument:
\begin{align}
 Q_\nu ( u ) \approx  \frac{ \Gamma \left( \frac{1}{2} + i \nu \right) \Gamma \left( - i \nu \right) }{\sqrt{2\pi} }
 \left(  \frac{u}{2}\right)^{\frac{1}{2} + i \nu} 
\end{align}
This leads to the collapsed limit
\begin{align}
k_s \langle \sigma^4 \rangle_s^{\rm pp} 
\approx \frac{ \sqrt{uv} }{2} \sum_{\mu_b = \pm \mu}
 \alpha_{\mu_1 \mu_2} \left( \frac{u}{2} \right)^{i \mu_1} \left( \frac{v}{2} \right)^{i \mu_2}
 \label{eqn:sig4_pp}
\end{align} 
plus $\mathcal{O} ( k_s^2 ) $ corrections, with coefficients
\begin{align}
\alpha_{\mu_1 \mu_2} =   \frac{ \alpha_{\mu_1} }{2 \pi}   \prod_{b=1}^2 \Gamma \left( \tfrac{1}{2} + i \mu_b \right)   \Gamma \left( - i \mu_b \right)  . 
\label{eqn:app_from_ap}
\end{align} 

The equal-time correlator $\langle \sigma_{\bfk_1} ( \tau ) \sigma_{\bfk_2} ( \tau  ) \sigma_{\bfk_3} (\tau ) \sigma_{\bfk_4} ( \tau ) \rangle' $ requires summing four in-in propagators with different time orderings, which in the late time limit $\tau \to 0$ produces \eqref{eqn:O_int} with
\begin{align}
 \alpha_\nu = \frac{ e^{+ \pi \nu} - e^{- \pi \nu} + i + i }{ 2 \sinh ( \pi \nu ) } \; . 
 \label{eqn:a_correlator}
\end{align} 
\eqref{eqn:sig4_pp} with \eqref{eqn:a_correlator} gives the result \eqref{eqn:seed_pp} quoted in the main text, with amplitudes
 \begin{align}
 \alpha  &= \frac{ 2 \pi i}{ \sin \left(  2 \pi i \nu \right) } ( 1 + \sin ( i \pi \nu ) ) \; ,   \label{eqn:pp_ab}  \\
 \beta &=  \frac{ \Gamma ( \frac{1}{2} + i \nu    )^2 \Gamma ( - i \nu  )^2 }{ \pi}  \left( 1 +  \sin ( i \pi \nu )   \right)  \; . \nonumber 
\end{align}
For contrast: the 4-particle dS $S$-matrix of \cite{Melville:2023kgd} requires only the time-ordered propagator \eqref{eqn:self-energy_form}, so is given by \eqref{eqn:O_int} with $2 \alpha_\nu = e^{\pi \nu}/\sinh ( \pi \nu)$, and the late-time wavefunction coefficient $\psi_4$ requires the bulk-to-bulk propagator, so is given by \eqref{eqn:O_int} with $2 \alpha_\nu = \left( e^{\pi \nu} -  f^-_\nu / f^+_\nu \right) /\sinh ( \pi \nu) $ and depends on the late-time limit of the mode functions \cite{Melville:2024ove}.  \\

\noindent {\bf Extension to higher point correlators.}
There is a straightforward generalisation of the above to the single-exchange part of any $n$-field observable.
Consider the interactions $\sigma^{n_1} \phi$ and $\sigma^{n_2} \phi$, which contribute to $n = n_1 + n_2$ field observables through the integral
\begin{align}
\langle \sigma^n \rangle_s = \int_{-\infty}^{+\infty} N_\nu   \, F_\nu^{j_{n_1}} (u) F_\nu^{j_{n_2}} (v) G_{\phi} ( \nu ) 
\end{align}
where now $u = \frac{| \bfk_1 + ... \bfk_{n_1} |}{k_1 + ... + k_{n_1} }$, $v = \frac{| \bfk_{n_1 + 1} + ... \bfk_{n} |}{k_{n_1 + 1} + ... + k_{n} }$ and the $F^{j_n}_\nu$ are $n$-particle generalisations of $f^{+}_\nu$, namely \eqref{eqn:F_def}. 
We again split this integral into EFT and particle production pieces, 
and find 
\begin{align}
\langle \sigma^n \rangle^{\rm EFT}_s  = \sum_{m,n=0}^{\infty} u^{2m + 1} \left( \frac{u}{v} \right)^n c_{mn}^{n_1 n_2}
\end{align} 
where
\begin{align}
 c_{mn}^{n_1 n_2} = \sum_{j=0}^m  \left. \frac{ (-1)^{n+j} \; c_{mnj}^{n_1 n_2} }{ \mu^2 - \nu^2 - \text{Re} \, \Sigma (\nu ) } \right|_{\nu = i (n + 2j + \frac{1}{2} )} \; . 
\end{align}
and the constants $c_{mnj}^{n_1 n_2}$ are given in \cite{Melville:2024ove}.
This is again universal (independent of whether we consider $S$-matrix, wavefunction or in-in correlator). since the odd (imaginary) part of $\Sigma ( \nu )$ vanishes at these values of $\nu$.
In the tree-level limit, $\text{Re} \, \Sigma (\nu )$ vanishes and this coincides with the result of \cite{Melville:2024ove}.
Again we see that positivity of $\Sigma ( i \alpha )$ constrains the shift in this EFT background due to the presence of heavy fields in the UV.   \\

\noindent {\bf Power spectrum.}
Finally, a remark about the power spectrum of $\phi$ (the equal-time in-in two-point function at late times). 
This may seem the natural observable to compute from a resummed propagator like \eqref{eqn:self-energy_form}. 
However, this function is uniquely fixed by the de Sitter symmetries. 
An equivalent statement is that the 2-particle $S$-matrix (i.e. the $1 \to 1$ scattering amplitude) is trivial, since it simply reflects the normalisation of the states. 
This is familiar on Minkowski, where the only on-shell information in $G(p)$ is the location of the pole in the complex $p^2$ plane, which we identify with the physical mass of the field, and its residue, which encodes the wavefunction normalisation of the field. 
The dS power spectrum is similarly insensitive to $\Sigma (\nu)$ beyond the pole location and residue. 
Once the mass and field have been renormalised, the remaining EFT coefficients do not contribute to the (on-shell) power spectrum. 
Instead, they show up in the propagation of off-shell $\phi$ particles: for instance in the 4-point exchange correlator computed above.

\bibliography{references}

\begin{thebibliography}{129}%
\makeatletter
\providecommand \@ifxundefined [1]{%
 \@ifx{#1\undefined}
}%
\providecommand \@ifnum [1]{%
 \ifnum #1\expandafter \@firstoftwo
 \else \expandafter \@secondoftwo
 \fi
}%
\providecommand \@ifx [1]{%
 \ifx #1\expandafter \@firstoftwo
 \else \expandafter \@secondoftwo
 \fi
}%
\providecommand \natexlab [1]{#1}%
\providecommand \enquote  [1]{``#1''}%
\providecommand \bibnamefont  [1]{#1}%
\providecommand \bibfnamefont [1]{#1}%
\providecommand \citenamefont [1]{#1}%
\providecommand \href@noop [0]{\@secondoftwo}%
\providecommand \href [0]{\begingroup \@sanitize@url \@href}%
\providecommand \@href[1]{\@@startlink{#1}\@@href}%
\providecommand \@@href[1]{\endgroup#1\@@endlink}%
\providecommand \@sanitize@url [0]{\catcode `\\12\catcode `\$12\catcode
  `\&12\catcode `\#12\catcode `\^12\catcode `\_12\catcode `\%12\relax}%
\providecommand \@@startlink[1]{}%
\providecommand \@@endlink[0]{}%
\providecommand \url  [0]{\begingroup\@sanitize@url \@url }%
\providecommand \@url [1]{\endgroup\@href {#1}{\urlprefix }}%
\providecommand \urlprefix  [0]{URL }%
\providecommand \Eprint [0]{\href }%
\providecommand \doibase [0]{http://dx.doi.org/}%
\providecommand \selectlanguage [0]{\@gobble}%
\providecommand \bibinfo  [0]{\@secondoftwo}%
\providecommand \bibfield  [0]{\@secondoftwo}%
\providecommand \translation [1]{[#1]}%
\providecommand \BibitemOpen [0]{}%
\providecommand \bibitemStop [0]{}%
\providecommand \bibitemNoStop [0]{.\EOS\space}%
\providecommand \EOS [0]{\spacefactor3000\relax}%
\providecommand \BibitemShut  [1]{\csname bibitem#1\endcsname}%
\let\auto@bib@innerbib\@empty
\bibitem [{\citenamefont {Isidori}\ \emph {et~al.}(2024)\citenamefont
  {Isidori}, \citenamefont {Wilsch},\ and\ \citenamefont
  {Wyler}}]{Isidori:2023pyp}%
  \BibitemOpen
  \bibfield  {author} {\bibinfo {author} {\bibfnamefont {G.}~\bibnamefont
  {Isidori}}, \bibinfo {author} {\bibfnamefont {F.}~\bibnamefont {Wilsch}}, \
  and\ \bibinfo {author} {\bibfnamefont {D.}~\bibnamefont {Wyler}},\ }\href
  {\doibase 10.1103/RevModPhys.96.015006} {\bibfield  {journal} {\bibinfo
  {journal} {Rev. Mod. Phys.}\ }\textbf {\bibinfo {volume} {96}},\ \bibinfo
  {pages} {015006} (\bibinfo {year} {2024})},\ \Eprint
  {http://arxiv.org/abs/2303.16922} {arXiv:2303.16922 [hep-ph]} \BibitemShut
  {NoStop}%
\bibitem [{\citenamefont {Arkani-Hamed}\ and\ \citenamefont
  {Maldacena}(2015)}]{Arkani-Hamed:2015bza}%
  \BibitemOpen
  \bibfield  {author} {\bibinfo {author} {\bibfnamefont {N.}~\bibnamefont
  {Arkani-Hamed}}\ and\ \bibinfo {author} {\bibfnamefont {J.}~\bibnamefont
  {Maldacena}},\ }\href@noop {} {\  (\bibinfo {year} {2015})},\ \Eprint
  {http://arxiv.org/abs/1503.08043} {arXiv:1503.08043 [hep-th]} \BibitemShut
  {NoStop}%
\bibitem [{\citenamefont {Baumann}\ \emph {et~al.}(2024)\citenamefont
  {Baumann}, \citenamefont {Green}, \citenamefont {Joyce}, \citenamefont
  {Pajer}, \citenamefont {Pimentel}, \citenamefont {Sleight},\ and\
  \citenamefont {Taronna}}]{Baumann:2022jpr}%
  \BibitemOpen
  \bibfield  {author} {\bibinfo {author} {\bibfnamefont {D.}~\bibnamefont
  {Baumann}}, \bibinfo {author} {\bibfnamefont {D.}~\bibnamefont {Green}},
  \bibinfo {author} {\bibfnamefont {A.}~\bibnamefont {Joyce}}, \bibinfo
  {author} {\bibfnamefont {E.}~\bibnamefont {Pajer}}, \bibinfo {author}
  {\bibfnamefont {G.~L.}\ \bibnamefont {Pimentel}}, \bibinfo {author}
  {\bibfnamefont {C.}~\bibnamefont {Sleight}}, \ and\ \bibinfo {author}
  {\bibfnamefont {M.}~\bibnamefont {Taronna}},\ }\href {\doibase
  10.21468/SciPostPhysCommRep.1} {\bibfield  {journal} {\bibinfo  {journal}
  {SciPost Phys. Comm. Rep.}\ }\textbf {\bibinfo {volume} {2024}},\ \bibinfo
  {pages} {1} (\bibinfo {year} {2024})},\ \Eprint
  {http://arxiv.org/abs/2203.08121} {arXiv:2203.08121 [hep-th]} \BibitemShut
  {NoStop}%
\bibitem [{\citenamefont {Ach{\'u}carro}\ \emph {et~al.}(2022)\citenamefont
  {Ach{\'u}carro} \emph {et~al.}}]{Achucarro:2022qrl}%
  \BibitemOpen
  \bibfield  {author} {\bibinfo {author} {\bibfnamefont {A.}~\bibnamefont
  {Ach{\'u}carro}} \emph {et~al.},\ }\href@noop {} {\  (\bibinfo {year}
  {2022})},\ \Eprint {http://arxiv.org/abs/2203.08128} {arXiv:2203.08128
  [astro-ph.CO]} \BibitemShut {NoStop}%
\bibitem [{\citenamefont {Baumann}\ \emph {et~al.}(2016)\citenamefont
  {Baumann}, \citenamefont {Green}, \citenamefont {Lee},\ and\ \citenamefont
  {Porto}}]{Baumann:2015nta}%
  \BibitemOpen
  \bibfield  {author} {\bibinfo {author} {\bibfnamefont {D.}~\bibnamefont
  {Baumann}}, \bibinfo {author} {\bibfnamefont {D.}~\bibnamefont {Green}},
  \bibinfo {author} {\bibfnamefont {H.}~\bibnamefont {Lee}}, \ and\ \bibinfo
  {author} {\bibfnamefont {R.~A.}\ \bibnamefont {Porto}},\ }\href {\doibase
  10.1103/PhysRevD.93.023523} {\bibfield  {journal} {\bibinfo  {journal} {Phys.
  Rev. D}\ }\textbf {\bibinfo {volume} {93}},\ \bibinfo {pages} {023523}
  (\bibinfo {year} {2016})},\ \Eprint {http://arxiv.org/abs/1502.07304}
  {arXiv:1502.07304 [hep-th]} \BibitemShut {NoStop}%
\bibitem [{\citenamefont {Grall}\ and\ \citenamefont
  {Melville}(2020)}]{Grall:2020tqc}%
  \BibitemOpen
  \bibfield  {author} {\bibinfo {author} {\bibfnamefont {T.}~\bibnamefont
  {Grall}}\ and\ \bibinfo {author} {\bibfnamefont {S.}~\bibnamefont
  {Melville}},\ }\href {\doibase 10.1088/1475-7516/2020/09/017} {\bibfield
  {journal} {\bibinfo  {journal} {JCAP}\ }\textbf {\bibinfo {volume} {09}},\
  \bibinfo {pages} {017} (\bibinfo {year} {2020})},\ \Eprint
  {http://arxiv.org/abs/2005.02366} {arXiv:2005.02366 [gr-qc]} \BibitemShut
  {NoStop}%
\bibitem [{\citenamefont {Grall}\ and\ \citenamefont
  {Melville}(2022)}]{Grall:2021xxm}%
  \BibitemOpen
  \bibfield  {author} {\bibinfo {author} {\bibfnamefont {T.}~\bibnamefont
  {Grall}}\ and\ \bibinfo {author} {\bibfnamefont {S.}~\bibnamefont
  {Melville}},\ }\href {\doibase 10.1103/PhysRevD.105.L121301} {\bibfield
  {journal} {\bibinfo  {journal} {Phys. Rev. D}\ }\textbf {\bibinfo {volume}
  {105}},\ \bibinfo {pages} {L121301} (\bibinfo {year} {2022})},\ \Eprint
  {http://arxiv.org/abs/2102.05683} {arXiv:2102.05683 [hep-th]} \BibitemShut
  {NoStop}%
\bibitem [{\citenamefont {Aoki}\ \emph {et~al.}(2021)\citenamefont {Aoki},
  \citenamefont {Mukohyama},\ and\ \citenamefont {Namba}}]{Aoki:2021ffc}%
  \BibitemOpen
  \bibfield  {author} {\bibinfo {author} {\bibfnamefont {K.}~\bibnamefont
  {Aoki}}, \bibinfo {author} {\bibfnamefont {S.}~\bibnamefont {Mukohyama}}, \
  and\ \bibinfo {author} {\bibfnamefont {R.}~\bibnamefont {Namba}},\ }\href
  {\doibase 10.1088/1475-7516/2021/10/079} {\bibfield  {journal} {\bibinfo
  {journal} {JCAP}\ }\textbf {\bibinfo {volume} {10}},\ \bibinfo {pages} {079}
  (\bibinfo {year} {2021})},\ \Eprint {http://arxiv.org/abs/2107.01755}
  {arXiv:2107.01755 [hep-th]} \BibitemShut {NoStop}%
\bibitem [{\citenamefont {Hui}\ \emph {et~al.}(2024)\citenamefont {Hui},
  \citenamefont {Kourkoulou}, \citenamefont {Nicolis}, \citenamefont {Podo},\
  and\ \citenamefont {Zhou}}]{Hui:2023pxc}%
  \BibitemOpen
  \bibfield  {author} {\bibinfo {author} {\bibfnamefont {L.}~\bibnamefont
  {Hui}}, \bibinfo {author} {\bibfnamefont {I.}~\bibnamefont {Kourkoulou}},
  \bibinfo {author} {\bibfnamefont {A.}~\bibnamefont {Nicolis}}, \bibinfo
  {author} {\bibfnamefont {A.}~\bibnamefont {Podo}}, \ and\ \bibinfo {author}
  {\bibfnamefont {S.}~\bibnamefont {Zhou}},\ }\href {\doibase
  10.1007/JHEP04(2024)145} {\bibfield  {journal} {\bibinfo  {journal} {JHEP}\
  }\textbf {\bibinfo {volume} {04}},\ \bibinfo {pages} {145} (\bibinfo {year}
  {2024})},\ \Eprint {http://arxiv.org/abs/2312.08440} {arXiv:2312.08440
  [hep-th]} \BibitemShut {NoStop}%
\bibitem [{\citenamefont {Creminelli}\ \emph
  {et~al.}(2024{\natexlab{a}})\citenamefont {Creminelli}, \citenamefont
  {Delladio}, \citenamefont {Janssen}, \citenamefont {Longo},\ and\
  \citenamefont {Senatore}}]{Creminelli:2023kze}%
  \BibitemOpen
  \bibfield  {author} {\bibinfo {author} {\bibfnamefont {P.}~\bibnamefont
  {Creminelli}}, \bibinfo {author} {\bibfnamefont {M.}~\bibnamefont
  {Delladio}}, \bibinfo {author} {\bibfnamefont {O.}~\bibnamefont {Janssen}},
  \bibinfo {author} {\bibfnamefont {A.}~\bibnamefont {Longo}}, \ and\ \bibinfo
  {author} {\bibfnamefont {L.}~\bibnamefont {Senatore}},\ }\href {\doibase
  10.1007/JHEP06(2024)201} {\bibfield  {journal} {\bibinfo  {journal} {JHEP}\
  }\textbf {\bibinfo {volume} {06}},\ \bibinfo {pages} {201} (\bibinfo {year}
  {2024}{\natexlab{a}})},\ \Eprint {http://arxiv.org/abs/2312.08441}
  {arXiv:2312.08441 [hep-th]} \BibitemShut {NoStop}%
\bibitem [{\citenamefont {Creminelli}\ \emph {et~al.}(2022)\citenamefont
  {Creminelli}, \citenamefont {Janssen},\ and\ \citenamefont
  {Senatore}}]{Creminelli:2022onn}%
  \BibitemOpen
  \bibfield  {author} {\bibinfo {author} {\bibfnamefont {P.}~\bibnamefont
  {Creminelli}}, \bibinfo {author} {\bibfnamefont {O.}~\bibnamefont {Janssen}},
  \ and\ \bibinfo {author} {\bibfnamefont {L.}~\bibnamefont {Senatore}},\
  }\href {\doibase 10.1007/JHEP09(2022)201} {\bibfield  {journal} {\bibinfo
  {journal} {JHEP}\ }\textbf {\bibinfo {volume} {09}},\ \bibinfo {pages} {201}
  (\bibinfo {year} {2022})},\ \Eprint {http://arxiv.org/abs/2207.14224}
  {arXiv:2207.14224 [hep-th]} \BibitemShut {NoStop}%
\bibitem [{\citenamefont {Serra}\ and\ \citenamefont
  {Trombetta}(2024)}]{Serra:2024tmz}%
  \BibitemOpen
  \bibfield  {author} {\bibinfo {author} {\bibfnamefont {F.}~\bibnamefont
  {Serra}}\ and\ \bibinfo {author} {\bibfnamefont {L.~G.}\ \bibnamefont
  {Trombetta}},\ }\href@noop {} {\  (\bibinfo {year} {2024})},\ \Eprint
  {http://arxiv.org/abs/2412.19745} {arXiv:2412.19745 [hep-th]} \BibitemShut
  {NoStop}%
\bibitem [{\citenamefont {Hui}\ \emph {et~al.}(2025)\citenamefont {Hui},
  \citenamefont {Nicolis}, \citenamefont {Podo},\ and\ \citenamefont
  {Zhou}}]{Hui:2025aja}%
  \BibitemOpen
  \bibfield  {author} {\bibinfo {author} {\bibfnamefont {L.}~\bibnamefont
  {Hui}}, \bibinfo {author} {\bibfnamefont {A.}~\bibnamefont {Nicolis}},
  \bibinfo {author} {\bibfnamefont {A.}~\bibnamefont {Podo}}, \ and\ \bibinfo
  {author} {\bibfnamefont {S.}~\bibnamefont {Zhou}},\ }\href {\doibase
  10.1007/JHEP07(2025)188} {\bibfield  {journal} {\bibinfo  {journal} {JHEP}\
  }\textbf {\bibinfo {volume} {07}},\ \bibinfo {pages} {188} (\bibinfo {year}
  {2025})},\ \Eprint {http://arxiv.org/abs/2502.04215} {arXiv:2502.04215
  [hep-th]} \BibitemShut {NoStop}%
\bibitem [{\citenamefont {Creminelli}\ \emph {et~al.}(2025)\citenamefont
  {Creminelli}, \citenamefont {Longo}, \citenamefont {Salehian},\ and\
  \citenamefont {Zahed}}]{Creminelli:2025rxj}%
  \BibitemOpen
  \bibfield  {author} {\bibinfo {author} {\bibfnamefont {P.}~\bibnamefont
  {Creminelli}}, \bibinfo {author} {\bibfnamefont {A.}~\bibnamefont {Longo}},
  \bibinfo {author} {\bibfnamefont {B.}~\bibnamefont {Salehian}}, \ and\
  \bibinfo {author} {\bibfnamefont {A.}~\bibnamefont {Zahed}},\ }\href@noop {}
  {\  (\bibinfo {year} {2025})},\ \Eprint {http://arxiv.org/abs/2512.10843}
  {arXiv:2512.10843 [hep-th]} \BibitemShut {NoStop}%
\bibitem [{\citenamefont {Heller}\ \emph {et~al.}(2023)\citenamefont {Heller},
  \citenamefont {Serantes}, \citenamefont {Spali{\'n}ski},\ and\ \citenamefont
  {Withers}}]{Heller:2022ejw}%
  \BibitemOpen
  \bibfield  {author} {\bibinfo {author} {\bibfnamefont {M.~P.}\ \bibnamefont
  {Heller}}, \bibinfo {author} {\bibfnamefont {A.}~\bibnamefont {Serantes}},
  \bibinfo {author} {\bibfnamefont {M.}~\bibnamefont {Spali{\'n}ski}}, \ and\
  \bibinfo {author} {\bibfnamefont {B.}~\bibnamefont {Withers}},\ }\href
  {\doibase 10.1103/PhysRevLett.130.261601} {\bibfield  {journal} {\bibinfo
  {journal} {Phys. Rev. Lett.}\ }\textbf {\bibinfo {volume} {130}},\ \bibinfo
  {pages} {261601} (\bibinfo {year} {2023})},\ \Eprint
  {http://arxiv.org/abs/2212.07434} {arXiv:2212.07434 [hep-th]} \BibitemShut
  {NoStop}%
\bibitem [{\citenamefont {Heller}\ \emph {et~al.}(2024)\citenamefont {Heller},
  \citenamefont {Serantes}, \citenamefont {Spali{\'n}ski},\ and\ \citenamefont
  {Withers}}]{Heller:2023jtd}%
  \BibitemOpen
  \bibfield  {author} {\bibinfo {author} {\bibfnamefont {M.~P.}\ \bibnamefont
  {Heller}}, \bibinfo {author} {\bibfnamefont {A.}~\bibnamefont {Serantes}},
  \bibinfo {author} {\bibfnamefont {M.}~\bibnamefont {Spali{\'n}ski}}, \ and\
  \bibinfo {author} {\bibfnamefont {B.}~\bibnamefont {Withers}},\ }\href
  {\doibase 10.1038/s41567-024-02635-5} {\bibfield  {journal} {\bibinfo
  {journal} {Nature Phys.}\ }\textbf {\bibinfo {volume} {20}},\ \bibinfo
  {pages} {1948} (\bibinfo {year} {2024})},\ \Eprint
  {http://arxiv.org/abs/2305.07703} {arXiv:2305.07703 [hep-th]} \BibitemShut
  {NoStop}%
\bibitem [{\citenamefont {Creminelli}\ \emph
  {et~al.}(2024{\natexlab{b}})\citenamefont {Creminelli}, \citenamefont
  {Janssen}, \citenamefont {Salehian},\ and\ \citenamefont
  {Senatore}}]{Creminelli:2024lhd}%
  \BibitemOpen
  \bibfield  {author} {\bibinfo {author} {\bibfnamefont {P.}~\bibnamefont
  {Creminelli}}, \bibinfo {author} {\bibfnamefont {O.}~\bibnamefont {Janssen}},
  \bibinfo {author} {\bibfnamefont {B.}~\bibnamefont {Salehian}}, \ and\
  \bibinfo {author} {\bibfnamefont {L.}~\bibnamefont {Senatore}},\ }\href
  {\doibase 10.1007/JHEP08(2024)066} {\bibfield  {journal} {\bibinfo  {journal}
  {JHEP}\ }\textbf {\bibinfo {volume} {08}},\ \bibinfo {pages} {066} (\bibinfo
  {year} {2024}{\natexlab{b}})},\ \Eprint {http://arxiv.org/abs/2405.09614}
  {arXiv:2405.09614 [hep-th]} \BibitemShut {NoStop}%
\bibitem [{\citenamefont {de~Rham}\ and\ \citenamefont
  {Tolley}(2020{\natexlab{a}})}]{deRham:2020zyh}%
  \BibitemOpen
  \bibfield  {author} {\bibinfo {author} {\bibfnamefont {C.}~\bibnamefont
  {de~Rham}}\ and\ \bibinfo {author} {\bibfnamefont {A.~J.}\ \bibnamefont
  {Tolley}},\ }\href {\doibase 10.1103/PhysRevD.102.084048} {\bibfield
  {journal} {\bibinfo  {journal} {Phys. Rev. D}\ }\textbf {\bibinfo {volume}
  {102}},\ \bibinfo {pages} {084048} (\bibinfo {year} {2020}{\natexlab{a}})},\
  \Eprint {http://arxiv.org/abs/2007.01847} {arXiv:2007.01847 [hep-th]}
  \BibitemShut {NoStop}%
\bibitem [{\citenamefont {Bittermann}\ \emph {et~al.}(2023)\citenamefont
  {Bittermann}, \citenamefont {McLoughlin},\ and\ \citenamefont
  {Rosen}}]{Bittermann:2022hhy}%
  \BibitemOpen
  \bibfield  {author} {\bibinfo {author} {\bibfnamefont {N.}~\bibnamefont
  {Bittermann}}, \bibinfo {author} {\bibfnamefont {D.}~\bibnamefont
  {McLoughlin}}, \ and\ \bibinfo {author} {\bibfnamefont {R.~A.}\ \bibnamefont
  {Rosen}},\ }\href {\doibase 10.1088/1361-6382/accc05} {\bibfield  {journal}
  {\bibinfo  {journal} {Class. Quant. Grav.}\ }\textbf {\bibinfo {volume}
  {40}},\ \bibinfo {pages} {115006} (\bibinfo {year} {2023})},\ \Eprint
  {http://arxiv.org/abs/2212.02559} {arXiv:2212.02559 [hep-th]} \BibitemShut
  {NoStop}%
\bibitem [{\citenamefont
  {Carrillo~Gonz{\'a}lez}(2024)}]{CarrilloGonzalez:2023emp}%
  \BibitemOpen
  \bibfield  {author} {\bibinfo {author} {\bibfnamefont {M.}~\bibnamefont
  {Carrillo~Gonz{\'a}lez}},\ }\href {\doibase 10.1103/PhysRevD.109.085008}
  {\bibfield  {journal} {\bibinfo  {journal} {Phys. Rev. D}\ }\textbf {\bibinfo
  {volume} {109}},\ \bibinfo {pages} {085008} (\bibinfo {year} {2024})},\
  \Eprint {http://arxiv.org/abs/2312.07651} {arXiv:2312.07651 [hep-th]}
  \BibitemShut {NoStop}%
\bibitem [{\citenamefont {Carrillo~Gonz{\'a}lez}\ and\ \citenamefont
  {C{\'e}spedes}(2025)}]{CarrilloGonzalez:2025fqq}%
  \BibitemOpen
  \bibfield  {author} {\bibinfo {author} {\bibfnamefont {M.}~\bibnamefont
  {Carrillo~Gonz{\'a}lez}}\ and\ \bibinfo {author} {\bibfnamefont
  {S.}~\bibnamefont {C{\'e}spedes}},\ }\href {\doibase
  10.1088/1475-7516/2025/08/071} {\bibfield  {journal} {\bibinfo  {journal}
  {JCAP}\ }\textbf {\bibinfo {volume} {08}},\ \bibinfo {pages} {071} (\bibinfo
  {year} {2025})},\ \Eprint {http://arxiv.org/abs/2502.19477} {arXiv:2502.19477
  [hep-th]} \BibitemShut {NoStop}%
\bibitem [{\citenamefont {McLoughlin}\ and\ \citenamefont
  {Rosen}(2025)}]{McLoughlin:2025shj}%
  \BibitemOpen
  \bibfield  {author} {\bibinfo {author} {\bibfnamefont {D.}~\bibnamefont
  {McLoughlin}}\ and\ \bibinfo {author} {\bibfnamefont {R.~A.}\ \bibnamefont
  {Rosen}},\ }\href@noop {} {\  (\bibinfo {year} {2025})},\ \Eprint
  {http://arxiv.org/abs/2502.19616} {arXiv:2502.19616 [hep-th]} \BibitemShut
  {NoStop}%
\bibitem [{\citenamefont {de~Rham}\ and\ \citenamefont
  {Tolley}(2020{\natexlab{b}})}]{deRham:2019ctd}%
  \BibitemOpen
  \bibfield  {author} {\bibinfo {author} {\bibfnamefont {C.}~\bibnamefont
  {de~Rham}}\ and\ \bibinfo {author} {\bibfnamefont {A.~J.}\ \bibnamefont
  {Tolley}},\ }\href {\doibase 10.1103/PhysRevD.101.063518} {\bibfield
  {journal} {\bibinfo  {journal} {Phys. Rev. D}\ }\textbf {\bibinfo {volume}
  {101}},\ \bibinfo {pages} {063518} (\bibinfo {year} {2020}{\natexlab{b}})},\
  \Eprint {http://arxiv.org/abs/1909.00881} {arXiv:1909.00881 [hep-th]}
  \BibitemShut {NoStop}%
\bibitem [{\citenamefont {Chen}\ \emph {et~al.}(2022)\citenamefont {Chen},
  \citenamefont {de~Rham}, \citenamefont {Margalit},\ and\ \citenamefont
  {Tolley}}]{Chen:2021bvg}%
  \BibitemOpen
  \bibfield  {author} {\bibinfo {author} {\bibfnamefont {C.~Y.~R.}\
  \bibnamefont {Chen}}, \bibinfo {author} {\bibfnamefont {C.}~\bibnamefont
  {de~Rham}}, \bibinfo {author} {\bibfnamefont {A.}~\bibnamefont {Margalit}}, \
  and\ \bibinfo {author} {\bibfnamefont {A.~J.}\ \bibnamefont {Tolley}},\
  }\href {\doibase 10.1007/JHEP03(2022)025} {\bibfield  {journal} {\bibinfo
  {journal} {JHEP}\ }\textbf {\bibinfo {volume} {03}},\ \bibinfo {pages} {025}
  (\bibinfo {year} {2022})},\ \Eprint {http://arxiv.org/abs/2112.05031}
  {arXiv:2112.05031 [hep-th]} \BibitemShut {NoStop}%
\bibitem [{\citenamefont {de~Rham}\ \emph {et~al.}(2022)\citenamefont
  {de~Rham}, \citenamefont {Tolley},\ and\ \citenamefont
  {Zhang}}]{deRham:2021bll}%
  \BibitemOpen
  \bibfield  {author} {\bibinfo {author} {\bibfnamefont {C.}~\bibnamefont
  {de~Rham}}, \bibinfo {author} {\bibfnamefont {A.~J.}\ \bibnamefont {Tolley}},
  \ and\ \bibinfo {author} {\bibfnamefont {J.}~\bibnamefont {Zhang}},\ }\href
  {\doibase 10.1103/PhysRevLett.128.131102} {\bibfield  {journal} {\bibinfo
  {journal} {Phys. Rev. Lett.}\ }\textbf {\bibinfo {volume} {128}},\ \bibinfo
  {pages} {131102} (\bibinfo {year} {2022})},\ \Eprint
  {http://arxiv.org/abs/2112.05054} {arXiv:2112.05054 [gr-qc]} \BibitemShut
  {NoStop}%
\bibitem [{\citenamefont {Bellazzini}\ \emph {et~al.}(2022)\citenamefont
  {Bellazzini}, \citenamefont {Isabella}, \citenamefont {Lewandowski},\ and\
  \citenamefont {Sgarlata}}]{Bellazzini:2021shn}%
  \BibitemOpen
  \bibfield  {author} {\bibinfo {author} {\bibfnamefont {B.}~\bibnamefont
  {Bellazzini}}, \bibinfo {author} {\bibfnamefont {G.}~\bibnamefont
  {Isabella}}, \bibinfo {author} {\bibfnamefont {M.}~\bibnamefont
  {Lewandowski}}, \ and\ \bibinfo {author} {\bibfnamefont {F.}~\bibnamefont
  {Sgarlata}},\ }\href {\doibase 10.1007/JHEP05(2022)154} {\bibfield  {journal}
  {\bibinfo  {journal} {JHEP}\ }\textbf {\bibinfo {volume} {05}},\ \bibinfo
  {pages} {154} (\bibinfo {year} {2022})},\ \Eprint
  {http://arxiv.org/abs/2108.05896} {arXiv:2108.05896 [hep-th]} \BibitemShut
  {NoStop}%
\bibitem [{\citenamefont {Carrillo~Gonzalez}\ \emph {et~al.}(2022)\citenamefont
  {Carrillo~Gonzalez}, \citenamefont {de~Rham}, \citenamefont {Pozsgay},\ and\
  \citenamefont {Tolley}}]{CarrilloGonzalez:2022fwg}%
  \BibitemOpen
  \bibfield  {author} {\bibinfo {author} {\bibfnamefont {M.}~\bibnamefont
  {Carrillo~Gonzalez}}, \bibinfo {author} {\bibfnamefont {C.}~\bibnamefont
  {de~Rham}}, \bibinfo {author} {\bibfnamefont {V.}~\bibnamefont {Pozsgay}}, \
  and\ \bibinfo {author} {\bibfnamefont {A.~J.}\ \bibnamefont {Tolley}},\
  }\href {\doibase 10.1103/PhysRevD.106.105018} {\bibfield  {journal} {\bibinfo
   {journal} {Phys. Rev. D}\ }\textbf {\bibinfo {volume} {106}},\ \bibinfo
  {pages} {105018} (\bibinfo {year} {2022})},\ \Eprint
  {http://arxiv.org/abs/2207.03491} {arXiv:2207.03491 [hep-th]} \BibitemShut
  {NoStop}%
\bibitem [{\citenamefont {Carrillo~Gonz{\'a}lez}\ \emph
  {et~al.}(2024)\citenamefont {Carrillo~Gonz{\'a}lez}, \citenamefont {de~Rham},
  \citenamefont {Jaitly}, \citenamefont {Pozsgay},\ and\ \citenamefont
  {Tokareva}}]{CarrilloGonzalez:2023cbf}%
  \BibitemOpen
  \bibfield  {author} {\bibinfo {author} {\bibfnamefont {M.}~\bibnamefont
  {Carrillo~Gonz{\'a}lez}}, \bibinfo {author} {\bibfnamefont {C.}~\bibnamefont
  {de~Rham}}, \bibinfo {author} {\bibfnamefont {S.}~\bibnamefont {Jaitly}},
  \bibinfo {author} {\bibfnamefont {V.}~\bibnamefont {Pozsgay}}, \ and\
  \bibinfo {author} {\bibfnamefont {A.}~\bibnamefont {Tokareva}},\ }\href
  {\doibase 10.1007/JHEP06(2024)146} {\bibfield  {journal} {\bibinfo  {journal}
  {JHEP}\ }\textbf {\bibinfo {volume} {06}},\ \bibinfo {pages} {146} (\bibinfo
  {year} {2024})},\ \Eprint {http://arxiv.org/abs/2307.04784} {arXiv:2307.04784
  [hep-th]} \BibitemShut {NoStop}%
\bibitem [{\citenamefont {Chen}\ \emph {et~al.}(2025)\citenamefont {Chen},
  \citenamefont {Margalit}, \citenamefont {de~Rham},\ and\ \citenamefont
  {Tolley}}]{Chen:2023rar}%
  \BibitemOpen
  \bibfield  {author} {\bibinfo {author} {\bibfnamefont {C.~Y.~R.}\
  \bibnamefont {Chen}}, \bibinfo {author} {\bibfnamefont {A.}~\bibnamefont
  {Margalit}}, \bibinfo {author} {\bibfnamefont {C.}~\bibnamefont {de~Rham}}, \
  and\ \bibinfo {author} {\bibfnamefont {A.~J.}\ \bibnamefont {Tolley}},\
  }\href {\doibase 10.1103/PhysRevD.111.024066} {\bibfield  {journal} {\bibinfo
   {journal} {Phys. Rev. D}\ }\textbf {\bibinfo {volume} {111}},\ \bibinfo
  {pages} {024066} (\bibinfo {year} {2025})},\ \Eprint
  {http://arxiv.org/abs/2309.04534} {arXiv:2309.04534 [hep-th]} \BibitemShut
  {NoStop}%
\bibitem [{\citenamefont {Melville}(2024)}]{Melville:2024zjq}%
  \BibitemOpen
  \bibfield  {author} {\bibinfo {author} {\bibfnamefont {S.}~\bibnamefont
  {Melville}},\ }\href {\doibase 10.1140/epjp/s13360-024-05520-5} {\bibfield
  {journal} {\bibinfo  {journal} {Eur. Phys. J. Plus}\ }\textbf {\bibinfo
  {volume} {139}},\ \bibinfo {pages} {725} (\bibinfo {year} {2024})},\ \Eprint
  {http://arxiv.org/abs/2401.05524} {arXiv:2401.05524 [gr-qc]} \BibitemShut
  {NoStop}%
\bibitem [{\citenamefont {Hogervorst}\ \emph {et~al.}(2023)\citenamefont
  {Hogervorst}, \citenamefont {Penedones},\ and\ \citenamefont
  {Vaziri}}]{Hogervorst:2021uvp}%
  \BibitemOpen
  \bibfield  {author} {\bibinfo {author} {\bibfnamefont {M.}~\bibnamefont
  {Hogervorst}}, \bibinfo {author} {\bibfnamefont {J.}~\bibnamefont
  {Penedones}}, \ and\ \bibinfo {author} {\bibfnamefont {K.~S.}\ \bibnamefont
  {Vaziri}},\ }\href {\doibase 10.1007/JHEP02(2023)162} {\bibfield  {journal}
  {\bibinfo  {journal} {JHEP}\ }\textbf {\bibinfo {volume} {02}},\ \bibinfo
  {pages} {162} (\bibinfo {year} {2023})},\ \Eprint
  {http://arxiv.org/abs/2107.13871} {arXiv:2107.13871 [hep-th]} \BibitemShut
  {NoStop}%
\bibitem [{\citenamefont {Green}\ \emph {et~al.}(2024)\citenamefont {Green},
  \citenamefont {Huang}, \citenamefont {Shen},\ and\ \citenamefont
  {Baumann}}]{Green:2023ids}%
  \BibitemOpen
  \bibfield  {author} {\bibinfo {author} {\bibfnamefont {D.}~\bibnamefont
  {Green}}, \bibinfo {author} {\bibfnamefont {Y.}~\bibnamefont {Huang}},
  \bibinfo {author} {\bibfnamefont {C.-H.}\ \bibnamefont {Shen}}, \ and\
  \bibinfo {author} {\bibfnamefont {D.}~\bibnamefont {Baumann}},\ }\href
  {\doibase 10.1007/JHEP04(2024)034} {\bibfield  {journal} {\bibinfo  {journal}
  {JHEP}\ }\textbf {\bibinfo {volume} {04}},\ \bibinfo {pages} {034} (\bibinfo
  {year} {2024})},\ \Eprint {http://arxiv.org/abs/2310.02490} {arXiv:2310.02490
  [hep-th]} \BibitemShut {NoStop}%
\bibitem [{\citenamefont {Chakraborty}\ \emph {et~al.}(2025)\citenamefont
  {Chakraborty}, \citenamefont {Cohen}, \citenamefont {Green},\ and\
  \citenamefont {Huang}}]{Chakraborty:2025mhh}%
  \BibitemOpen
  \bibfield  {author} {\bibinfo {author} {\bibfnamefont {P.}~\bibnamefont
  {Chakraborty}}, \bibinfo {author} {\bibfnamefont {T.}~\bibnamefont {Cohen}},
  \bibinfo {author} {\bibfnamefont {D.}~\bibnamefont {Green}}, \ and\ \bibinfo
  {author} {\bibfnamefont {Y.}~\bibnamefont {Huang}},\ }\href@noop {} {\
  (\bibinfo {year} {2025})},\ \Eprint {http://arxiv.org/abs/2508.08359}
  {arXiv:2508.08359 [hep-th]} \BibitemShut {NoStop}%
\bibitem [{\citenamefont {de~Rham}\ \emph {et~al.}(2025)\citenamefont
  {de~Rham}, \citenamefont {Jazayeri},\ and\ \citenamefont
  {Tolley}}]{deRham:2025mjh}%
  \BibitemOpen
  \bibfield  {author} {\bibinfo {author} {\bibfnamefont {C.}~\bibnamefont
  {de~Rham}}, \bibinfo {author} {\bibfnamefont {S.}~\bibnamefont {Jazayeri}}, \
  and\ \bibinfo {author} {\bibfnamefont {A.~J.}\ \bibnamefont {Tolley}},\
  }\href {\doibase 10.1103/q6rq-sj9t} {\bibfield  {journal} {\bibinfo
  {journal} {Phys. Rev. D}\ }\textbf {\bibinfo {volume} {112}},\ \bibinfo
  {pages} {083531} (\bibinfo {year} {2025})},\ \Eprint
  {http://arxiv.org/abs/2506.19198} {arXiv:2506.19198 [hep-th]} \BibitemShut
  {NoStop}%
\bibitem [{\citenamefont {Arkani-Hamed}\ \emph {et~al.}(2021)\citenamefont
  {Arkani-Hamed}, \citenamefont {Huang},\ and\ \citenamefont
  {Huang}}]{Arkani-Hamed:2020blm}%
  \BibitemOpen
  \bibfield  {author} {\bibinfo {author} {\bibfnamefont {N.}~\bibnamefont
  {Arkani-Hamed}}, \bibinfo {author} {\bibfnamefont {T.-C.}\ \bibnamefont
  {Huang}}, \ and\ \bibinfo {author} {\bibfnamefont {Y.-t.}\ \bibnamefont
  {Huang}},\ }\href {\doibase 10.1007/JHEP05(2021)259} {\bibfield  {journal}
  {\bibinfo  {journal} {JHEP}\ }\textbf {\bibinfo {volume} {05}},\ \bibinfo
  {pages} {259} (\bibinfo {year} {2021})},\ \Eprint
  {http://arxiv.org/abs/2012.15849} {arXiv:2012.15849 [hep-th]} \BibitemShut
  {NoStop}%
\bibitem [{\citenamefont {Chiang}\ \emph {et~al.}(2022)\citenamefont {Chiang},
  \citenamefont {Huang}, \citenamefont {Li}, \citenamefont {Rodina},\ and\
  \citenamefont {Weng}}]{Chiang:2021ziz}%
  \BibitemOpen
  \bibfield  {author} {\bibinfo {author} {\bibfnamefont {L.-Y.}\ \bibnamefont
  {Chiang}}, \bibinfo {author} {\bibfnamefont {Y.-t.}\ \bibnamefont {Huang}},
  \bibinfo {author} {\bibfnamefont {W.}~\bibnamefont {Li}}, \bibinfo {author}
  {\bibfnamefont {L.}~\bibnamefont {Rodina}}, \ and\ \bibinfo {author}
  {\bibfnamefont {H.-C.}\ \bibnamefont {Weng}},\ }\href {\doibase
  10.1007/JHEP03(2022)063} {\bibfield  {journal} {\bibinfo  {journal} {JHEP}\
  }\textbf {\bibinfo {volume} {03}},\ \bibinfo {pages} {063} (\bibinfo {year}
  {2022})},\ \Eprint {http://arxiv.org/abs/2105.02862} {arXiv:2105.02862
  [hep-th]} \BibitemShut {NoStop}%
\bibitem [{\citenamefont {Arkani-Hamed}\ and\ \citenamefont
  {Trnka}(2014)}]{Arkani-Hamed:2013jha}%
  \BibitemOpen
  \bibfield  {author} {\bibinfo {author} {\bibfnamefont {N.}~\bibnamefont
  {Arkani-Hamed}}\ and\ \bibinfo {author} {\bibfnamefont {J.}~\bibnamefont
  {Trnka}},\ }\href {\doibase 10.1007/JHEP10(2014)030} {\bibfield  {journal}
  {\bibinfo  {journal} {JHEP}\ }\textbf {\bibinfo {volume} {10}},\ \bibinfo
  {pages} {030} (\bibinfo {year} {2014})},\ \Eprint
  {http://arxiv.org/abs/1312.2007} {arXiv:1312.2007 [hep-th]} \BibitemShut
  {NoStop}%
\bibitem [{\citenamefont {Arkani-Hamed}\ \emph {et~al.}(2017)\citenamefont
  {Arkani-Hamed}, \citenamefont {Benincasa},\ and\ \citenamefont
  {Postnikov}}]{Arkani-Hamed:2017fdk}%
  \BibitemOpen
  \bibfield  {author} {\bibinfo {author} {\bibfnamefont {N.}~\bibnamefont
  {Arkani-Hamed}}, \bibinfo {author} {\bibfnamefont {P.}~\bibnamefont
  {Benincasa}}, \ and\ \bibinfo {author} {\bibfnamefont {A.}~\bibnamefont
  {Postnikov}},\ }\href@noop {} {\  (\bibinfo {year} {2017})},\ \Eprint
  {http://arxiv.org/abs/1709.02813} {arXiv:1709.02813 [hep-th]} \BibitemShut
  {NoStop}%
\bibitem [{\citenamefont {Arkani-Hamed}\ \emph {et~al.}(2025)\citenamefont
  {Arkani-Hamed}, \citenamefont {Figueiredo},\ and\ \citenamefont
  {Vaz{\~a}o}}]{Arkani-Hamed:2024jbp}%
  \BibitemOpen
  \bibfield  {author} {\bibinfo {author} {\bibfnamefont {N.}~\bibnamefont
  {Arkani-Hamed}}, \bibinfo {author} {\bibfnamefont {C.}~\bibnamefont
  {Figueiredo}}, \ and\ \bibinfo {author} {\bibfnamefont {F.}~\bibnamefont
  {Vaz{\~a}o}},\ }\href {\doibase 10.1007/JHEP11(2025)029} {\bibfield
  {journal} {\bibinfo  {journal} {JHEP}\ }\textbf {\bibinfo {volume} {11}},\
  \bibinfo {pages} {029} (\bibinfo {year} {2025})},\ \Eprint
  {http://arxiv.org/abs/2412.19881} {arXiv:2412.19881 [hep-th]} \BibitemShut
  {NoStop}%
\bibitem [{Note1()}]{Note1}%
  \BibitemOpen
  \bibinfo {note} {Explicitly, our mode functions are \begin {align} f_\mu ^+ (
  k \tau ) &= \protect \frac {\protect \sqrt {\pi }}{2 i} e^{+ \pi \mu /2} H_{i
  \mu }^{(2)} ( - k \tau ) \protect \tmspace +\thickmuskip {.2777em} , \\
  f^-_\mu ( k \tau ) &=\protect \frac {i \protect \sqrt {\pi }}{2} e^{- \pi \mu
  /2} H_{i \mu }^{(1)} ( - k \tau ) \protect \tmspace +\thickmuskip {.2777em} .
  \end {align} and obey $f^+_\mu = (f^-_\mu )^*$.}\BibitemShut {Stop}%
\bibitem [{\citenamefont {Di~Pietro}\ \emph {et~al.}(2022)\citenamefont
  {Di~Pietro}, \citenamefont {Gorbenko},\ and\ \citenamefont
  {Komatsu}}]{DiPietro:2021sjt}%
  \BibitemOpen
  \bibfield  {author} {\bibinfo {author} {\bibfnamefont {L.}~\bibnamefont
  {Di~Pietro}}, \bibinfo {author} {\bibfnamefont {V.}~\bibnamefont {Gorbenko}},
  \ and\ \bibinfo {author} {\bibfnamefont {S.}~\bibnamefont {Komatsu}},\ }\href
  {\doibase 10.1007/JHEP03(2022)023} {\bibfield  {journal} {\bibinfo  {journal}
  {JHEP}\ }\textbf {\bibinfo {volume} {03}},\ \bibinfo {pages} {023} (\bibinfo
  {year} {2022})},\ \Eprint {http://arxiv.org/abs/2108.01695} {arXiv:2108.01695
  [hep-th]} \BibitemShut {NoStop}%
\bibitem [{\citenamefont {Melville}\ and\ \citenamefont
  {Pimentel}(2024{\natexlab{a}})}]{Melville:2024ove}%
  \BibitemOpen
  \bibfield  {author} {\bibinfo {author} {\bibfnamefont {S.}~\bibnamefont
  {Melville}}\ and\ \bibinfo {author} {\bibfnamefont {G.~L.}\ \bibnamefont
  {Pimentel}},\ }\href {\doibase 10.1007/JHEP08(2024)211} {\bibfield  {journal}
  {\bibinfo  {journal} {JHEP}\ }\textbf {\bibinfo {volume} {08}},\ \bibinfo
  {pages} {211} (\bibinfo {year} {2024}{\natexlab{a}})},\ \Eprint
  {http://arxiv.org/abs/2404.05712} {arXiv:2404.05712 [hep-th]} \BibitemShut
  {NoStop}%
\bibitem [{Note2()}]{Note2}%
  \BibitemOpen
  \bibinfo {note} {Concretely, consider a diagrammatic expansion of the in-in
  path integral in which each vertex is labelled by $\pm $ according to its
  branch of the integration contour. At leading order, this gives $G = \Pi
  ^{++} + \DOTSB \sum@ \slimits@ _{\alpha , \beta } \Pi ^{+\alpha } \Sigma
  ^{\alpha \beta } \Pi ^{\beta +} + ... $, where $\Pi ^{\alpha \beta }$ are the
  free theory propagators for $\phi $ with different operator orderings.
  Comparing with the leading order expansion of \protect \textup {\hbox
  {\mathsurround \z@ \protect \normalfont (\ignorespaces \ref
  {eqn:self-energy_form}\unskip \@@italiccorr )}} leads to \protect \textup
  {\hbox {\mathsurround \z@ \protect \normalfont (\ignorespaces \ref
  {eqn:Sigma_def}\unskip \@@italiccorr )}}. Some combinatorics then shows that
  \protect \textup {\hbox {\mathsurround \z@ \protect \normalfont
  (\ignorespaces \ref {eqn:self-energy_form}\unskip \@@italiccorr )}} will
  match the in-in sum over vertices at every order in $\Sigma $.}\BibitemShut
  {Stop}%
\bibitem [{Note3()}]{Note3}%
  \BibitemOpen
  \bibinfo {note} {Here we refer to the causality condition that the analytic
  structure of $G(\nu )$, which encodes the time-ordering, remains the same in
  the interacting theory: namely an even combination of $e^{\pm \pi \nu }$
  terms which have singularities in the upper/lower half planes
  respectively.}\BibitemShut {Stop}%
\bibitem [{\citenamefont {Marolf}\ and\ \citenamefont
  {Morrison}(2010)}]{Marolf:2010zp}%
  \BibitemOpen
  \bibfield  {author} {\bibinfo {author} {\bibfnamefont {D.}~\bibnamefont
  {Marolf}}\ and\ \bibinfo {author} {\bibfnamefont {I.~A.}\ \bibnamefont
  {Morrison}},\ }\href {\doibase 10.1103/PhysRevD.82.105032} {\bibfield
  {journal} {\bibinfo  {journal} {Phys. Rev. D}\ }\textbf {\bibinfo {volume}
  {82}},\ \bibinfo {pages} {105032} (\bibinfo {year} {2010})},\ \Eprint
  {http://arxiv.org/abs/1006.0035} {arXiv:1006.0035 [gr-qc]} \BibitemShut
  {NoStop}%
\bibitem [{\citenamefont {Higuchi}\ \emph {et~al.}(2011)\citenamefont
  {Higuchi}, \citenamefont {Marolf},\ and\ \citenamefont
  {Morrison}}]{Higuchi:2010xt}%
  \BibitemOpen
  \bibfield  {author} {\bibinfo {author} {\bibfnamefont {A.}~\bibnamefont
  {Higuchi}}, \bibinfo {author} {\bibfnamefont {D.}~\bibnamefont {Marolf}}, \
  and\ \bibinfo {author} {\bibfnamefont {I.~A.}\ \bibnamefont {Morrison}},\
  }\href {\doibase 10.1103/PhysRevD.83.084029} {\bibfield  {journal} {\bibinfo
  {journal} {Phys. Rev. D}\ }\textbf {\bibinfo {volume} {83}},\ \bibinfo
  {pages} {084029} (\bibinfo {year} {2011})},\ \Eprint
  {http://arxiv.org/abs/1012.3415} {arXiv:1012.3415 [gr-qc]} \BibitemShut
  {NoStop}%
\bibitem [{\citenamefont {Sleight}\ and\ \citenamefont
  {Taronna}(2021{\natexlab{a}})}]{Sleight:2020obc}%
  \BibitemOpen
  \bibfield  {author} {\bibinfo {author} {\bibfnamefont {C.}~\bibnamefont
  {Sleight}}\ and\ \bibinfo {author} {\bibfnamefont {M.}~\bibnamefont
  {Taronna}},\ }\href {\doibase 10.1103/PhysRevD.104.L081902} {\bibfield
  {journal} {\bibinfo  {journal} {Phys. Rev. D}\ }\textbf {\bibinfo {volume}
  {104}},\ \bibinfo {pages} {L081902} (\bibinfo {year} {2021}{\natexlab{a}})},\
  \Eprint {http://arxiv.org/abs/2007.09993} {arXiv:2007.09993 [hep-th]}
  \BibitemShut {NoStop}%
\bibitem [{\citenamefont {Sleight}\ and\ \citenamefont
  {Taronna}(2021{\natexlab{b}})}]{Sleight:2021plv}%
  \BibitemOpen
  \bibfield  {author} {\bibinfo {author} {\bibfnamefont {C.}~\bibnamefont
  {Sleight}}\ and\ \bibinfo {author} {\bibfnamefont {M.}~\bibnamefont
  {Taronna}},\ }\href {\doibase 10.1007/JHEP12(2021)074} {\bibfield  {journal}
  {\bibinfo  {journal} {JHEP}\ }\textbf {\bibinfo {volume} {12}},\ \bibinfo
  {pages} {074} (\bibinfo {year} {2021}{\natexlab{b}})},\ \Eprint
  {http://arxiv.org/abs/2109.02725} {arXiv:2109.02725 [hep-th]} \BibitemShut
  {NoStop}%
\bibitem [{\citenamefont {Abhishek}\ \emph {et~al.}(2025)\citenamefont
  {Abhishek}, \citenamefont {Sleight},\ and\ \citenamefont
  {Taronna}}]{MdAbhishek:2025dhx}%
  \BibitemOpen
  \bibfield  {author} {\bibinfo {author} {\bibfnamefont {M.}~\bibnamefont
  {Abhishek}}, \bibinfo {author} {\bibfnamefont {C.}~\bibnamefont {Sleight}}, \
  and\ \bibinfo {author} {\bibfnamefont {M.}~\bibnamefont {Taronna}},\
  }\href@noop {} {\  (\bibinfo {year} {2025})},\ \Eprint
  {http://arxiv.org/abs/2509.09536} {arXiv:2509.09536 [hep-th]} \BibitemShut
  {NoStop}%
\bibitem [{\citenamefont {Collins}\ \emph {et~al.}(2013)\citenamefont
  {Collins}, \citenamefont {Holman},\ and\ \citenamefont
  {Ross}}]{Collins:2012nq}%
  \BibitemOpen
  \bibfield  {author} {\bibinfo {author} {\bibfnamefont {H.}~\bibnamefont
  {Collins}}, \bibinfo {author} {\bibfnamefont {R.}~\bibnamefont {Holman}}, \
  and\ \bibinfo {author} {\bibfnamefont {A.}~\bibnamefont {Ross}},\ }\href
  {\doibase 10.1007/JHEP02(2013)108} {\bibfield  {journal} {\bibinfo  {journal}
  {JHEP}\ }\textbf {\bibinfo {volume} {02}},\ \bibinfo {pages} {108} (\bibinfo
  {year} {2013})},\ \Eprint {http://arxiv.org/abs/1208.3255} {arXiv:1208.3255
  [hep-th]} \BibitemShut {NoStop}%
\bibitem [{\citenamefont {Burgess}\ \emph {et~al.}(2016)\citenamefont
  {Burgess}, \citenamefont {Holman},\ and\ \citenamefont
  {Tasinato}}]{Burgess:2015ajz}%
  \BibitemOpen
  \bibfield  {author} {\bibinfo {author} {\bibfnamefont {C.~P.}\ \bibnamefont
  {Burgess}}, \bibinfo {author} {\bibfnamefont {R.}~\bibnamefont {Holman}}, \
  and\ \bibinfo {author} {\bibfnamefont {G.}~\bibnamefont {Tasinato}},\ }\href
  {\doibase 10.1007/JHEP01(2016)153} {\bibfield  {journal} {\bibinfo  {journal}
  {JHEP}\ }\textbf {\bibinfo {volume} {01}},\ \bibinfo {pages} {153} (\bibinfo
  {year} {2016})},\ \Eprint {http://arxiv.org/abs/1512.00169} {arXiv:1512.00169
  [gr-qc]} \BibitemShut {NoStop}%
\bibitem [{\citenamefont {Hongo}\ \emph {et~al.}(2019)\citenamefont {Hongo},
  \citenamefont {Kim}, \citenamefont {Noumi},\ and\ \citenamefont
  {Ota}}]{Hongo:2018ant}%
  \BibitemOpen
  \bibfield  {author} {\bibinfo {author} {\bibfnamefont {M.}~\bibnamefont
  {Hongo}}, \bibinfo {author} {\bibfnamefont {S.}~\bibnamefont {Kim}}, \bibinfo
  {author} {\bibfnamefont {T.}~\bibnamefont {Noumi}}, \ and\ \bibinfo {author}
  {\bibfnamefont {A.}~\bibnamefont {Ota}},\ }\href {\doibase
  10.1007/JHEP02(2019)131} {\bibfield  {journal} {\bibinfo  {journal} {JHEP}\
  }\textbf {\bibinfo {volume} {02}},\ \bibinfo {pages} {131} (\bibinfo {year}
  {2019})},\ \Eprint {http://arxiv.org/abs/1805.06240} {arXiv:1805.06240
  [hep-th]} \BibitemShut {NoStop}%
\bibitem [{\citenamefont {Burgess}\ and\ \citenamefont
  {Kaplanek}(2022)}]{Burgess:2022rdo}%
  \BibitemOpen
  \bibfield  {author} {\bibinfo {author} {\bibfnamefont {C.~P.}\ \bibnamefont
  {Burgess}}\ and\ \bibinfo {author} {\bibfnamefont {G.}~\bibnamefont
  {Kaplanek}},\ }\href {\doibase 10.1007/978-981-19-3079-9\_7-1} {\  (\bibinfo
  {year} {2022}),\ 10.1007/978-981-19-3079-9\_7-1},\ \Eprint
  {http://arxiv.org/abs/2212.09157} {arXiv:2212.09157 [hep-th]} \BibitemShut
  {NoStop}%
\bibitem [{\citenamefont {Burgess}\ \emph {et~al.}(2023)\citenamefont
  {Burgess}, \citenamefont {Holman}, \citenamefont {Kaplanek}, \citenamefont
  {Martin},\ and\ \citenamefont {Vennin}}]{Burgess:2022nwu}%
  \BibitemOpen
  \bibfield  {author} {\bibinfo {author} {\bibfnamefont {C.~P.}\ \bibnamefont
  {Burgess}}, \bibinfo {author} {\bibfnamefont {R.}~\bibnamefont {Holman}},
  \bibinfo {author} {\bibfnamefont {G.}~\bibnamefont {Kaplanek}}, \bibinfo
  {author} {\bibfnamefont {J.}~\bibnamefont {Martin}}, \ and\ \bibinfo {author}
  {\bibfnamefont {V.}~\bibnamefont {Vennin}},\ }\href {\doibase
  10.1088/1475-7516/2023/07/022} {\bibfield  {journal} {\bibinfo  {journal}
  {JCAP}\ }\textbf {\bibinfo {volume} {07}},\ \bibinfo {pages} {022} (\bibinfo
  {year} {2023})},\ \Eprint {http://arxiv.org/abs/2211.11046} {arXiv:2211.11046
  [hep-th]} \BibitemShut {NoStop}%
\bibitem [{\citenamefont {Salcedo}\ \emph {et~al.}(2024)\citenamefont
  {Salcedo}, \citenamefont {Colas},\ and\ \citenamefont
  {Pajer}}]{Salcedo:2024smn}%
  \BibitemOpen
  \bibfield  {author} {\bibinfo {author} {\bibfnamefont {S.~A.}\ \bibnamefont
  {Salcedo}}, \bibinfo {author} {\bibfnamefont {T.}~\bibnamefont {Colas}}, \
  and\ \bibinfo {author} {\bibfnamefont {E.}~\bibnamefont {Pajer}},\ }\href
  {\doibase 10.1007/JHEP10(2024)248} {\bibfield  {journal} {\bibinfo  {journal}
  {JHEP}\ }\textbf {\bibinfo {volume} {10}},\ \bibinfo {pages} {248} (\bibinfo
  {year} {2024})},\ \Eprint {http://arxiv.org/abs/2404.15416} {arXiv:2404.15416
  [hep-th]} \BibitemShut {NoStop}%
\bibitem [{\citenamefont {Salcedo}\ \emph
  {et~al.}(2025{\natexlab{a}})\citenamefont {Salcedo}, \citenamefont {Colas},\
  and\ \citenamefont {Pajer}}]{Salcedo:2024nex}%
  \BibitemOpen
  \bibfield  {author} {\bibinfo {author} {\bibfnamefont {S.~A.}\ \bibnamefont
  {Salcedo}}, \bibinfo {author} {\bibfnamefont {T.}~\bibnamefont {Colas}}, \
  and\ \bibinfo {author} {\bibfnamefont {E.}~\bibnamefont {Pajer}},\ }\href
  {\doibase 10.1007/JHEP03(2025)138} {\bibfield  {journal} {\bibinfo  {journal}
  {JHEP}\ }\textbf {\bibinfo {volume} {03}},\ \bibinfo {pages} {138} (\bibinfo
  {year} {2025}{\natexlab{a}})},\ \Eprint {http://arxiv.org/abs/2412.12299}
  {arXiv:2412.12299 [hep-th]} \BibitemShut {NoStop}%
\bibitem [{\citenamefont {Salcedo}\ \emph
  {et~al.}(2025{\natexlab{b}})\citenamefont {Salcedo}, \citenamefont {Colas},
  \citenamefont {Dufner},\ and\ \citenamefont {Pajer}}]{Salcedo:2025ezu}%
  \BibitemOpen
  \bibfield  {author} {\bibinfo {author} {\bibfnamefont {S.~A.}\ \bibnamefont
  {Salcedo}}, \bibinfo {author} {\bibfnamefont {T.}~\bibnamefont {Colas}},
  \bibinfo {author} {\bibfnamefont {L.}~\bibnamefont {Dufner}}, \ and\ \bibinfo
  {author} {\bibfnamefont {E.}~\bibnamefont {Pajer}},\ }\href@noop {} {\
  (\bibinfo {year} {2025}{\natexlab{b}})},\ \Eprint
  {http://arxiv.org/abs/2507.03103} {arXiv:2507.03103 [hep-th]} \BibitemShut
  {NoStop}%
\bibitem [{\citenamefont {Colas}\ \emph {et~al.}(2025)\citenamefont {Colas},
  \citenamefont {Qin},\ and\ \citenamefont {Tong}}]{Colas:2025ind}%
  \BibitemOpen
  \bibfield  {author} {\bibinfo {author} {\bibfnamefont {T.}~\bibnamefont
  {Colas}}, \bibinfo {author} {\bibfnamefont {Z.}~\bibnamefont {Qin}}, \ and\
  \bibinfo {author} {\bibfnamefont {X.}~\bibnamefont {Tong}},\ }\href@noop {}
  {\  (\bibinfo {year} {2025})},\ \Eprint {http://arxiv.org/abs/2512.07941}
  {arXiv:2512.07941 [hep-th]} \BibitemShut {NoStop}%
\bibitem [{\citenamefont {Salcedo}\ \emph {et~al.}(2023)\citenamefont
  {Salcedo}, \citenamefont {Lee}, \citenamefont {Melville},\ and\ \citenamefont
  {Pajer}}]{Salcedo:2022aal}%
  \BibitemOpen
  \bibfield  {author} {\bibinfo {author} {\bibfnamefont {S.~A.}\ \bibnamefont
  {Salcedo}}, \bibinfo {author} {\bibfnamefont {M.~H.~G.}\ \bibnamefont {Lee}},
  \bibinfo {author} {\bibfnamefont {S.}~\bibnamefont {Melville}}, \ and\
  \bibinfo {author} {\bibfnamefont {E.}~\bibnamefont {Pajer}},\ }\href
  {\doibase 10.1007/JHEP06(2023)020} {\bibfield  {journal} {\bibinfo  {journal}
  {JHEP}\ }\textbf {\bibinfo {volume} {06}},\ \bibinfo {pages} {020} (\bibinfo
  {year} {2023})},\ \Eprint {http://arxiv.org/abs/2212.08009} {arXiv:2212.08009
  [hep-th]} \BibitemShut {NoStop}%
\bibitem [{\citenamefont {Green}\ and\ \citenamefont
  {Sun}(2025)}]{Green:2024cmx}%
  \BibitemOpen
  \bibfield  {author} {\bibinfo {author} {\bibfnamefont {D.}~\bibnamefont
  {Green}}\ and\ \bibinfo {author} {\bibfnamefont {G.}~\bibnamefont {Sun}},\
  }\href {\doibase 10.1007/JHEP04(2025)166} {\bibfield  {journal} {\bibinfo
  {journal} {JHEP}\ }\textbf {\bibinfo {volume} {04}},\ \bibinfo {pages} {166}
  (\bibinfo {year} {2025})},\ \Eprint {http://arxiv.org/abs/2412.02739}
  {arXiv:2412.02739 [hep-th]} \BibitemShut {NoStop}%
\bibitem [{\citenamefont {Kawaguchi}\ \emph {et~al.}(2024)\citenamefont
  {Kawaguchi}, \citenamefont {Tsujikawa},\ and\ \citenamefont
  {Yamada}}]{Kawaguchi:2024lsw}%
  \BibitemOpen
  \bibfield  {author} {\bibinfo {author} {\bibfnamefont {R.}~\bibnamefont
  {Kawaguchi}}, \bibinfo {author} {\bibfnamefont {S.}~\bibnamefont
  {Tsujikawa}}, \ and\ \bibinfo {author} {\bibfnamefont {Y.}~\bibnamefont
  {Yamada}},\ }\href {\doibase 10.1016/j.physletb.2024.138962} {\bibfield
  {journal} {\bibinfo  {journal} {Phys. Lett. B}\ }\textbf {\bibinfo {volume}
  {856}},\ \bibinfo {pages} {138962} (\bibinfo {year} {2024})},\ \Eprint
  {http://arxiv.org/abs/2403.16022} {arXiv:2403.16022 [hep-th]} \BibitemShut
  {NoStop}%
\bibitem [{\citenamefont {Duaso~Pueyo}\ \emph {et~al.}(2025)\citenamefont
  {Duaso~Pueyo}, \citenamefont {Goodhew}, \citenamefont {McCulloch},\ and\
  \citenamefont {Pajer}}]{DuasoPueyo:2025lmq}%
  \BibitemOpen
  \bibfield  {author} {\bibinfo {author} {\bibfnamefont {C.}~\bibnamefont
  {Duaso~Pueyo}}, \bibinfo {author} {\bibfnamefont {H.}~\bibnamefont
  {Goodhew}}, \bibinfo {author} {\bibfnamefont {C.}~\bibnamefont {McCulloch}},
  \ and\ \bibinfo {author} {\bibfnamefont {E.}~\bibnamefont {Pajer}},\
  }\href@noop {} {\  (\bibinfo {year} {2025})},\ \Eprint
  {http://arxiv.org/abs/2505.17820} {arXiv:2505.17820 [hep-th]} \BibitemShut
  {NoStop}%
\bibitem [{\citenamefont {Eden}\ \emph {et~al.}(1966)\citenamefont {Eden},
  \citenamefont {Landshoff}, \citenamefont {Olive},\ and\ \citenamefont
  {Polkinghorne}}]{Eden:1966dnq}%
  \BibitemOpen
  \bibfield  {author} {\bibinfo {author} {\bibfnamefont {R.~J.}\ \bibnamefont
  {Eden}}, \bibinfo {author} {\bibfnamefont {P.~V.}\ \bibnamefont {Landshoff}},
  \bibinfo {author} {\bibfnamefont {D.~I.}\ \bibnamefont {Olive}}, \ and\
  \bibinfo {author} {\bibfnamefont {J.~C.}\ \bibnamefont {Polkinghorne}},\
  }\href@noop {} {\emph {\bibinfo {title} {{The analytic S-matrix}}}}\
  (\bibinfo  {publisher} {Cambridge Univ. Press},\ \bibinfo {address}
  {Cambridge},\ \bibinfo {year} {1966})\BibitemShut {NoStop}%
\bibitem [{\citenamefont {de~Rham}\ \emph {et~al.}(2017)\citenamefont
  {de~Rham}, \citenamefont {Melville}, \citenamefont {Tolley},\ and\
  \citenamefont {Zhou}}]{deRham:2017avq}%
  \BibitemOpen
  \bibfield  {author} {\bibinfo {author} {\bibfnamefont {C.}~\bibnamefont
  {de~Rham}}, \bibinfo {author} {\bibfnamefont {S.}~\bibnamefont {Melville}},
  \bibinfo {author} {\bibfnamefont {A.~J.}\ \bibnamefont {Tolley}}, \ and\
  \bibinfo {author} {\bibfnamefont {S.-Y.}\ \bibnamefont {Zhou}},\ }\href
  {\doibase 10.1103/PhysRevD.96.081702} {\bibfield  {journal} {\bibinfo
  {journal} {Phys. Rev. D}\ }\textbf {\bibinfo {volume} {96}},\ \bibinfo
  {pages} {081702} (\bibinfo {year} {2017})},\ \Eprint
  {http://arxiv.org/abs/1702.06134} {arXiv:1702.06134 [hep-th]} \BibitemShut
  {NoStop}%
\bibitem [{\citenamefont {Adams}\ \emph {et~al.}(2006)\citenamefont {Adams},
  \citenamefont {Arkani-Hamed}, \citenamefont {Dubovsky}, \citenamefont
  {Nicolis},\ and\ \citenamefont {Rattazzi}}]{Adams:2006sv}%
  \BibitemOpen
  \bibfield  {author} {\bibinfo {author} {\bibfnamefont {A.}~\bibnamefont
  {Adams}}, \bibinfo {author} {\bibfnamefont {N.}~\bibnamefont {Arkani-Hamed}},
  \bibinfo {author} {\bibfnamefont {S.}~\bibnamefont {Dubovsky}}, \bibinfo
  {author} {\bibfnamefont {A.}~\bibnamefont {Nicolis}}, \ and\ \bibinfo
  {author} {\bibfnamefont {R.}~\bibnamefont {Rattazzi}},\ }\href {\doibase
  10.1088/1126-6708/2006/10/014} {\bibfield  {journal} {\bibinfo  {journal}
  {JHEP}\ }\textbf {\bibinfo {volume} {10}},\ \bibinfo {pages} {014} (\bibinfo
  {year} {2006})},\ \Eprint {http://arxiv.org/abs/hep-th/0602178}
  {arXiv:hep-th/0602178} \BibitemShut {NoStop}%
\bibitem [{\citenamefont {Bellazzini}\ \emph {et~al.}(2021)\citenamefont
  {Bellazzini}, \citenamefont {Elias~Mir{\'o}}, \citenamefont {Rattazzi},
  \citenamefont {Riembau},\ and\ \citenamefont {Riva}}]{Bellazzini:2020cot}%
  \BibitemOpen
  \bibfield  {author} {\bibinfo {author} {\bibfnamefont {B.}~\bibnamefont
  {Bellazzini}}, \bibinfo {author} {\bibfnamefont {J.}~\bibnamefont
  {Elias~Mir{\'o}}}, \bibinfo {author} {\bibfnamefont {R.}~\bibnamefont
  {Rattazzi}}, \bibinfo {author} {\bibfnamefont {M.}~\bibnamefont {Riembau}}, \
  and\ \bibinfo {author} {\bibfnamefont {F.}~\bibnamefont {Riva}},\ }\href
  {\doibase 10.1103/PhysRevD.104.036006} {\bibfield  {journal} {\bibinfo
  {journal} {Phys. Rev. D}\ }\textbf {\bibinfo {volume} {104}},\ \bibinfo
  {pages} {036006} (\bibinfo {year} {2021})},\ \Eprint
  {http://arxiv.org/abs/2011.00037} {arXiv:2011.00037 [hep-th]} \BibitemShut
  {NoStop}%
\bibitem [{Note4()}]{Note4}%
  \BibitemOpen
  \bibinfo {note} {Notice that $g_2 = 0$ would require $\rho = 0$ for all
  energies above $\Lambda $, i.e. that the theory is free. These bounds are
  strict inequalities when applied to theories with non-trivial
  interactions.}\BibitemShut {Stop}%
\bibitem [{\citenamefont {Bros}(1991)}]{Bros:1990cu}%
  \BibitemOpen
  \bibfield  {author} {\bibinfo {author} {\bibfnamefont {J.}~\bibnamefont
  {Bros}},\ }\href {\doibase 10.1016/0920-5632(91)90119-Y} {\bibfield
  {journal} {\bibinfo  {journal} {Nucl. Phys. B Proc. Suppl.}\ }\textbf
  {\bibinfo {volume} {18}},\ \bibinfo {pages} {22} (\bibinfo {year}
  {1991})}\BibitemShut {NoStop}%
\bibitem [{\citenamefont {Bros}\ and\ \citenamefont
  {Moschella}(1996)}]{Bros:1995js}%
  \BibitemOpen
  \bibfield  {author} {\bibinfo {author} {\bibfnamefont {J.}~\bibnamefont
  {Bros}}\ and\ \bibinfo {author} {\bibfnamefont {U.}~\bibnamefont
  {Moschella}},\ }\href {\doibase 10.1142/S0129055X96000123} {\bibfield
  {journal} {\bibinfo  {journal} {Rev. Math. Phys.}\ }\textbf {\bibinfo
  {volume} {8}},\ \bibinfo {pages} {327} (\bibinfo {year} {1996})},\ \Eprint
  {http://arxiv.org/abs/gr-qc/9511019} {arXiv:gr-qc/9511019} \BibitemShut
  {NoStop}%
\bibitem [{\citenamefont {Hollands}(2012)}]{Hollands:2011we}%
  \BibitemOpen
  \bibfield  {author} {\bibinfo {author} {\bibfnamefont {S.}~\bibnamefont
  {Hollands}},\ }\href {\doibase 10.1007/s00023-011-0140-1} {\bibfield
  {journal} {\bibinfo  {journal} {Annales Henri Poincare}\ }\textbf {\bibinfo
  {volume} {13}},\ \bibinfo {pages} {1039} (\bibinfo {year} {2012})},\ \Eprint
  {http://arxiv.org/abs/1105.1996} {arXiv:1105.1996 [gr-qc]} \BibitemShut
  {NoStop}%
\bibitem [{\citenamefont {Marolf}\ \emph {et~al.}(2013)\citenamefont {Marolf},
  \citenamefont {Morrison},\ and\ \citenamefont {Srednicki}}]{Marolf:2012kh}%
  \BibitemOpen
  \bibfield  {author} {\bibinfo {author} {\bibfnamefont {D.}~\bibnamefont
  {Marolf}}, \bibinfo {author} {\bibfnamefont {I.~A.}\ \bibnamefont
  {Morrison}}, \ and\ \bibinfo {author} {\bibfnamefont {M.}~\bibnamefont
  {Srednicki}},\ }\href {\doibase 10.1088/0264-9381/30/15/155023} {\bibfield
  {journal} {\bibinfo  {journal} {Class. Quant. Grav.}\ }\textbf {\bibinfo
  {volume} {30}},\ \bibinfo {pages} {155023} (\bibinfo {year} {2013})},\
  \Eprint {http://arxiv.org/abs/1209.6039} {arXiv:1209.6039 [hep-th]}
  \BibitemShut {NoStop}%
\bibitem [{\citenamefont {Penedones}\ \emph {et~al.}(2025)\citenamefont
  {Penedones}, \citenamefont {Salehi~Vaziri},\ and\ \citenamefont
  {Sun}}]{Penedones:2023uqc}%
  \BibitemOpen
  \bibfield  {author} {\bibinfo {author} {\bibfnamefont {J.}~\bibnamefont
  {Penedones}}, \bibinfo {author} {\bibfnamefont {K.}~\bibnamefont
  {Salehi~Vaziri}}, \ and\ \bibinfo {author} {\bibfnamefont {Z.}~\bibnamefont
  {Sun}},\ }\href {\doibase 10.1103/PhysRevD.111.045001} {\bibfield  {journal}
  {\bibinfo  {journal} {Phys. Rev. D}\ }\textbf {\bibinfo {volume} {111}},\
  \bibinfo {pages} {045001} (\bibinfo {year} {2025})},\ \Eprint
  {http://arxiv.org/abs/2301.04146} {arXiv:2301.04146 [hep-th]} \BibitemShut
  {NoStop}%
\bibitem [{\citenamefont {Loparco}\ \emph {et~al.}(2023)\citenamefont
  {Loparco}, \citenamefont {Penedones}, \citenamefont {Salehi~Vaziri},\ and\
  \citenamefont {Sun}}]{Loparco:2023rug}%
  \BibitemOpen
  \bibfield  {author} {\bibinfo {author} {\bibfnamefont {M.}~\bibnamefont
  {Loparco}}, \bibinfo {author} {\bibfnamefont {J.}~\bibnamefont {Penedones}},
  \bibinfo {author} {\bibfnamefont {K.}~\bibnamefont {Salehi~Vaziri}}, \ and\
  \bibinfo {author} {\bibfnamefont {Z.}~\bibnamefont {Sun}},\ }\href {\doibase
  10.1007/JHEP12(2023)159} {\bibfield  {journal} {\bibinfo  {journal} {JHEP}\
  }\textbf {\bibinfo {volume} {12}},\ \bibinfo {pages} {159} (\bibinfo {year}
  {2023})},\ \Eprint {http://arxiv.org/abs/2306.00090} {arXiv:2306.00090
  [hep-th]} \BibitemShut {NoStop}%
\bibitem [{Note5()}]{Note5}%
  \BibitemOpen
  \bibinfo {note} {Alternatively, notice that the analytic structure of $\Sigma
  $ implied by causality means we can express it as $\protect \frac {1}{2\pi i}
  \DOTSI \ointop \ilimits@ _\nu d\mu ' \protect \tmspace +\thinmuskip {.1667em}
  2 \mu ' \Sigma (\mu ') / ( \mu '^2 - \nu ^2 - i \epsilon )$ around the pole
  in the lower half of the complex plane, and then deform this contour so that
  it runs over $\DOTSI \intop \ilimits@ _{-\infty }^{+\infty } d \mu '$ and
  then change variables to $\mu '^2$ to produce \protect \textup {\hbox
  {\mathsurround \z@ \protect \normalfont (\ignorespaces \ref
  {eqn:Sigma_KL_dS}\unskip \@@italiccorr )}}.}\BibitemShut {Stop}%
\bibitem [{\citenamefont {Dubovsky}\ \emph {et~al.}(2008)\citenamefont
  {Dubovsky}, \citenamefont {Nicolis}, \citenamefont {Trincherini},\ and\
  \citenamefont {Villadoro}}]{Dubovsky:2007ac}%
  \BibitemOpen
  \bibfield  {author} {\bibinfo {author} {\bibfnamefont {S.}~\bibnamefont
  {Dubovsky}}, \bibinfo {author} {\bibfnamefont {A.}~\bibnamefont {Nicolis}},
  \bibinfo {author} {\bibfnamefont {E.}~\bibnamefont {Trincherini}}, \ and\
  \bibinfo {author} {\bibfnamefont {G.}~\bibnamefont {Villadoro}},\ }\href
  {\doibase 10.1103/PhysRevD.77.084016} {\bibfield  {journal} {\bibinfo
  {journal} {Phys. Rev. D}\ }\textbf {\bibinfo {volume} {77}},\ \bibinfo
  {pages} {084016} (\bibinfo {year} {2008})},\ \Eprint
  {http://arxiv.org/abs/0709.1483} {arXiv:0709.1483 [hep-th]} \BibitemShut
  {NoStop}%
\bibitem [{\citenamefont {{Nachtmann}}(1968)}]{Nachtmann:1968}%
  \BibitemOpen
  \bibfield  {author} {\bibinfo {author} {\bibfnamefont {O.}~\bibnamefont
  {{Nachtmann}}},\ }\href
  {https://ui.adsabs.harvard.edu/abs/1968OAWMN.176..363N} {\bibfield  {journal}
  {\bibinfo  {journal} {Oesterreichische Akademie Wissenschaften Mathematisch
  naturwissenschaftliche Klasse Sitzungsberichte Abteilung}\ }\textbf {\bibinfo
  {volume} {176}},\ \bibinfo {pages} {363} (\bibinfo {year}
  {1968})}\BibitemShut {NoStop}%
\bibitem [{Note6()}]{Note6}%
  \BibitemOpen
  \bibinfo {note} {Note that a low-energy observer can only determine $g_n
  (\Lambda )$ from the EFT coefficients when $\Lambda $ is below threshold, in
  this case $\Lambda < 2 \mu $. In Figure \ref {fig:d3_island} we have plotted
  all values of $\Lambda $ to illustrate that our bounds are always satisfied,
  regardless of the EFT series \protect \textup {\hbox {\mathsurround \z@
  \protect \normalfont (\ignorespaces \ref {eqn:g_from_c}\unskip \@@italiccorr
  )}} converging.}\BibitemShut {Stop}%
\bibitem [{\citenamefont {Fitzpatrick}\ \emph {et~al.}(2011)\citenamefont
  {Fitzpatrick}, \citenamefont {Katz}, \citenamefont {Poland},\ and\
  \citenamefont {Simmons-Duffin}}]{Fitzpatrick:2010zm}%
  \BibitemOpen
  \bibfield  {author} {\bibinfo {author} {\bibfnamefont {A.~L.}\ \bibnamefont
  {Fitzpatrick}}, \bibinfo {author} {\bibfnamefont {E.}~\bibnamefont {Katz}},
  \bibinfo {author} {\bibfnamefont {D.}~\bibnamefont {Poland}}, \ and\ \bibinfo
  {author} {\bibfnamefont {D.}~\bibnamefont {Simmons-Duffin}},\ }\href
  {\doibase 10.1007/JHEP07(2011)023} {\bibfield  {journal} {\bibinfo  {journal}
  {JHEP}\ }\textbf {\bibinfo {volume} {07}},\ \bibinfo {pages} {023} (\bibinfo
  {year} {2011})},\ \Eprint {http://arxiv.org/abs/1007.2412} {arXiv:1007.2412
  [hep-th]} \BibitemShut {NoStop}%
\bibitem [{\citenamefont {Fitzpatrick}\ and\ \citenamefont
  {Kaplan}(2012)}]{Fitzpatrick:2011hu}%
  \BibitemOpen
  \bibfield  {author} {\bibinfo {author} {\bibfnamefont {A.~L.}\ \bibnamefont
  {Fitzpatrick}}\ and\ \bibinfo {author} {\bibfnamefont {J.}~\bibnamefont
  {Kaplan}},\ }\href {\doibase 10.1007/JHEP10(2012)127} {\bibfield  {journal}
  {\bibinfo  {journal} {JHEP}\ }\textbf {\bibinfo {volume} {10}},\ \bibinfo
  {pages} {127} (\bibinfo {year} {2012})},\ \Eprint
  {http://arxiv.org/abs/1111.6972} {arXiv:1111.6972 [hep-th]} \BibitemShut
  {NoStop}%
\bibitem [{\citenamefont {Alishahiha}\ \emph {et~al.}(2004)\citenamefont
  {Alishahiha}, \citenamefont {Silverstein},\ and\ \citenamefont
  {Tong}}]{Alishahiha:2004eh}%
  \BibitemOpen
  \bibfield  {author} {\bibinfo {author} {\bibfnamefont {M.}~\bibnamefont
  {Alishahiha}}, \bibinfo {author} {\bibfnamefont {E.}~\bibnamefont
  {Silverstein}}, \ and\ \bibinfo {author} {\bibfnamefont {D.}~\bibnamefont
  {Tong}},\ }\href {\doibase 10.1103/PhysRevD.70.123505} {\bibfield  {journal}
  {\bibinfo  {journal} {Phys. Rev.}\ }\textbf {\bibinfo {volume} {D70}},\
  \bibinfo {pages} {123505} (\bibinfo {year} {2004})},\ \Eprint
  {http://arxiv.org/abs/hep-th/0404084} {arXiv:hep-th/0404084} \BibitemShut
  {NoStop}%
\bibitem [{\citenamefont {Tolley}\ and\ \citenamefont
  {Wyman}(2010)}]{Tolley:2009fg}%
  \BibitemOpen
  \bibfield  {author} {\bibinfo {author} {\bibfnamefont {A.~J.}\ \bibnamefont
  {Tolley}}\ and\ \bibinfo {author} {\bibfnamefont {M.}~\bibnamefont {Wyman}},\
  }\href {\doibase 10.1103/PhysRevD.81.043502} {\bibfield  {journal} {\bibinfo
  {journal} {Phys. Rev.}\ }\textbf {\bibinfo {volume} {D81}},\ \bibinfo {pages}
  {043502} (\bibinfo {year} {2010})},\ \Eprint {http://arxiv.org/abs/0910.1853}
  {arXiv:0910.1853 [hep-th]} \BibitemShut {NoStop}%
\bibitem [{\citenamefont {Baumann}\ and\ \citenamefont
  {Green}(2011)}]{Baumann:2011su}%
  \BibitemOpen
  \bibfield  {author} {\bibinfo {author} {\bibfnamefont {D.}~\bibnamefont
  {Baumann}}\ and\ \bibinfo {author} {\bibfnamefont {D.}~\bibnamefont
  {Green}},\ }\href {\doibase 10.1088/1475-7516/2011/09/014} {\bibfield
  {journal} {\bibinfo  {journal} {JCAP}\ }\textbf {\bibinfo {volume} {1109}},\
  \bibinfo {pages} {014} (\bibinfo {year} {2011})},\ \Eprint
  {http://arxiv.org/abs/1102.5343} {arXiv:1102.5343 [hep-th]} \BibitemShut
  {NoStop}%
\bibitem [{\citenamefont {Achucarro}\ \emph {et~al.}(2011)\citenamefont
  {Achucarro}, \citenamefont {Gong}, \citenamefont {Hardeman}, \citenamefont
  {Palma},\ and\ \citenamefont {Patil}}]{Achucarro:2010da}%
  \BibitemOpen
  \bibfield  {author} {\bibinfo {author} {\bibfnamefont {A.}~\bibnamefont
  {Achucarro}}, \bibinfo {author} {\bibfnamefont {J.-O.}\ \bibnamefont {Gong}},
  \bibinfo {author} {\bibfnamefont {S.}~\bibnamefont {Hardeman}}, \bibinfo
  {author} {\bibfnamefont {G.~A.}\ \bibnamefont {Palma}}, \ and\ \bibinfo
  {author} {\bibfnamefont {S.~P.}\ \bibnamefont {Patil}},\ }\href {\doibase
  10.1088/1475-7516/2011/01/030} {\bibfield  {journal} {\bibinfo  {journal}
  {JCAP}\ }\textbf {\bibinfo {volume} {1101}},\ \bibinfo {pages} {030}
  (\bibinfo {year} {2011})},\ \Eprint {http://arxiv.org/abs/1010.3693}
  {arXiv:1010.3693 [hep-ph]} \BibitemShut {NoStop}%
\bibitem [{\citenamefont {Flauger}\ \emph {et~al.}(2017)\citenamefont
  {Flauger}, \citenamefont {Mirbabayi}, \citenamefont {Senatore},\ and\
  \citenamefont {Silverstein}}]{Flauger:2016idt}%
  \BibitemOpen
  \bibfield  {author} {\bibinfo {author} {\bibfnamefont {R.}~\bibnamefont
  {Flauger}}, \bibinfo {author} {\bibfnamefont {M.}~\bibnamefont {Mirbabayi}},
  \bibinfo {author} {\bibfnamefont {L.}~\bibnamefont {Senatore}}, \ and\
  \bibinfo {author} {\bibfnamefont {E.}~\bibnamefont {Silverstein}},\ }\href
  {\doibase 10.1088/1475-7516/2017/10/058} {\bibfield  {journal} {\bibinfo
  {journal} {JCAP}\ }\textbf {\bibinfo {volume} {1710}},\ \bibinfo {pages}
  {058} (\bibinfo {year} {2017})},\ \Eprint {http://arxiv.org/abs/1606.00513}
  {arXiv:1606.00513 [hep-th]} \BibitemShut {NoStop}%
\bibitem [{\citenamefont {Davighi}\ \emph {et~al.}(2022)\citenamefont
  {Davighi}, \citenamefont {Melville},\ and\ \citenamefont
  {You}}]{Davighi:2021osh}%
  \BibitemOpen
  \bibfield  {author} {\bibinfo {author} {\bibfnamefont {J.}~\bibnamefont
  {Davighi}}, \bibinfo {author} {\bibfnamefont {S.}~\bibnamefont {Melville}}, \
  and\ \bibinfo {author} {\bibfnamefont {T.}~\bibnamefont {You}},\ }\href
  {\doibase 10.1007/JHEP02(2022)167} {\bibfield  {journal} {\bibinfo  {journal}
  {JHEP}\ }\textbf {\bibinfo {volume} {02}},\ \bibinfo {pages} {167} (\bibinfo
  {year} {2022})},\ \Eprint {http://arxiv.org/abs/2108.06334} {arXiv:2108.06334
  [hep-th]} \BibitemShut {NoStop}%
\bibitem [{Note7()}]{Note7}%
  \BibitemOpen
  \bibinfo {note} {This is analoguous to scattering with space-like momenta on
  Minkowski, for which the $i \epsilon $ becomes unimportant because there is
  no operator ordering ambiguity.}\BibitemShut {Stop}%
\bibitem [{\citenamefont {Caron-Huot}\ and\ \citenamefont
  {Van~Duong}(2021)}]{Caron-Huot:2020cmc}%
  \BibitemOpen
  \bibfield  {author} {\bibinfo {author} {\bibfnamefont {S.}~\bibnamefont
  {Caron-Huot}}\ and\ \bibinfo {author} {\bibfnamefont {V.}~\bibnamefont
  {Van~Duong}},\ }\href {\doibase 10.1007/JHEP05(2021)280} {\bibfield
  {journal} {\bibinfo  {journal} {JHEP}\ }\textbf {\bibinfo {volume} {05}},\
  \bibinfo {pages} {280} (\bibinfo {year} {2021})},\ \Eprint
  {http://arxiv.org/abs/2011.02957} {arXiv:2011.02957 [hep-th]} \BibitemShut
  {NoStop}%
\bibitem [{Note8()}]{Note8}%
  \BibitemOpen
  \bibinfo {note} {\protect \textup {\hbox {\mathsurround \z@ \protect
  \normalfont (\ignorespaces \ref {eqn:rho_phi_to_rho_J}\unskip \@@italiccorr
  )}} makes it clear that $\rho _J$ and $\Sigma $ are related to $\rho _\phi $
  and \protect {$\delimiter "426830A \phi \phi \delimiter "526930B $} by an
  amputation of two propagators that closely resembles LSZ reduction: $\Sigma $
  is therefore closer in spirit to an $S$-matrix, and is the natural object for
  EFT matching.}\BibitemShut {Stop}%
\bibitem [{Note9()}]{Note9}%
  \BibitemOpen
  \bibinfo {note} {For the equal-mass case considered above, these
  singularities are cancelled by zeroes of $\protect \qopname \relax o{sinh}(
  \pi \nu )$ and the cut-off is instead set by singularities at $| \nu |^2 = 4
  \mu ^2 + d^2/4$).}\BibitemShut {Stop}%
\bibitem [{\citenamefont {Lee}\ and\ \citenamefont {Melville}(pear)}]{us}%
  \BibitemOpen
  \bibfield  {author} {\bibinfo {author} {\bibfnamefont {M.~H.~G.}\
  \bibnamefont {Lee}}\ and\ \bibinfo {author} {\bibfnamefont {S.}~\bibnamefont
  {Melville}},\ }\href@noop {} {\bibfield  {journal} {\bibinfo  {journal} {{\it
  Propagators, poles and positivity for de Sitter scalar fields}}\ } (\bibinfo
  {year} {to appear})}\BibitemShut {NoStop}%
\bibitem [{Note10()}]{Note10}%
  \BibitemOpen
  \bibinfo {note} {See for instance \cite {Melville:2023kgd} for a description
  of how to ``square'' an in-out $S$-matrix on de Sitter into an in-in
  Bunch-Davies correlator.}\BibitemShut {Stop}%
\bibitem [{\citenamefont {Kruczenski}\ \emph {et~al.}(2022)\citenamefont
  {Kruczenski}, \citenamefont {Penedones},\ and\ \citenamefont {van
  Rees}}]{Kruczenski:2022lot}%
  \BibitemOpen
  \bibfield  {author} {\bibinfo {author} {\bibfnamefont {M.}~\bibnamefont
  {Kruczenski}}, \bibinfo {author} {\bibfnamefont {J.}~\bibnamefont
  {Penedones}}, \ and\ \bibinfo {author} {\bibfnamefont {B.~C.}\ \bibnamefont
  {van Rees}},\ }\href@noop {} {\  (\bibinfo {year} {2022})},\ \Eprint
  {http://arxiv.org/abs/2203.02421} {arXiv:2203.02421 [hep-th]} \BibitemShut
  {NoStop}%
\bibitem [{\citenamefont {Melville}\ and\ \citenamefont
  {Pajer}(2021)}]{Melville:2021lst}%
  \BibitemOpen
  \bibfield  {author} {\bibinfo {author} {\bibfnamefont {S.}~\bibnamefont
  {Melville}}\ and\ \bibinfo {author} {\bibfnamefont {E.}~\bibnamefont
  {Pajer}},\ }\href {\doibase 10.1007/JHEP05(2021)249} {\bibfield  {journal}
  {\bibinfo  {journal} {JHEP}\ }\textbf {\bibinfo {volume} {05}},\ \bibinfo
  {pages} {249} (\bibinfo {year} {2021})},\ \Eprint
  {http://arxiv.org/abs/2103.09832} {arXiv:2103.09832 [hep-th]} \BibitemShut
  {NoStop}%
\bibitem [{\citenamefont {Goodhew}\ \emph {et~al.}(2021)\citenamefont
  {Goodhew}, \citenamefont {Jazayeri}, \citenamefont {Lee},\ and\ \citenamefont
  {Pajer}}]{Goodhew:2021oqg}%
  \BibitemOpen
  \bibfield  {author} {\bibinfo {author} {\bibfnamefont {H.}~\bibnamefont
  {Goodhew}}, \bibinfo {author} {\bibfnamefont {S.}~\bibnamefont {Jazayeri}},
  \bibinfo {author} {\bibfnamefont {M.~H.~G.}\ \bibnamefont {Lee}}, \ and\
  \bibinfo {author} {\bibfnamefont {E.}~\bibnamefont {Pajer}},\ }\href
  {\doibase 10.1088/1475-7516/2021/08/003} {\bibfield  {journal} {\bibinfo
  {journal} {JCAP}\ }\textbf {\bibinfo {volume} {08}},\ \bibinfo {pages} {003}
  (\bibinfo {year} {2021})},\ \Eprint {http://arxiv.org/abs/2104.06587}
  {arXiv:2104.06587 [hep-th]} \BibitemShut {NoStop}%
\bibitem [{\citenamefont {Tong}\ \emph {et~al.}(2022)\citenamefont {Tong},
  \citenamefont {Wang},\ and\ \citenamefont {Zhu}}]{Tong:2021wai}%
  \BibitemOpen
  \bibfield  {author} {\bibinfo {author} {\bibfnamefont {X.}~\bibnamefont
  {Tong}}, \bibinfo {author} {\bibfnamefont {Y.}~\bibnamefont {Wang}}, \ and\
  \bibinfo {author} {\bibfnamefont {Y.}~\bibnamefont {Zhu}},\ }\href {\doibase
  10.1007/JHEP03(2022)181} {\bibfield  {journal} {\bibinfo  {journal} {JHEP}\
  }\textbf {\bibinfo {volume} {03}},\ \bibinfo {pages} {181} (\bibinfo {year}
  {2022})},\ \Eprint {http://arxiv.org/abs/2112.03448} {arXiv:2112.03448
  [hep-th]} \BibitemShut {NoStop}%
\bibitem [{\citenamefont {Agui~Salcedo}\ and\ \citenamefont
  {Melville}(2023)}]{AguiSalcedo:2023nds}%
  \BibitemOpen
  \bibfield  {author} {\bibinfo {author} {\bibfnamefont {S.}~\bibnamefont
  {Agui~Salcedo}}\ and\ \bibinfo {author} {\bibfnamefont {S.}~\bibnamefont
  {Melville}},\ }\href {\doibase 10.1007/JHEP12(2023)076} {\bibfield  {journal}
  {\bibinfo  {journal} {JHEP}\ }\textbf {\bibinfo {volume} {12}},\ \bibinfo
  {pages} {076} (\bibinfo {year} {2023})},\ \Eprint
  {http://arxiv.org/abs/2308.00680} {arXiv:2308.00680 [hep-th]} \BibitemShut
  {NoStop}%
\bibitem [{\citenamefont {Qin}\ and\ \citenamefont
  {Xianyu}(2023)}]{Qin:2023bjk}%
  \BibitemOpen
  \bibfield  {author} {\bibinfo {author} {\bibfnamefont {Z.}~\bibnamefont
  {Qin}}\ and\ \bibinfo {author} {\bibfnamefont {Z.-Z.}\ \bibnamefont
  {Xianyu}},\ }\href {\doibase 10.1007/JHEP09(2023)116} {\bibfield  {journal}
  {\bibinfo  {journal} {JHEP}\ }\textbf {\bibinfo {volume} {09}},\ \bibinfo
  {pages} {116} (\bibinfo {year} {2023})},\ \Eprint
  {http://arxiv.org/abs/2304.13295} {arXiv:2304.13295 [hep-th]} \BibitemShut
  {NoStop}%
\bibitem [{\citenamefont {Ema}\ and\ \citenamefont
  {Mukaida}(2024)}]{Ema:2024hkj}%
  \BibitemOpen
  \bibfield  {author} {\bibinfo {author} {\bibfnamefont {Y.}~\bibnamefont
  {Ema}}\ and\ \bibinfo {author} {\bibfnamefont {K.}~\bibnamefont {Mukaida}},\
  }\href {\doibase 10.1007/JHEP12(2024)194} {\bibfield  {journal} {\bibinfo
  {journal} {JHEP}\ }\textbf {\bibinfo {volume} {12}},\ \bibinfo {pages} {194}
  (\bibinfo {year} {2024})},\ \Eprint {http://arxiv.org/abs/2409.07521}
  {arXiv:2409.07521 [hep-th]} \BibitemShut {NoStop}%
\bibitem [{\citenamefont {Baumann}\ \emph {et~al.}(2022)\citenamefont
  {Baumann}, \citenamefont {Chen}, \citenamefont {Duaso~Pueyo}, \citenamefont
  {Joyce}, \citenamefont {Lee},\ and\ \citenamefont
  {Pimentel}}]{Baumann:2021fxj}%
  \BibitemOpen
  \bibfield  {author} {\bibinfo {author} {\bibfnamefont {D.}~\bibnamefont
  {Baumann}}, \bibinfo {author} {\bibfnamefont {W.-M.}\ \bibnamefont {Chen}},
  \bibinfo {author} {\bibfnamefont {C.}~\bibnamefont {Duaso~Pueyo}}, \bibinfo
  {author} {\bibfnamefont {A.}~\bibnamefont {Joyce}}, \bibinfo {author}
  {\bibfnamefont {H.}~\bibnamefont {Lee}}, \ and\ \bibinfo {author}
  {\bibfnamefont {G.~L.}\ \bibnamefont {Pimentel}},\ }\href {\doibase
  10.1007/JHEP09(2022)010} {\bibfield  {journal} {\bibinfo  {journal} {JHEP}\
  }\textbf {\bibinfo {volume} {09}},\ \bibinfo {pages} {010} (\bibinfo {year}
  {2022})},\ \Eprint {http://arxiv.org/abs/2106.05294} {arXiv:2106.05294
  [hep-th]} \BibitemShut {NoStop}%
\bibitem [{\citenamefont {Jazayeri}\ \emph {et~al.}(2021)\citenamefont
  {Jazayeri}, \citenamefont {Pajer},\ and\ \citenamefont
  {Stefanyszyn}}]{Jazayeri:2021fvk}%
  \BibitemOpen
  \bibfield  {author} {\bibinfo {author} {\bibfnamefont {S.}~\bibnamefont
  {Jazayeri}}, \bibinfo {author} {\bibfnamefont {E.}~\bibnamefont {Pajer}}, \
  and\ \bibinfo {author} {\bibfnamefont {D.}~\bibnamefont {Stefanyszyn}},\
  }\href {\doibase 10.1007/JHEP10(2021)065} {\bibfield  {journal} {\bibinfo
  {journal} {JHEP}\ }\textbf {\bibinfo {volume} {10}},\ \bibinfo {pages} {065}
  (\bibinfo {year} {2021})},\ \Eprint {http://arxiv.org/abs/2103.08649}
  {arXiv:2103.08649 [hep-th]} \BibitemShut {NoStop}%
\bibitem [{\citenamefont {Meltzer}(2021{\natexlab{a}})}]{Meltzer:2021zin}%
  \BibitemOpen
  \bibfield  {author} {\bibinfo {author} {\bibfnamefont {D.}~\bibnamefont
  {Meltzer}},\ }\href {\doibase 10.1088/1475-7516/2021/12/018} {\bibfield
  {journal} {\bibinfo  {journal} {JCAP}\ }\textbf {\bibinfo {volume} {12}},\
  \bibinfo {pages} {018} (\bibinfo {year} {2021}{\natexlab{a}})},\ \Eprint
  {http://arxiv.org/abs/2107.10266} {arXiv:2107.10266 [hep-th]} \BibitemShut
  {NoStop}%
\bibitem [{\citenamefont {Chowdhury}\ \emph
  {et~al.}(2025{\natexlab{a}})\citenamefont {Chowdhury}, \citenamefont
  {Lipstein}, \citenamefont {Mei}, \citenamefont {Sachs},\ and\ \citenamefont
  {Vanhove}}]{Chowdhury:2023arc}%
  \BibitemOpen
  \bibfield  {author} {\bibinfo {author} {\bibfnamefont {C.}~\bibnamefont
  {Chowdhury}}, \bibinfo {author} {\bibfnamefont {A.}~\bibnamefont {Lipstein}},
  \bibinfo {author} {\bibfnamefont {J.}~\bibnamefont {Mei}}, \bibinfo {author}
  {\bibfnamefont {I.}~\bibnamefont {Sachs}}, \ and\ \bibinfo {author}
  {\bibfnamefont {P.}~\bibnamefont {Vanhove}},\ }\href {\doibase
  10.1007/JHEP03(2025)007} {\bibfield  {journal} {\bibinfo  {journal} {JHEP}\
  }\textbf {\bibinfo {volume} {03}},\ \bibinfo {pages} {007} (\bibinfo {year}
  {2025}{\natexlab{a}})},\ \Eprint {http://arxiv.org/abs/2312.13803}
  {arXiv:2312.13803 [hep-th]} \BibitemShut {NoStop}%
\bibitem [{\citenamefont {Chowdhury}\ \emph
  {et~al.}(2025{\natexlab{b}})\citenamefont {Chowdhury}, \citenamefont
  {Lipstein}, \citenamefont {Marshall}, \citenamefont {Mei},\ and\
  \citenamefont {Sachs}}]{Chowdhury:2025ohm}%
  \BibitemOpen
  \bibfield  {author} {\bibinfo {author} {\bibfnamefont {C.}~\bibnamefont
  {Chowdhury}}, \bibinfo {author} {\bibfnamefont {A.}~\bibnamefont {Lipstein}},
  \bibinfo {author} {\bibfnamefont {J.}~\bibnamefont {Marshall}}, \bibinfo
  {author} {\bibfnamefont {J.}~\bibnamefont {Mei}}, \ and\ \bibinfo {author}
  {\bibfnamefont {I.}~\bibnamefont {Sachs}},\ }\href@noop {} {\  (\bibinfo
  {year} {2025}{\natexlab{b}})},\ \Eprint {http://arxiv.org/abs/2503.10598}
  {arXiv:2503.10598 [hep-th]} \BibitemShut {NoStop}%
\bibitem [{\citenamefont {Jazayeri}\ \emph {et~al.}(2025)\citenamefont
  {Jazayeri}, \citenamefont {Tong},\ and\ \citenamefont
  {Zhu}}]{Jazayeri:2025vlv}%
  \BibitemOpen
  \bibfield  {author} {\bibinfo {author} {\bibfnamefont {S.}~\bibnamefont
  {Jazayeri}}, \bibinfo {author} {\bibfnamefont {X.}~\bibnamefont {Tong}}, \
  and\ \bibinfo {author} {\bibfnamefont {Y.}~\bibnamefont {Zhu}},\ }\href@noop
  {} {\  (\bibinfo {year} {2025})},\ \Eprint {http://arxiv.org/abs/2511.00152}
  {arXiv:2511.00152 [hep-th]} \BibitemShut {NoStop}%
\bibitem [{\citenamefont {Meltzer}(2021{\natexlab{b}})}]{Meltzer:2021bmb}%
  \BibitemOpen
  \bibfield  {author} {\bibinfo {author} {\bibfnamefont {D.}~\bibnamefont
  {Meltzer}},\ }\href {\doibase 10.1007/JHEP05(2021)098} {\bibfield  {journal}
  {\bibinfo  {journal} {JHEP}\ }\textbf {\bibinfo {volume} {05}},\ \bibinfo
  {pages} {098} (\bibinfo {year} {2021}{\natexlab{b}})},\ \Eprint
  {http://arxiv.org/abs/2103.15839} {arXiv:2103.15839 [hep-th]} \BibitemShut
  {NoStop}%
\bibitem [{\citenamefont {Kristiano}\ \emph {et~al.}(2025)\citenamefont
  {Kristiano}, \citenamefont {Namba}, \citenamefont {Naruko}, \citenamefont
  {Saito},\ and\ \citenamefont {Yamauchi}}]{Kristiano:2025cod}%
  \BibitemOpen
  \bibfield  {author} {\bibinfo {author} {\bibfnamefont {J.}~\bibnamefont
  {Kristiano}}, \bibinfo {author} {\bibfnamefont {R.}~\bibnamefont {Namba}},
  \bibinfo {author} {\bibfnamefont {A.}~\bibnamefont {Naruko}}, \bibinfo
  {author} {\bibfnamefont {R.}~\bibnamefont {Saito}}, \ and\ \bibinfo {author}
  {\bibfnamefont {D.}~\bibnamefont {Yamauchi}},\ }\href@noop {} {\  (\bibinfo
  {year} {2025})},\ \Eprint {http://arxiv.org/abs/2511.22623} {arXiv:2511.22623
  [hep-th]} \BibitemShut {NoStop}%
\bibitem [{\citenamefont {Cheung}\ \emph {et~al.}(2008)\citenamefont {Cheung},
  \citenamefont {Creminelli}, \citenamefont {Fitzpatrick}, \citenamefont
  {Kaplan},\ and\ \citenamefont {Senatore}}]{Cheung:2007st}%
  \BibitemOpen
  \bibfield  {author} {\bibinfo {author} {\bibfnamefont {C.}~\bibnamefont
  {Cheung}}, \bibinfo {author} {\bibfnamefont {P.}~\bibnamefont {Creminelli}},
  \bibinfo {author} {\bibfnamefont {A.~L.}\ \bibnamefont {Fitzpatrick}},
  \bibinfo {author} {\bibfnamefont {J.}~\bibnamefont {Kaplan}}, \ and\ \bibinfo
  {author} {\bibfnamefont {L.}~\bibnamefont {Senatore}},\ }\href {\doibase
  10.1088/1126-6708/2008/03/014} {\bibfield  {journal} {\bibinfo  {journal}
  {JHEP}\ }\textbf {\bibinfo {volume} {03}},\ \bibinfo {pages} {014} (\bibinfo
  {year} {2008})},\ \Eprint {http://arxiv.org/abs/0709.0293} {arXiv:0709.0293
  [hep-th]} \BibitemShut {NoStop}%
\bibitem [{\citenamefont {Senatore}\ and\ \citenamefont
  {Zaldarriaga}(2012)}]{Senatore:2010wk}%
  \BibitemOpen
  \bibfield  {author} {\bibinfo {author} {\bibfnamefont {L.}~\bibnamefont
  {Senatore}}\ and\ \bibinfo {author} {\bibfnamefont {M.}~\bibnamefont
  {Zaldarriaga}},\ }\href {\doibase 10.1007/JHEP04(2012)024} {\bibfield
  {journal} {\bibinfo  {journal} {JHEP}\ }\textbf {\bibinfo {volume} {04}},\
  \bibinfo {pages} {024} (\bibinfo {year} {2012})},\ \Eprint
  {http://arxiv.org/abs/1009.2093} {arXiv:1009.2093 [hep-th]} \BibitemShut
  {NoStop}%
\bibitem [{\citenamefont {Noumi}\ \emph {et~al.}(2013)\citenamefont {Noumi},
  \citenamefont {Yamaguchi},\ and\ \citenamefont {Yokoyama}}]{Noumi:2012vr}%
  \BibitemOpen
  \bibfield  {author} {\bibinfo {author} {\bibfnamefont {T.}~\bibnamefont
  {Noumi}}, \bibinfo {author} {\bibfnamefont {M.}~\bibnamefont {Yamaguchi}}, \
  and\ \bibinfo {author} {\bibfnamefont {D.}~\bibnamefont {Yokoyama}},\ }\href
  {\doibase 10.1007/JHEP06(2013)051} {\bibfield  {journal} {\bibinfo  {journal}
  {JHEP}\ }\textbf {\bibinfo {volume} {06}},\ \bibinfo {pages} {051} (\bibinfo
  {year} {2013})},\ \Eprint {http://arxiv.org/abs/1211.1624} {arXiv:1211.1624
  [hep-th]} \BibitemShut {NoStop}%
\bibitem [{\citenamefont {Green}\ and\ \citenamefont
  {Pajer}(2020)}]{Green:2020ebl}%
  \BibitemOpen
  \bibfield  {author} {\bibinfo {author} {\bibfnamefont {D.}~\bibnamefont
  {Green}}\ and\ \bibinfo {author} {\bibfnamefont {E.}~\bibnamefont {Pajer}},\
  }\href {\doibase 10.1088/1475-7516/2020/09/032} {\bibfield  {journal}
  {\bibinfo  {journal} {JCAP}\ }\textbf {\bibinfo {volume} {09}},\ \bibinfo
  {pages} {032} (\bibinfo {year} {2020})},\ \Eprint
  {http://arxiv.org/abs/2004.09587} {arXiv:2004.09587 [hep-th]} \BibitemShut
  {NoStop}%
\bibitem [{\citenamefont {Melville}\ and\ \citenamefont
  {Noller}(2020)}]{Melville:2019wyy}%
  \BibitemOpen
  \bibfield  {author} {\bibinfo {author} {\bibfnamefont {S.}~\bibnamefont
  {Melville}}\ and\ \bibinfo {author} {\bibfnamefont {J.}~\bibnamefont
  {Noller}},\ }\href {\doibase 10.1103/PhysRevD.101.021502} {\bibfield
  {journal} {\bibinfo  {journal} {Phys. Rev. D}\ }\textbf {\bibinfo {volume}
  {101}},\ \bibinfo {pages} {021502} (\bibinfo {year} {2020})},\ \bibinfo
  {note} {[Erratum: Phys.Rev.D 102, 049902 (2020)]},\ \Eprint
  {http://arxiv.org/abs/1904.05874} {arXiv:1904.05874 [astro-ph.CO]}
  \BibitemShut {NoStop}%
\bibitem [{\citenamefont {Ye}\ and\ \citenamefont {Piao}(2020)}]{Ye:2019oxx}%
  \BibitemOpen
  \bibfield  {author} {\bibinfo {author} {\bibfnamefont {G.}~\bibnamefont
  {Ye}}\ and\ \bibinfo {author} {\bibfnamefont {Y.-S.}\ \bibnamefont {Piao}},\
  }\href {\doibase 10.1140/epjc/s10052-020-7973-z} {\bibfield  {journal}
  {\bibinfo  {journal} {Eur. Phys. J. C}\ }\textbf {\bibinfo {volume} {80}},\
  \bibinfo {pages} {421} (\bibinfo {year} {2020})},\ \Eprint
  {http://arxiv.org/abs/1908.08644} {arXiv:1908.08644 [hep-th]} \BibitemShut
  {NoStop}%
\bibitem [{\citenamefont {Kennedy}\ and\ \citenamefont
  {Lombriser}(2020)}]{Kennedy:2020ehn}%
  \BibitemOpen
  \bibfield  {author} {\bibinfo {author} {\bibfnamefont {J.}~\bibnamefont
  {Kennedy}}\ and\ \bibinfo {author} {\bibfnamefont {L.}~\bibnamefont
  {Lombriser}},\ }\href {\doibase 10.1103/PhysRevD.102.044062} {\bibfield
  {journal} {\bibinfo  {journal} {Phys. Rev. D}\ }\textbf {\bibinfo {volume}
  {102}},\ \bibinfo {pages} {044062} (\bibinfo {year} {2020})},\ \Eprint
  {http://arxiv.org/abs/2003.05318} {arXiv:2003.05318 [gr-qc]} \BibitemShut
  {NoStop}%
\bibitem [{\citenamefont {Tokuda}\ \emph {et~al.}(2020)\citenamefont {Tokuda},
  \citenamefont {Aoki},\ and\ \citenamefont {Hirano}}]{Tokuda:2020mlf}%
  \BibitemOpen
  \bibfield  {author} {\bibinfo {author} {\bibfnamefont {J.}~\bibnamefont
  {Tokuda}}, \bibinfo {author} {\bibfnamefont {K.}~\bibnamefont {Aoki}}, \ and\
  \bibinfo {author} {\bibfnamefont {S.}~\bibnamefont {Hirano}},\ }\href
  {\doibase 10.1007/JHEP11(2020)054} {\bibfield  {journal} {\bibinfo  {journal}
  {JHEP}\ }\textbf {\bibinfo {volume} {11}},\ \bibinfo {pages} {054} (\bibinfo
  {year} {2020})},\ \Eprint {http://arxiv.org/abs/2007.15009} {arXiv:2007.15009
  [hep-th]} \BibitemShut {NoStop}%
\bibitem [{\citenamefont {de~Rham}\ \emph {et~al.}(2021)\citenamefont
  {de~Rham}, \citenamefont {Melville},\ and\ \citenamefont
  {Noller}}]{deRham:2021fpu}%
  \BibitemOpen
  \bibfield  {author} {\bibinfo {author} {\bibfnamefont {C.}~\bibnamefont
  {de~Rham}}, \bibinfo {author} {\bibfnamefont {S.}~\bibnamefont {Melville}}, \
  and\ \bibinfo {author} {\bibfnamefont {J.}~\bibnamefont {Noller}},\ }\href
  {\doibase 10.1088/1475-7516/2021/08/018} {\bibfield  {journal} {\bibinfo
  {journal} {JCAP}\ }\textbf {\bibinfo {volume} {08}},\ \bibinfo {pages} {018}
  (\bibinfo {year} {2021})},\ \Eprint {http://arxiv.org/abs/2103.06855}
  {arXiv:2103.06855 [astro-ph.CO]} \BibitemShut {NoStop}%
\bibitem [{\citenamefont {Melville}\ and\ \citenamefont
  {Noller}(2022)}]{Melville:2022ykg}%
  \BibitemOpen
  \bibfield  {author} {\bibinfo {author} {\bibfnamefont {S.}~\bibnamefont
  {Melville}}\ and\ \bibinfo {author} {\bibfnamefont {J.}~\bibnamefont
  {Noller}},\ }\href {\doibase 10.1088/1475-7516/2022/06/031} {\bibfield
  {journal} {\bibinfo  {journal} {JCAP}\ }\textbf {\bibinfo {volume} {06}},\
  \bibinfo {pages} {031} (\bibinfo {year} {2022})},\ \Eprint
  {http://arxiv.org/abs/2202.01222} {arXiv:2202.01222 [hep-th]} \BibitemShut
  {NoStop}%
\bibitem [{\citenamefont {Xu}\ and\ \citenamefont {Zhou}(2023)}]{Xu:2023lpq}%
  \BibitemOpen
  \bibfield  {author} {\bibinfo {author} {\bibfnamefont {H.}~\bibnamefont
  {Xu}}\ and\ \bibinfo {author} {\bibfnamefont {S.-Y.}\ \bibnamefont {Zhou}},\
  }\href {\doibase 10.1088/1475-7516/2023/11/076} {\bibfield  {journal}
  {\bibinfo  {journal} {JCAP}\ }\textbf {\bibinfo {volume} {11}},\ \bibinfo
  {pages} {076} (\bibinfo {year} {2023})},\ \Eprint
  {http://arxiv.org/abs/2306.06639} {arXiv:2306.06639 [hep-th]} \BibitemShut
  {NoStop}%
\bibitem [{\citenamefont {Arkani-Hamed}\ \emph {et~al.}(2020)\citenamefont
  {Arkani-Hamed}, \citenamefont {Baumann}, \citenamefont {Lee},\ and\
  \citenamefont {Pimentel}}]{Arkani-Hamed:2018kmz}%
  \BibitemOpen
  \bibfield  {author} {\bibinfo {author} {\bibfnamefont {N.}~\bibnamefont
  {Arkani-Hamed}}, \bibinfo {author} {\bibfnamefont {D.}~\bibnamefont
  {Baumann}}, \bibinfo {author} {\bibfnamefont {H.}~\bibnamefont {Lee}}, \ and\
  \bibinfo {author} {\bibfnamefont {G.~L.}\ \bibnamefont {Pimentel}},\ }\href
  {\doibase 10.1007/JHEP04(2020)105} {\bibfield  {journal} {\bibinfo  {journal}
  {JHEP}\ }\textbf {\bibinfo {volume} {04}},\ \bibinfo {pages} {105} (\bibinfo
  {year} {2020})},\ \Eprint {http://arxiv.org/abs/1811.00024} {arXiv:1811.00024
  [hep-th]} \BibitemShut {NoStop}%
\bibitem [{\citenamefont {Baumann}\ \emph {et~al.}(2020)\citenamefont
  {Baumann}, \citenamefont {Duaso~Pueyo}, \citenamefont {Joyce}, \citenamefont
  {Lee},\ and\ \citenamefont {Pimentel}}]{Baumann:2019oyu}%
  \BibitemOpen
  \bibfield  {author} {\bibinfo {author} {\bibfnamefont {D.}~\bibnamefont
  {Baumann}}, \bibinfo {author} {\bibfnamefont {C.}~\bibnamefont
  {Duaso~Pueyo}}, \bibinfo {author} {\bibfnamefont {A.}~\bibnamefont {Joyce}},
  \bibinfo {author} {\bibfnamefont {H.}~\bibnamefont {Lee}}, \ and\ \bibinfo
  {author} {\bibfnamefont {G.~L.}\ \bibnamefont {Pimentel}},\ }\href {\doibase
  10.1007/JHEP12(2020)204} {\bibfield  {journal} {\bibinfo  {journal} {JHEP}\
  }\textbf {\bibinfo {volume} {12}},\ \bibinfo {pages} {204} (\bibinfo {year}
  {2020})},\ \Eprint {http://arxiv.org/abs/1910.14051} {arXiv:1910.14051
  [hep-th]} \BibitemShut {NoStop}%
\bibitem [{Note11()}]{Note11}%
  \BibitemOpen
  \bibinfo {note} {Explicitly, they are given by: \begin {align} F_\nu ^j (1/z)
  = \protect \frac {1}{\protect \sqrt {2}} \Gamma \left ( j \pm i \nu \right )
  P_{i \nu - \protect \frac {1}{2}}^{\protect \frac {1}{2} - j} (z) \left (
  \protect \sqrt {z^2 - 1 } \right )^{ \protect \frac {1}{2} - j } \label
  {eqn:F_def} \end {align} and are specified by a half-integer index $j =
  \protect \frac {d-2}{2} ( n -1 )$ that depends on the number of spatial
  dimensions and the number of particles at each vertex. $j=1/2$ for \protect
  {$\delimiter "426830A \sigma ^4 \delimiter "526930B _s$} in
  $d=3$.}\BibitemShut {Stop}%
\bibitem [{\citenamefont {Lee}\ \emph {et~al.}(2016)\citenamefont {Lee},
  \citenamefont {Baumann},\ and\ \citenamefont {Pimentel}}]{Lee:2016vti}%
  \BibitemOpen
  \bibfield  {author} {\bibinfo {author} {\bibfnamefont {H.}~\bibnamefont
  {Lee}}, \bibinfo {author} {\bibfnamefont {D.}~\bibnamefont {Baumann}}, \ and\
  \bibinfo {author} {\bibfnamefont {G.~L.}\ \bibnamefont {Pimentel}},\ }\href
  {\doibase 10.1007/JHEP12(2016)040} {\bibfield  {journal} {\bibinfo  {journal}
  {JHEP}\ }\textbf {\bibinfo {volume} {12}},\ \bibinfo {pages} {040} (\bibinfo
  {year} {2016})},\ \Eprint {http://arxiv.org/abs/1607.03735} {arXiv:1607.03735
  [hep-th]} \BibitemShut {NoStop}%
\bibitem [{\citenamefont {Chen}\ \emph
  {et~al.}(2017{\natexlab{a}})\citenamefont {Chen}, \citenamefont {Wang},\ and\
  \citenamefont {Xianyu}}]{Chen:2016uwp}%
  \BibitemOpen
  \bibfield  {author} {\bibinfo {author} {\bibfnamefont {X.}~\bibnamefont
  {Chen}}, \bibinfo {author} {\bibfnamefont {Y.}~\bibnamefont {Wang}}, \ and\
  \bibinfo {author} {\bibfnamefont {Z.-Z.}\ \bibnamefont {Xianyu}},\ }\href
  {\doibase 10.1103/PhysRevLett.118.261302} {\bibfield  {journal} {\bibinfo
  {journal} {Phys. Rev. Lett.}\ }\textbf {\bibinfo {volume} {118}},\ \bibinfo
  {pages} {261302} (\bibinfo {year} {2017}{\natexlab{a}})},\ \Eprint
  {http://arxiv.org/abs/1610.06597} {arXiv:1610.06597 [hep-th]} \BibitemShut
  {NoStop}%
\bibitem [{\citenamefont {Chen}\ \emph
  {et~al.}(2017{\natexlab{b}})\citenamefont {Chen}, \citenamefont {Wang},\ and\
  \citenamefont {Xianyu}}]{Chen:2016hrz}%
  \BibitemOpen
  \bibfield  {author} {\bibinfo {author} {\bibfnamefont {X.}~\bibnamefont
  {Chen}}, \bibinfo {author} {\bibfnamefont {Y.}~\bibnamefont {Wang}}, \ and\
  \bibinfo {author} {\bibfnamefont {Z.-Z.}\ \bibnamefont {Xianyu}},\ }\href
  {\doibase 10.1007/JHEP04(2017)058} {\bibfield  {journal} {\bibinfo  {journal}
  {JHEP}\ }\textbf {\bibinfo {volume} {04}},\ \bibinfo {pages} {058} (\bibinfo
  {year} {2017}{\natexlab{b}})},\ \Eprint {http://arxiv.org/abs/1612.08122}
  {arXiv:1612.08122 [hep-th]} \BibitemShut {NoStop}%
\bibitem [{\citenamefont {Wang}\ and\ \citenamefont
  {Xianyu}(2020)}]{Wang:2019gbi}%
  \BibitemOpen
  \bibfield  {author} {\bibinfo {author} {\bibfnamefont {L.-T.}\ \bibnamefont
  {Wang}}\ and\ \bibinfo {author} {\bibfnamefont {Z.-Z.}\ \bibnamefont
  {Xianyu}},\ }\href {\doibase 10.1007/JHEP02(2020)044} {\bibfield  {journal}
  {\bibinfo  {journal} {JHEP}\ }\textbf {\bibinfo {volume} {02}},\ \bibinfo
  {pages} {044} (\bibinfo {year} {2020})},\ \Eprint
  {http://arxiv.org/abs/1910.12876} {arXiv:1910.12876 [hep-ph]} \BibitemShut
  {NoStop}%
\bibitem [{\citenamefont {Kumar}\ and\ \citenamefont
  {Sundrum}(2020)}]{Kumar:2019ebj}%
  \BibitemOpen
  \bibfield  {author} {\bibinfo {author} {\bibfnamefont {S.}~\bibnamefont
  {Kumar}}\ and\ \bibinfo {author} {\bibfnamefont {R.}~\bibnamefont
  {Sundrum}},\ }\href {\doibase 10.1007/JHEP04(2020)077} {\bibfield  {journal}
  {\bibinfo  {journal} {JHEP}\ }\textbf {\bibinfo {volume} {04}},\ \bibinfo
  {pages} {077} (\bibinfo {year} {2020})},\ \Eprint
  {http://arxiv.org/abs/1908.11378} {arXiv:1908.11378 [hep-ph]} \BibitemShut
  {NoStop}%
\bibitem [{\citenamefont {Bodas}\ \emph {et~al.}(2021)\citenamefont {Bodas},
  \citenamefont {Kumar},\ and\ \citenamefont {Sundrum}}]{Bodas:2020yho}%
  \BibitemOpen
  \bibfield  {author} {\bibinfo {author} {\bibfnamefont {A.}~\bibnamefont
  {Bodas}}, \bibinfo {author} {\bibfnamefont {S.}~\bibnamefont {Kumar}}, \ and\
  \bibinfo {author} {\bibfnamefont {R.}~\bibnamefont {Sundrum}},\ }\href
  {\doibase 10.1007/JHEP02(2021)079} {\bibfield  {journal} {\bibinfo  {journal}
  {JHEP}\ }\textbf {\bibinfo {volume} {02}},\ \bibinfo {pages} {079} (\bibinfo
  {year} {2021})},\ \Eprint {http://arxiv.org/abs/2010.04727} {arXiv:2010.04727
  [hep-ph]} \BibitemShut {NoStop}%
\bibitem [{\citenamefont {Pinol}\ \emph {et~al.}(2023)\citenamefont {Pinol},
  \citenamefont {Aoki}, \citenamefont {Renaux-Petel},\ and\ \citenamefont
  {Yamaguchi}}]{Pinol:2021aun}%
  \BibitemOpen
  \bibfield  {author} {\bibinfo {author} {\bibfnamefont {L.}~\bibnamefont
  {Pinol}}, \bibinfo {author} {\bibfnamefont {S.}~\bibnamefont {Aoki}},
  \bibinfo {author} {\bibfnamefont {S.}~\bibnamefont {Renaux-Petel}}, \ and\
  \bibinfo {author} {\bibfnamefont {M.}~\bibnamefont {Yamaguchi}},\ }\href
  {\doibase 10.1103/PhysRevD.107.L021301} {\bibfield  {journal} {\bibinfo
  {journal} {Phys. Rev. D}\ }\textbf {\bibinfo {volume} {107}},\ \bibinfo
  {pages} {L021301} (\bibinfo {year} {2023})},\ \Eprint
  {http://arxiv.org/abs/2112.05710} {arXiv:2112.05710 [hep-th]} \BibitemShut
  {NoStop}%
\bibitem [{\citenamefont {Tong}\ and\ \citenamefont
  {Xianyu}(2022)}]{Tong:2022cdz}%
  \BibitemOpen
  \bibfield  {author} {\bibinfo {author} {\bibfnamefont {X.}~\bibnamefont
  {Tong}}\ and\ \bibinfo {author} {\bibfnamefont {Z.-Z.}\ \bibnamefont
  {Xianyu}},\ }\href {\doibase 10.1007/JHEP10(2022)194} {\bibfield  {journal}
  {\bibinfo  {journal} {JHEP}\ }\textbf {\bibinfo {volume} {10}},\ \bibinfo
  {pages} {194} (\bibinfo {year} {2022})},\ \Eprint
  {http://arxiv.org/abs/2203.06349} {arXiv:2203.06349 [hep-ph]} \BibitemShut
  {NoStop}%
\bibitem [{\citenamefont {Melville}\ and\ \citenamefont
  {Pimentel}(2024{\natexlab{b}})}]{Melville:2023kgd}%
  \BibitemOpen
  \bibfield  {author} {\bibinfo {author} {\bibfnamefont {S.}~\bibnamefont
  {Melville}}\ and\ \bibinfo {author} {\bibfnamefont {G.~L.}\ \bibnamefont
  {Pimentel}},\ }\href {\doibase 10.1103/PhysRevD.110.103530} {\bibfield
  {journal} {\bibinfo  {journal} {Phys. Rev. D}\ }\textbf {\bibinfo {volume}
  {110}},\ \bibinfo {pages} {103530} (\bibinfo {year} {2024}{\natexlab{b}})},\
  \Eprint {http://arxiv.org/abs/2309.07092} {arXiv:2309.07092 [hep-th]}
  \BibitemShut {NoStop}%
\end{thebibliography}%

\end{document}